\pdfoutput=1
\documentclass[12pt]{article}
\usepackage{jheppub}
\usepackage{latexsym}

\usepackage{hyperref}
\usepackage{amssymb}
\usepackage{amsmath}
\usepackage{bbm}
 \usepackage{epsfig}
 \usepackage{graphicx}

\newcommand{\be}{\begin{equation}}
\newcommand{\ee}{\end{equation}}
\newcommand{\ba}{\begin{eqnarray}}
\newcommand{\ea}{\end{eqnarray}}
\newcommand{\baa}{\begin{array}}
\newcommand{\eaa}{\end{array}}
\newcommand{\bi}{\begin{itemize}}
\newcommand{\ei}{\end{itemize}}
\newcommand{\edoc}{\end{document}}

\newcommand{\nn}{\nonumber \\}
\newcommand{\nr}[1]{(\ref{#1})}
\newcommand{\la}[1]{\label{#1}}

\newcommand{\rmi}[1]{{\mbox{\scriptsize #1}}}
\newcommand{\fr}[2]{{\frac{#1}{#2}\,}}
\newcommand{\fra}[2]{{\textstyle{\frac{#1}{#2}\,}}}

\newcommand{\hn}{\hat n}
\newcommand{\tn}{\tilde n}
\newcommand{\nt}{\tilde n}
\newcommand{\ws}{w}
\newcommand{\Azs}{\Phi}
\newcommand{\veff}{V_\rmi{eff}}
\newcommand{\lend}{\lambda_\rmi{end}}

\newcommand{\Ks}{K}
\def\ads#1{{AdS$_{#1}$}}
\def\sp{, \quad}
\def\l{\lambda}
\newcommand{\f}{{\kappa}} 

 \def\e{\epsilon}
 \def\p{\partial}

\def\tr{{\rm tr}}

\def\CL{{\cal L}}

\def\CO{{\cal O}}
\def\LUV{{\CL_\rmi{UV}}}
\newcommand{\dd}{\mathrm{d}}

\title{A holographic model for QCD in the Veneziano limit at finite temperature and density}
\author[a,b]{T. Alho}
\author[c]{M.J\"arvinen}
\author[b]{K. Kajantie}
\author[c,d,e]{E. Kiritsis}
\author[c]{C. Rosen}
\author[f,b]{K. Tuominen}
\affiliation[a]{Department of Physics, P.O.Box 35, FI-40014 University of Jyv\"askyl\"a, Finland}
\affiliation[b]{Helsinki Institute of Physics, P.O.Box 64, FI-00014 University of Helsinki, Finland}
\affiliation[c]{\href{http://hep.physics.uoc.gr}{Crete Center for Theoretical Physics},
Department of Physics, University of Crete, 71003 Heraklion, Greece.}
\affiliation[d]{\href{http://www.apc.univ-paris7.fr/APC_CS/}{APC},  Univ Paris Diderot, Sorbonne Paris Cit\'e,
UMR 7164 CNRS, F-75205 Paris, France.}

\affiliation[e]{\href{http://wwwth.cern.ch/}{Theory Group, Physics Department, CERN}, CH-1211, Geneva 23, Switzerland}

\affiliation[f]{Department of Physics, P.O.Box 64, FI-00014 University of Helsinki, Finland}
\emailAdd{timo.s.alho@jyu.fi}
\emailAdd{mjarvine@physics.uoc.gr}
\emailAdd{keijo.kajantie@helsinki.fi}
\emailAdd{hep.physics.uoc.gr/$\sim$kiritsis/ }
\emailAdd{rosen@physics.uoc.gr}
\emailAdd{kimmo.i.tuominen@helsinki.fi}

\preprint{\\ \hfill CCTP-2013-19\\ \hfill CCQCN-2013-6 \\\hfill HIP-2013-20/TH \\ \hfill CERN-PH-TH//2013-320}

\abstract{A holographic model
of QCD in the limit of
large number of colors, $N_c$,  and massless fermion flavors, $N_f$, but constant ratio
$x_f=N_f/N_c$ is analyzed at finite temperature and chemical potential.
The five dimensional gravity model contains three bulk fields:
a scalar dilaton sourcing ${\rm Tr}F^2$, a scalar tachyon dual to $\bar qq$ and a 
4-vector dual to the baryon current $\bar q \gamma^{\mu} q$.
The main result is the  $\mu,T$ phase diagram of the holographic theory.
 A first order
deconfining transition along $T_h(\mu)$ and a chiral transition at $T_\chi(\mu)>T_h(\mu)$ are found.
The chiral transition is of second order for all $\mu$.
The dependence of thermodynamical quantities including the speed of sound and
susceptibilities on the chemical potential and temperature is computed.
A new quantum critical regime is found at zero temperature and finite chemical potential. 
It is controlled by an AdS$_2\times \mathbbm{R}^3$ geometry and displays semi-local criticality.

}

\keywords{Gauge/ gravity duality, holography, QCD phase diagrams}

\begin{document}
\maketitle

\section{Introduction}

The phase diagram of QCD, as a function of temperature $T$ and chemical potential $\mu$,
corresponding to baryon density or some other conserved charge like isospin,
displays a rich structure \cite{Kogut:2004su}. 
Particularly interesting and important features of the phase diagram  are the nature of the chiral phase transition, the location of the chiral critical point and its properties. All these have been extensively studied both with effective chiral models
\cite{Scavenius:2000qd,Kahara:2008yg} and other approaches reviewed e.g. in \cite{Stephanov:2004wx}, and holography \cite{Myers:2009ij, DeWolfe:2010he,DeWolfe:2011ts}. Since first principle lattice methods \cite{Kaczmarek:2011zz, Karsch:2010hm, Endrodi:2011gv} are currently still limited to small values of $\mu/T$, the model studies provide important complement. However, the location of the critical point is very dependent on the details of the models \cite{Stephanov:2004wx}.

In addition to temperature and density, typically also external perturbations of the chiral
symmetry, i.e. finite quark masses, are present and provide further dimensions to the phase
diagrams. For example, in the limit of vanishing quark masses,  the finite temperature phase
transition of two flavor QCD at zero chemical potential is expected to be of second order.
At low temperatures,  the chiral transition at finite $\mu$ is expected to be of first order 
\cite{Scavenius:2000qd}.
At intermediate temperatures and densities the first and second order transition lines were conjectured to 
meet at a tricritical point. The finite quark mass softens the singularity at the
second order phase transition which becomes a smooth crossover. The line of first
order transitions is unaffected in the presence of small external perturbation,
and the tricritical point becomes the critical endpoint for this line. The fate of the existence of
the critical point of QCD in the $(\mu, T)$-plane at the physical value of quark masses is ultimately determined by the form of the critical manifold in the  multidimensional space of parameters $\mu,T,m_q$ \cite{deForcrand:2008zi}. One can imagine several possibilities to occur.  Indeed, it can be that the existence of a critical point near the chiral limit implies that the critical point does not exist at physical masses. Depending on the shape of the critical manifold, a variety of other possibilities can be imagined.

Effective field theories utilizing holographic methods, motivated by the AdS/CFT correspondence \cite{Maldacena:1997re,Witten:1998qj,Gubser:1998bc}, have become a major tool in the analysis of strongly coupled theories both in elementary particle and condensed matter physics
\cite{Hartnoll:2009sz,Herzog:2009xv}.
A class of bottom up models for QCD-like theories, which captures the entire renormalization group evolution of the corresponding quantum field theory from weak to strong coupling has been developed in \cite{Gursoy:2007cb,Gursoy:2007er,Gursoy:2010fj}. 
A particular application of this framework is the determination of the vacuum and finite temperature phase diagrams of the associated quantum gauge theories \cite{Gursoy:2007er,Gursoy:2008za,Gursoy:2008bu,Gursoy:2009jd,Jarvinen:2009fe,Alanen:2009na,Alanen:2010tg,Alanen:2011hh}.
The framework has been extended to account for the dynamics of chiral symmetry breaking in the presence of flavors \cite{Bigazzi:2005md,Casero:2007ae,Iatrakis:2010zf,Iatrakis:2010jb,Iatrakis:2011ht,Arean:2013tja}. 
In order to consider effects coming from the backreaction of flavor to color, holographic models with dynamics close to that of QCD in the 
Veneziano limit were explored and developed
\cite{Jarvinen:2011qe,Alho:2012mh,Arean:2012mq,Arean:2013tja}. In this work we consider adding finite chemical potential\footnote{For a different effort in that direction, see \cite{Stoffers:2010sp}.}
in order to determine the phase diagram in the $(T,\mu)$-plane by computing the pressure
$p(T,\mu;m_q=0)$ in the phases where chiral symmetry is intact or spontaneously broken.

Concretely, we consider the holographic model for equilibrium QCD with $N_f$
massless quarks at the limit  $N_f\to\infty$, $N_c\to\infty$  and fixed ratio $x_f=N_f/N_c$. 
For a thorough discussion of the fundamentals of this type of bottom-up holographic model
for QCD in the Veneziano limit (V-QCD)
at zero or finite $T$ but zero density,
we refer to \cite{Jarvinen:2011qe, Alho:2012mh}. Here we only outline the features arising when
we allow also finite density and chemical potential in V-QCD.
According to the holographic dictums to add baryon density we must turn on a source for the five-dimensional gauge field 
$A_a$.
The dynamics of the baryon number gauge field $A_{a}$ is determined by its appearance in the tachyon DBI action, which can be schematically written as
\be
\sqrt{-\det{(g_{ab}+\f\, \p_a\tau\p_b\tau+\ws\, F_{ab})}}.
\ee
Here $\f$ and $\ws$ are couplings, $F_{ab}=\p_aA_b-\p_bA_a$, and $\tau$ is the tachyon, sourcing $\bar qq$. To turn on a uniform constant density, the Ansatz
$A_a=\Phi(z)\delta_{a 0}$ should be made, where
$z$ is the coordinate of the 5th dimension and
the only non-zero component of $F_{ab}$ is $F_{z0}=\p_z\Phi(z)$.
The action contains only the derivative of $\Azs$ and
the finite density arises as the integration constant $\nt$ of the equation of motion of the
cyclic configuration space
coordinate $\Azs$. 

The three bulk fields
$\l,\Azs,\tau$ correspond to the three arguments in $p(T,\mu;m_q)$, and we will consider only the 
case $m_q=0$ in this paper and denote the pressure simply by $p(T,\mu)$.
As in \cite{Alho:2012mh} we find that there are two types of $m_q=0$ solutions:
those with vanishing tachyon (chirally symmetric) and those with nonzero tachyon (breaking chiral symmetry spontaneously).
To determine the pressure, the strategy is therefore to find black hole solutions with one or two scalar hair (corresponding to the dilaton and tachyon scalars) and a non-trivial charge density. Such solutions, when they exist, compete also with finite temperature but zero charge solutions without a black hole. The reason is that these zero charge solutions always have a constant $\Phi=\mu$ and therefore correspond to saddle points with finite chemical potential but zero charge density. Such solutions are expected to dominate at small enough temperature and chemical potential, and we identify them with the ``hadron gas" vacuum phase with zero pressure. 
Increasing the charge density, we have the possibility of a trivial or non-trivial tachyon field. 
The latter possibility describes a ``deconfined" but chirality breaking plasma, while the former corresponds to chirally symmetric plasma. 
To determine which of these two dominates, one solves numerically for the coupled equations of motion of the fields, and 
finds pressures $p_s(T,\mu)$ and $p_b(T,\mu)$
corresponding, respectively, to the solutions with intact or spontaneously broken chiral symmetry.
Equality of pressures, temperatures and chemical potentials then defines the phase boundary
on the $T,\mu$ plane. 

\begin{figure}[!htb]

\centering

\includegraphics[width=0.6\textwidth]{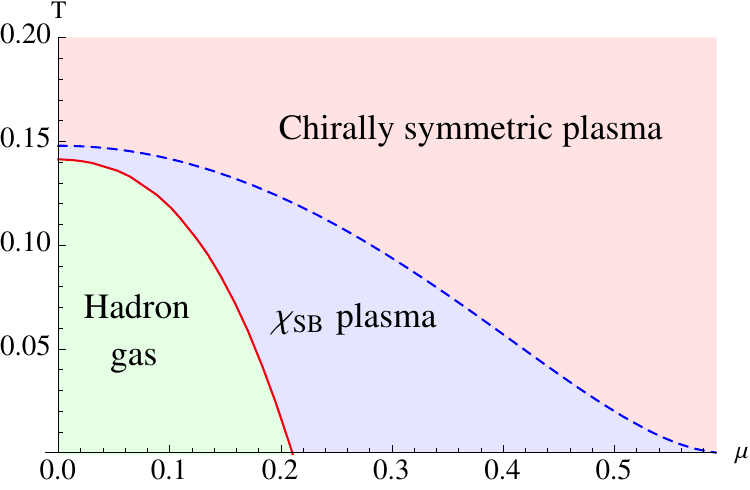}

\caption{\small Chemical potential dependence of transition temperatures of the
deconfining ($T_h(\mu)$) and chiral ($T_\chi(\mu)$) transitions at $m_q=0$. The dashed line
corresponds to a second order phase transition while the solid lines corresponds to a first order transition.
If finite quark mass is turned on, the second order transitions become smooth crossovers. The $T=0$ lines in the $\chi_{SB}$ plasma phase as well as the chirally symmetric phase correspond to a new quantum critical semilocal phase at finite density.
}
\label{Ttrans}
\end{figure}

The main outcome of this work is the phase diagram
shown in Fig.~\ref{Ttrans} which was obtained for the theory with $x_f={N_f\over N_f}=1$, 
namely for the same number of massless flavors and colors.
The chiral transition is of second order for all $\mu$, with the transition line ending at zero temperature, $\mu \approx 0.6$. For larger $\mu$, the system is always in the chirally symmetric deconfined phase.
There is also a tentative deconfining transition at $T_h(\mu)$ between the chirality breaking plasma and the ``hadron gas" phase discussed above. 
This phase boundary is  
determined by the condition $p_b(T,\mu)=p_\rmi{low}=0$.
\footnote{It is well known that in the presence of flavor there is no order parameter for deconfinement: confined phases can be continuously connected to Coulomb and Higgs phases.
However, at large $N_c$ the pressure itself can be considered as an order parameter for deconfinement.
The confined phase has $p\sim {\cal O}(1)$, while deconfined phase has $p\sim {\cal O}(N_c^2)$. 
When we talk about confined and deconfined phases we have this definition in mind.}

To motivate this in the field theory, note that the degrees of freedom of the low
temperature phase are the Goldstone bosons of the spontaneously broken chiral symmetry,
and their number is $\sim N_f^2$. On the other hand, the number of degrees of freedom in
the high-temperature phase is $\sim 2N_c^2+\fra72 N_cN_f$. As we consider only the case $x_f=1$
we obtain $p_{\rmi{low}}/p_{\rmi{high}}\sim 2/11\sim 0$. The relative
weight of the low-temperature degrees of freedom grows with $x_f$, and ultimately at some $x_{c}\simeq 4$,
in terms of the free energy, they become indistinguishable from the high temperature ones.
This signifies the quantum phase transition from a
confining gauge theory to the one whose long-distance behavior at zero temperature is
governed by a nontrivial and stable infrared fixed point. We leave the study
of the finite temperature and density phases in the limit $x_f\rightarrow 4$ for a further investigation.

Except for the lack of a critical point, all these features of this phase diagram agree on the general expectations.
However, for the phase diagram of QCD at low temperatures there is a surprise: There exists a new quantum critical regime at $T=0$, 
with exotic properties which realize the symmetries of the associated geometry, that is 
AdS$_2 \times {\mathbbm{R}}^3$. 
The presence of the AdS$_2 \times {\mathbbm{R}}^3$ geometry in the holographic solution 
indicates that there is a scaling symmetry of the time direction which does not act in the spatial directions. Such symmetries have been called semilocal.
While this is an unexpected symmetry in a field theory at finite density, it is natural and generic 
in the holographic context \cite{Gouteraux:2012yr}, and appears even in simple black holes as the 
Reissner-Nordstr\"om black hole \cite{Liu:2009dm}.
This new scaling region exists on the $T=0$  segment of the chirality breaking plasma as well as on 
the $T=0$ line of the chirally symmetric plasma. The physics in this critical regime is similar to that of a theory with zero speed of light: all spatial points decouple in the IR.

It is well known that such AdS$_2$ solutions are highly unstable as AdS$_2$ has a rather restrictive 
Breitenlohner-Freedman bound. The instabilities associated to the fields we consider can be 
understood in terms of the physics of the phase diagram and we describe them in detail in section \ref{stab}.
However, there can be further instabilities associated with other operators which we have 
not included here. It is possible 
that such quantum critical points play an important role in the appearance of color 
superconductivity and color flavor locking at high density.

There are many technical obstacles one has to cross before obtaining the final numerical results for the phase diagram:
First, to find the relevant charged black-hole solutions one has to guarantee that the metric function $f(z)$ vanishes
at the horizon $z=z_h$. As the horizon is a singular point of the equations, the numerical evolution must start close to the horizon with the appropriate boundary conditions.
 Second, the UV quark mass will be fixed to zero in order to have exact chiral symmetry.
To implement this, we must solve the entire coupled set of equations of motion and tune the boundary conditions
 so that
the leading term of the tachyon field at small $z$ (near the boundary) is $\sim z^3$, instead of a linear one
corresponding to a finite quark mass. This requires high numerical precision in the solution of the non-linear equations of motion.
The third difficulty is that the quantity to be computed  is a function of two variables, $p(T,\mu)$. The numerics is
correspondingly parametrised by two parameters, the value of the dilaton at the horizon
$\l_h$ and the integration constant $\nt$. These parameters cannot be continuous ones, but one can
determine, say, $T(\l_h,\nt)$ as a function of $\l_h$ for fixed values of $\nt$, and vice versa.
Proceeding in this way one obtains $p(T,\mu)$ on two grids on the $T,\mu$
plane (see Fig.~\ref{gridfig}).
Fourth and final issue is that one has to guarantee, using
scaling properties of the equations of motion, that all the
physically dimensionful quantities are expressed in the same units.

All of these considerations make the numerical problem at hand challenging. In this paper we focus on
the details of introducing the chemical potential to the model, limit ourselves to one set of potentials
chosen from \cite{Alho:2012mh} and to one value $x_f=1$. This allows us to show that the method works,
produces interesting results and motivates further studies. We have released the numerical code which has been used to compute the results presented in this paper \cite{TSAcode}.

In Section \ref{act} we specify the model and give the equations of motion
and their scaling properties. In Section \ref{constScalars} we find all solutions with constant scalars as they are 
critical end points of flows. They correspond to AdS$_5$ and AdS$_2$ geometries and we analyze their RG stability.
In Section~\ref{num} we discuss the horizon expansion
required for initialising numerical solution and the physical values of the parameters
$\l_h,\,\nt$ of numerical integration. The main numerical results for the pressures,
the transition temperatures $T_h(\mu)$, $T_\chi(\mu)$ and sound velocity
are shown and discussed in Section~\ref{sectp}. The Appendices contain a detailed
discussion of the numerical solutions, examples of the computed values of $T$ and
$\mu$ and a detailed presentation of the chiral phase transition line on the
plane of numerical parameters $\l_h,\,\nt$.

\section{Action and the equations of motion}\la{act}

\subsection{Definition of the action}
The action of the model for vanishing chemical potential has been discussed thoroughly in
\cite{Jarvinen:2011qe,Alho:2012mh,Arean:2012mq}. We focus here on the additional terms needed to describe the finite
baryon density. The action of the model is, in standard notation \cite{Alho:2012mh},
\be
S={1\over16\pi G_5}\int d^5x\,L,
\la{S}
\ee
where the Lagrangian is
\ba
&&\sqrt{-g}\biggl[ R+\left[-\fra43g^{\mu\nu}\partial_\mu\phi\partial_\nu\phi+V_g(\lambda)\right]
-\, V_f(\lambda,\tau)
\sqrt{-\det\left[ g_{ab}+\f(\l)\partial_a\tau\partial_b\tau+\ws(\l) 
F_{ab}\right]}\biggr]
\nonumber\\
&=&
b^5\biggl[-{f\over b^2}\left(8{\ddot b\over b}+4{\dot b^2\over b^2}
+8{\dot b\over b}{\dot f\over f}+{\ddot f\over f}+\fr43\dot\phi^2\right)+V_g(\lambda)
-\, V_f(\lambda,\tau)
\sqrt{1+{f\f\over b^2} \dot\tau^2-{\ws^2\over b^4}
\dot \Azs^2}\biggr].
\label{lagrang}
\ea
The metric Ansatz is
\be
\dd s^2=b^2(z)\left[-f(z)\dd t^2+\dd {\bf x}^2+{\dd z^2\over f(z)}\right],
\quad b(z)=e^{A(z)} \underset{z\to0}{\longrightarrow}{\CL_\rmi{UV}\over z},\quad f(0)=1.
\la{bg}
\ee
The functions $b$ and $f$ of the metric, the dilaton $\l=e^\phi$, the tachyon $\tau$ and the bulk
density $\Azs$ depend only on the extra dimensional coordinate $z$.
The Gibbons-Hawking counterterm
is implied.

The Lagrangian \nr{lagrang} is parametrized in terms of the potentials $V_g(\lambda)$,
$V_f(\lambda,\tau)$, $\f (\lambda)$ and $w(\lambda)$ which are chosen to satisfy two 
basic requirements. First,
in the ultraviolet, i.e. in the weak coupling limit, the model should reproduce the known
perturbative behaviors of the corresponding field theory. Second, in the deep infrared the
model should lead to the generation of a dynamical wall shielding the singular behavior as
$\lambda\rightarrow \infty$, which is responsible for confinement in the absence of the tachyon.

In the numerical study in this article we will take
the gauge coupling constant function
$\ws$ in the DBI action to be proportional to the other model function $\f$ as\footnote{This choice is at the boundary of allowed choices as indicated in \cite{Arean:2012mq}.}
\be
\ws(\l) 
=\CL_A^2\f(\l).
\la{defellA}
\ee
Here the scale $\CL_A\sim \CL_\rmi{UV}$ appears in order to match the dimensions correctly.
It can be formally eliminated from the Lagrangian~\eqref{lagrang} by rescaling $\Phi$.
In Appendix~\ref{appD} we shall find that $\CL_A\approx \CL_\rmi{UV}(x_f=0)=1$.
Note that if one expands the Lagrangian \nr{lagrang} in $F_{ab}$ and writes it in the form
$-\fra1{4e^2}F^2$, one can identify a dimensionless coupling
\be
\gamma^2=e^2\LUV^2={\LUV^2\over V_f\ws^2}={\LUV^2\over V_f\CL_A^4\f^2}.
\la{defgam}
\ee

Explicitly, the potentials are \cite{Alho:2012mh}
\be
V_g(\lambda)={12\over\CL_0^2}\biggl[1+{88\lambda\over27}+{4619\lambda^2
\over 729}{\sqrt{1+\ln(1+\lambda)}\over(1+\lambda)^{2/3}}\biggr],
\quad
V_f (\lambda, \tau) = x_f V_{f0} (\lambda) e^{-\fra32 \tau^2/\CL_\rmi{UV}^2},
\label{VfSB}
\ee
where the function $V_{f0}(\lambda)$ is given by
\ba\label{Vf0SB}
\hspace{-1cm}V_{f0}&=&{12\over\CL_\rmi{UV}^2x_f}\biggl[{\CL_\rmi{UV}^2\over\CL_0^2}
-1+{8\over27}\biggl(11{\CL_\rmi{UV}^2\over\CL_0^2}-11+2x_f\biggr)\lambda\nn
&&\hspace{-1cm}+{1\over729}\biggl(4619{\CL_\rmi{UV}^2\over\CL_0^2}-4619+1714x_f-92x_f^2\biggr)\lambda^2\biggr]
\equiv W_0+W_1\l+W_2\l^2.
\ea
The scale $\CL_\rmi{UV}$ has a nontrivial dependence on $x_f$,
$\CL_\rmi{UV}=\CL_0(1+\fra74 x_f)^{1/3}$, which is determined by matching the pressure
to the Stefan-Boltzmann limit at $\mu=0$ \cite{Alho:2012mh}. The function $\f (\lambda)$ is given by
\be
 \kappa(\l)= {[1+\ln(1+\l)]^{\bar\mu}\over[1+\fra34(\fra{115-16x_f}{27}+\bar\mu)\l]^{4/3}}.
 \label{akappa}
\ee

The numerical factors appearing in \nr{VfSB} and \nr{Vf0SB} simply provide the equivalence with the
known perturbative behavior in the weak coupling limit. This matching is obtained via the definition
\be
\beta(\l)={d\l\over db/b},
\ee
and recalling that the 2-loop beta function for the coupling  $\l=N_cg^2(\mu)/(8\pi^2)$ of the boundary theory is
\be
\beta(\l)={d\l(\mu)\over d\ln\mu}=-b_0\l^2-b_1\l^3,\quad
\quad b_0=\fra13(11-2x_f),\quad b_1=\fra16(34-13x_f)
\ee
in the Veneziano limit. Analogously, the numerical factors appearing in $\f(\lambda)$ in Eq. \nr{akappa} are obtained by first defining
\be
\gamma(\l)={d\ln\tau\over d\ln b}+1,
\ee
and then relating to the quark mass anomalous dimension with the scheme independent
coefficient $\gamma_0$ defined by
\be
\gamma(\l)={d\ln m\over d\ln\mu}=-\gamma_0\l+\cdots,\quad \gamma_0=\fr32={9b_0\over2(11-2x_f)}.
\ee
The actual numerical value of the quark mass (and the condensate $\langle\bar{q}q\rangle$)
is fixed by the UV expansion of the tachyon (remembering that the energy dimension of $\tau$ is $-1$):
\be
\tau(z)/\CL_\rmi{UV}
=m_qz\,(-\ln\Lambda z)^{-\frac{\gamma_0}{b_0}}
+\langle\bar qq\rangle z^3\,(-\ln\Lambda z)^{\frac{\gamma_0}{b_0}}.
\la{tauUV}
\ee
To have exact chiral symmetry one must find solutions for which $m_q=0$, and achieving this,  is one of
the technically most demanding tasks of this model (for details, see Appendix \ref{scalings}).

The behavior of the potentials at large values of the fields $\l$ and $\tau$
is determined by requirements of a confining spectrum and breaking of the chiral symmetry in the
deep infrared \cite{Gursoy:2007er,Gursoy:2008za,Gursoy:2009jd,Jarvinen:2011qe}. To fix the last remaining parameter we choose $\bar\mu
=-\fra12$. 
This choice, according to \cite{Alho:2012mh}, leads to regular thermodynamics at zero chemical potential.

With these definitions, the numerical results in this paper are given for the potentials
\nr{VfSB}-\nr{akappa}
and for $x_f=1$ case only. Of course the above choice for the potentials and $\f$ is not
unique but other possibilities exist as discussed in \cite{Jarvinen:2011qe,Arean:2012mq}. The definitions
presented above are taken in this paper to provide for a benchmark study of this model,
and focused analyses of other potentials and other values of
$x_f$, in particular approaching the conformal region at $x_f\approx4$, are left for future studies.

As a final remark here, we emphasize that the duality between classical gravity and field theory can
be derived in the string theory framework only in the strong coupling limit. In our case, the matching to the
scheme  independent perturbative results in the weak coupling limit has to be regarded as a model
assumption, to be judged on the basis of its consequences.
Among these, an immediate and important one is that one can describe thermodynamics
up to arbitrarily high $T$ and $\mu$ and identify solid known behaviors.
Actually it is quite nontrivial that this matching can be carried out and the correct running
of the quark mass and the condensate implemented using the DBI action.
The model is then an effective theory extending weakly coupled results at large $T,\,\mu$
to the strongly coupled domain.

\subsection{$\Azs$ equation of motion}
The fermionic part of the action, given by
\be
L_f[\tau,\dot\tau,\dot \Azs]=
V_f(\l,\tau)b^5\sqrt{1+{f\f\over b^2} \dot\tau^2-{\ws^2\over b^4}
\dot \Azs^2}\,,
\ee
depends only on $\dot \Azs$ so that $\Azs$ is a cyclic coordinate.
Since both $L_f$ and $\dot \Azs$ have energy dimension 2, we have a dimensionless
constant of integration $\hat n$:
\be
{\partial L_f\over\partial\dot \Azs}={-
bV_f\ws^2\dot \Azs\over
\sqrt{1+{f\f\over b^2} \dot\tau^2-{\ws^2\over b^4}
\dot \Azs^2}}=\hat n.
\la{A0eom}
\ee
From this one solves 
\be
\dot \Azs=-{\hn b^2\over\ws}\sqrt{\biggl(1+{f\f\over b^2} \dot\tau^2\biggr)
{1 \over \hat n^2+(
b^3\ws V_f)^2}}\equiv
-{\hn \over b V_f \ws^2 }\sqrt{\biggl(1+{f\f\over b^2} \dot\tau^2\biggr){1 \over 1+\Ks}},
\la{A0value}
\ee
where we have also introduced the dimensionless density factor
\be
\Ks(z)={\hn^2\over
b^6\ws^2V_f^2}={\hn^2\over \CL_A^4
b^6\f^2V_f^2}. 
\la{defGz}
\ee
The factor $\Ks$ defined above contains the density effects in this holographic model and will appear
repeatedly in what follows.

After the bulk fields $\l$ and $\tau$ have been determined from their equations of motion and the
Einstein's equations,
$\Azs(z)$ can be computed by integrating Eq.~\nr{A0value}:
\be
\Azs(z)=\mu+\int_0^{z} dz \, \dot \Azs(z)
\la{A0int}
\ee
with the constraint that the field $\Azs$ vanishes at the horizon $z=z_h$,
\be
\Azs(z_h)=0=\mu+\int_0^{z_h} dz \, \dot \Azs(z),
\la{A0cond}
\ee
from which $\mu$ is determined.

\subsection{Equations of motion for other bulk fields}
Using the previous results for $\Azs$, differential equations for $b,\l,f,\tau$ can be derived.
They are for $b(z)$
\be
3{\ddot b\over b}+6{\dot b^2\over b^2}+3{\dot b\over b}{\dot f\over f}
-{b^2\over f}V_g+{b^2\over f}V_f\left(1+\fr12 \fr{f\f}{b^2}\dot\tau^2\right)\sqrt{{1+\Ks\over1+\fra{f\f}{b^2}\dot\tau^2}}
=0,
\la{eq1}
\ee
for $\l(z)$
\ba
&&{\ddot\l\over\l}-{\dot\l^2\over\l^2}+3{\dot b\over b}{\dot\l\over\l}+
{\dot f\over f}{\dot\l\over\l}+\fr38{b^2\over f}\l{\p V_g\over\p\l}\nn
&&-\fr38{1\over \sqrt{1+\Ks}}\biggl\{{V_f\over\sqrt{1+{f\f\over b^2}\dot\tau^2}}
\l\f'\biggl[\fra12\dot\tau^2(1-\Ks)-{b^2\over f\f}\Ks\biggr]
\biggr.
\nn&&\biggl.
\hspace{1cm}+{b^2\over f}
\sqrt{1+{f\f\over b^2}\dot\tau^2}\,\l\p_\l V_f\biggr\}=0,
\la{eq2}
\ea
for $f(z)$
\be
\ddot f+{3\dot b\over b}\dot f- {\hn^2\over b\ws}\sqrt{{1+\fra{f\f}{b^2}\dot\tau^2\over \hn^2+b^6\ws^2V_f^2}}=0,
\la{eq3}
\ee
and for $\tau(z)$
\ba
&&
\left(1+\Ks\right)
\ddot\tau
-\biggl({b^2\over f\f}+\dot\tau^2\biggr){\partial\ln V_f\over\partial\tau}
\nonumber\\&&
+{f\f\over 2b^2}\biggl[{d\ln b^8f \f\over dz}+2\dot\l{\partial\ln V_f\over\partial\l}+
\Ks 
\biggl({d\ln (b^{2}f/\f)\over dz}\biggr)\biggr]\dot\tau^3
\nonumber\\&&
+\biggl[{d\ln b^3f\f\over dz}+\dot\l{\partial\ln V_f\over\partial\l}+
\Ks 
\biggl({d\ln f\over dz}\biggr)\biggr]\dot\tau
=0.
\la{eq4}
\ea
For $\l$ we also have the first order equation
\be
12 \frac{\dot b^2}{b^2} + 3 \frac{\dot b }{b} \frac{\dot f}{f} - \frac{4}{3} \frac{\dot \l^2}{ \l^2} = 
\frac{b^2}{f}\left(V_g-V_f\sqrt{\frac{1+K}{1+\frac{f\kappa}{b^2}\dot \tau^2}}\right).
\label{1orderell}
\ee

It turns out useful to define the quantity
\be
\veff (\lambda,\tau)=V_g(\l)-V_f(\l,\tau)\sqrt{1+{\hn^2\over b^6\ws^2V_f^2}}.
\la{1stdefveff}
\ee
Using this in $\tau=0$ case the equations can be written in more compact form as follows: First we have from the above definition 
\be
V_\rmi{eff}(\l,\tau=0)=V_g(\l)-\sqrt{V_f^2(\l,0)+{\hn^2\over b^6w(\l)^2}}.
\la{defvef}
\ee
Treating this as a function of $\lambda$ and $b$, the three remaining equations of motion are then
\ba
&&{\ddot b\over b}-2{\dot b^2\over b^2}+\fr49{\dot\l^2\over \l^2}=0, \la{eq10}\\
&&{\ddot\l\over\l}-{\dot\l^2\over\l^2}+3{\dot b\over b}{\dot\l\over\l}+
{\dot f\over f}{\dot\l\over\l}+{3b^2\over 8f}\l\partial_\l V_\rmi{eff}=0, \la{eq20}\\
&&\ddot f+{3\dot b\over b}\dot f-\fr13b^3\partial_b V_\rmi{eff}=0. \la{eq30}
\ea

The energy unit of the solutions is determined by fixing the small-$z$ behavior of the
dilaton to the perturbatively known field theory behavior, i.e. $b_0\l(z)=-1/\ln(\Lambda_0 z)$ with
$\Lambda_0=1$. To do this accurately enough,
one has to go to extremely small values of $z$ and it is better to use $\ln(z)$ or
actually $\ln b$ as the coordinate.
For numerics we thus write the equations in the $A=\ln b$ basis changing $z$ to $A$ via the
relation
\be
q(A)=e^A{dz\over dA},
\quad 
e^{-A}{dA\over dz}={\dot b\over b^2}.
\la{defq}
\ee
Then we have for $q(A)$, primes denoting derivatives with respect to $A$,
\be
12-6\frac{q^\prime}{q}+\frac{4}{3}\frac{\lambda^{\prime 2}}{\lambda^2}+3\frac{f^\prime}{f}
= \frac{q^2}{f}\left(V_g-V_f\sqrt{1+{f\f\tau^{\prime 2}\over q^2}}
\sqrt{1+\Ks}\right).
\label{A1}
\ee
The remaining equations of motion are for $\l(A)$
\ba
&&{\l''\over\l}-{\l'^2\over\l^2}+\biggl(4-{q'\over q}\biggr){\l'\over\l}+
{f'\over f}{\l'\over\l}+\fr38{q^2\over f}\l{\p V_g\over\p\l}\la{A2}\\
&&-\fr38{\l\over \sqrt{1+\Ks}}\biggl\{{1\over\sqrt{1+{f\f\over q^2}\tau^{\prime 2}}}
\f'(\l)V_f\,\biggl[\fra12\tau^{\prime 2}(1-\Ks)-{q^2\over f\f}\Ks\biggr]+
{q^2\over f}\sqrt{1+{f\f\over q^2}\tau^{\prime 2}}\,\p_\l V_f\biggr\}=0,\nonumber
\ea
for $f(A)$
\be
f^{\prime\prime}+(4-\frac{q^\prime}{q})f^\prime
= q^2{\hn^2\over\ws}e^{-3A}\sqrt{{1+f\f\tau^{\prime 2}/q^2\over
\hn^2+(e^{3A}\ws V_f)^2}}
=- q\hn e^{-4A}\Azs'\la{A3}
\ee
and for $\tau(A)$
\ba
&&\hspace{-1.cm}
\left(1+\Ks\right) 
\tau''
-\biggl({q^2\over f\f}+\tau^{\prime 2}\biggr){\partial\ln V_f\over\partial\tau}
\nonumber\\&&\hspace{-1.cm}
+{f\f\over 2q^2}\biggl[8+{d\ln f \f\over dA}+2\l'{\partial\ln V_f\over\partial\l}+
\Ks \biggl(2+{d\ln f/\f\over dA}\biggr)\biggr]\tau^{\prime 3}
\nonumber\\&&\hspace{-1.cm}
+\biggl[4-{q'\over q}+{d\ln f\f\over dA}+\l'{\partial\ln V_f\over\partial\l}+
\Ks \biggl(1-{q'\over q}+{f'\over f}\biggr)\biggr]\tau^\prime
=0.
\la{A4}
\ea
Here
\be
\Ks=\Ks(A) ={\hn^2\over e^{6A}V_f^2\ws^2}.
\la{GA}
\ee
This has the formally notable consequence that the $A$-equations are not
autonomous; there is explicit $A$ dependence.
The consequence of this will become explicit when we consider the scaling properties
of the solutions in the following section; see Eq.~\nr{Ascaling}.

The equation \nr{eq3} for $f$ can be integrated once:
\be
\dot f(z)={1\over b^3(z)}\biggl[C_1+\int_0^z
du
{\hn^2b^2\over \ws}\sqrt{{1+\fra{f\f}{b^2}\dot\tau^2\over \hn^2+b^6\ws^2V_f^2}}
\biggr]={1\over b^3}[C_1+\hn(\mu-\Azs(z))],
\ee
using \nr{A0value}.
Then $f(z)$ is obtained by one more integration, with integration constants determined
by $f(0)=1$, $f(z_h)=0$.
Actually we are
most interested in the charged black hole temperature, for which one obtains
\be
4\pi T=-\dot f(z_h)
={1-\hn\int_0^{z_h}du{\Azs(u)\over b^3(u)}\over
b^3(z_h)\int_0^{z_h}{du\over b^3(u)}}.
\la{chargedtemp}
\ee

\subsection{Scaling properties of equations of motion}\la{sectscalings}
Numerical solutions have to be transformed to the required standard form by using scaling properties
of the equations. A thorough discussion is given in Appendix \ref{scalings}, and
we summarize the main points in the following. The quantities which are not mentioned will remain
unchanged and all bulk fields are taken to be either functions of $z$
or $A$.

For the $z$ equations \eqref{eq1}--\eqref{eq4} one performs the following scalings:
\bi
\item
The boundary value of $f(z)$ must be set to 1 so that the boundary metric is pure
AdS, $f(0)=1$. This is achieved by scaling
\be
f\to {f\over f_0},\quad f_0\equiv f(0).
\ee
In order to keep $b^2/f$ and $\Ks(z)$ in \nr{defGz} invariant, this requires that further
\be
b\to {b\over\sqrt{f_0}},\quad
\hn\to {\hn \over f_0^{3/2}},
\quad \dot \Azs\to {1\over f_0}\dot \Azs.
\la{fscalz}
\ee
Note that also the integration constant $\hn$ is scaled.

\item
The unit of energy can be changed by $z\to\Lambda z$, together with
\be
b\to{b\over\Lambda},
\quad
\hn\to{\hn\over\Lambda^3},
\quad
\dot \Azs\to {1\over\Lambda^2}\dot \Azs,
\ee
which leave the equations of motion invariant.
\ei

For the $A$ equations \nr{A1}-\nr{A4} the corresponding scalings are:
\bi
\item
Scaling of $f$ to $f_0=f(\infty)=1$ requires that $q^2/f$ be constant, so that
\be
f\to {f\over f_0},\quad q\to {q\over\sqrt{f_0}}.
\la{fscalA}
\ee
Note that the density factor $\Ks(A)$ in \nr{GA} is not affected by this scaling.
\item
The scaling corresponding to $z\to\Lambda z$ is
\be
A\to A-\ln\Lambda, \quad b=e^A\to b{1\over\Lambda}.
\la{Ascaling}
\ee
The invariance of the density factor $\Ks(A)$ and Eq. \nr{A3} then demand that
\be
\hn\to{\hn\over\Lambda^3}
,\quad \Azs'\to{1\over\Lambda}\Azs'.
\la{Lascaling}
\ee
\ei

\section{Constant Scalar Solutions and IR Stability}\la{constScalars}

To gain intuition on what to expect at zero temperature and finite chemical potential 
we now consider some special solutions of the equations of motion derived in Sec.\,\ref{act}. 
We need to determine the fixed point solutions with translational symmetry since 
flows between different such solutions categorize the various RG flows 
of the boundary theory. In general the fixed point solutions with translational symmetry are 
 \ads{p} solutions either with fixed scalars or hyperscaling violating solutions when the scalars run off to infinity, \cite{Gouteraux:2012yr,Gouteraux:2011ce}.

We have not found hyperscaling violating asymptotics in this theory.
The other remaining scaling solutions must then have constant scalars.
These solutions will be  the non-linear generalization of \ads{5} Reissner-Nordstr\"om black hole (the so-called DBI black hole), and solutions with scaling \ads{2} regions in the IR, at extremality. 

To search for these, we turn to the equations of motion and make the following replacements:
\begin{equation}
\lambda(z)\to\lambda_0 ,\qquad \tau(z)\to\tau_0 \qquad \mathrm{and}\qquad  V_f,\,V_g,\,\kappa,\,w\to V_f^0, \, V_g^0,\,\kappa_0,\, w_0
\end{equation}
where a zero sub- or superscript indicates the constant value of the appropriate quantity in the fixed point and  
\begin{equation}
V^0_f\equiv V_f(\lambda,\tau)\Big|_{\lambda = \lambda_0,\,\tau=\tau_0}
\sp
\partial_{\lambda }V^0 \equiv \partial_{\lambda} V_f(\lambda,\tau) \Big|_{\lambda = \lambda_0,\,\tau=\tau_0},\,\,{\rm{etc.}}
\end{equation}
The two classes of solutions are distinguished by whether the scale factor $A$ is constant or not.
If it is constant we obtain \ads{2} 
type solutions while if it is non-trivial we obtain \ads{5} type solutions.
 
\subsection{\ads{5} and the DBI Black-Hole Solution}

For constant scalars many of the equations become quite simple, and often can be decoupled. For example, the equation governing the warp factor, $A=\log b$ is just
\begin{equation}\label{eq:Aeq}
{A}^{\prime\prime}(z)-{A^\prime(z)}^2=0,
\end{equation}
which has two independent  solutions,  $A(z)=-\log z$ or $A$ constant. The first  matches the \ads{5} result in these coordinates. This is the solution one anticipates as a UV fixed point in the dual theory. It will turn out to be the charged DBI black hole, which becomes the AdS$_5$ Reissner-Nordstr\"om solution in the limit of small gauge coupling. We can systematically insert this solution into the remaining equations of motion.

The Maxwell equation in this limit reads
\begin{equation}
{\Phi^\prime}(z)\equiv\mathcal{E}(z) = - \frac{\hat{n}\,z}{V_f^0 w_0^2\sqrt{1+\frac{\hat{n}^2 z^6}{V_f^0{}^2 w_0^2}}}
\end{equation}
and from this we obtain the behavior of the blackening function $f$ in the uniform scalar background. This function is described by the equation of motion
\begin{equation}\label{eq:heom}
f^{\prime\prime}(z)-\frac{3}{z}{f^\prime}(z)= \frac{\hat{n}^2}{V_f^0w_0^2\sqrt{1+\frac{\hat{n}^2 z^6}{V_f^0{}^2w_0^2}}}z^4.
\end{equation}
The solutions of the correponding homogeneous equation give the standard blackening 
for the AdS black hole in five dimensions. The general solution of the inhomogeneous equation 
therefore takes the form
\begin{equation}\label{eq:hsolcs}
f(z) = c_0-\frac{z^4}{z_0^4}+Q(z),
\end{equation}
familiar for charged black holes. Here $c_0$ and $z_0$ are integration constants. The function $Q$, 
which carries the information about the electric source for the black hole, can be computed by 
integrating (\ref{eq:heom}) twice. It turns out to be (see also Eq.~\nr{mures1})
\begin{equation}
Q(z) = -\frac{1}{12}V_f^0\sqrt{1+\frac{\hat{n}^2 z^6}{V_f^0{}^2w_0^2}}+\frac{1}{8}
\frac{\hat{n}^2}{V_f^0w_0^2}z^6\,{}_2 F_1(\frac{1}{3},\frac{1}{2},\frac{4}{3};-\frac{\hat{n}^2 z^6}{V_f^0{}^2w_0^2}),
\end{equation}
where ${}_2 F_1$ is the 
hypergeometric function. For consistency, the blackening function must be compatible with the 
constraint equation (\ref{1orderell}), given by
\begin{equation}
{f^\prime(z)}-\frac{4}{z}f(z)=\frac{1}{3z}\left(V_f^0\sqrt{1+\frac{\hat{n}^2 z^6}{V_f^0{}^2w_0^2}}-V_g^0\right).
\end{equation}
Note that this equation effectively governs the constant term in (\ref{eq:hsolcs}), or equivalently the near boundary value of the blackening function. To leading order in $\hat{n}$ the solution consistent with the above constraint is\footnote{Notice that when computing the full RG flow we have chosen to normalize $f$ to one in the UV. We have the freedom to do this if a constant term is also included in the UV solution for $A$.}
\begin{equation} \label{fads5}
f(z) = \frac{1}{12}\big( V_g^0-V_f^0\big)-\frac{z^4}{z_0^4}+\mathcal{O}(\hat{n}^2)
\end{equation}
which is exactly the form one would anticipate in \ads{5} with charged branes.

The remaining equations of motion, those for the dilaton and tachyon, contain algebraic constraints for various parameters of the theory. Specifically, the dilaton equation implies
\begin{equation}\label{eq:dilcon}
\lambda_0 = 0 \qquad \mathrm{or}\qquad \partial_\lambda V_g^0=\partial_\lambda V_f^0 = \partial_\lambda  w_0 =0
\end{equation}
while the tachyon equation needs either
\begin{equation}
\partial_\tau V_f^0 = \partial_\tau  w_0=0 \qquad \mathrm{or} \qquad V_f^0\kappa_0 = \infty
\end{equation}
in order to be satisfied. 
These constraints have a simple interpretation. The set of equations
\begin{equation}
\partial_\lambda V_g^0 = \partial_\lambda V_f^0 = \partial_\tau V_f^0 = 0
\end{equation}
are simply the requirement that all the potentials are extremized  at the appropriate value of $(\lambda_0,\tau_0)$. Evidently, the same must be true for the gauge kinetic function $w(\lambda,\tau)$.

For the V-QCD potentials of interest, specifically those from Section \ref{act}, it turns out that the extremization condition in \ref{eq:dilcon} can never be realized and the only possibility is the vanishing dilaton, $\lambda_0=0$. The gauge kinetic function $w$ and the flavor potential $V_f$ 
are of the general form of Eq. (\ref{defellA}) and Eq. (\ref{VfSB}), respectively:
\begin{equation}\label{eq:Vfnw}
V_f= x_fv_f(\lambda)\,e^{-a(\lambda)\tau^2} \qquad \mathrm{and}\qquad w = w(\lambda)
\end{equation}
so $\partial_\tau w = 0$ and the tachyon constraint reduces to
\begin{equation}\label{eq:Vfext}
0 = \partial_\tau V_f^0 \sim -2x_f \tau_0\, a(\lambda_0) v_f(\lambda_0) e^{-a(\lambda_0)\tau_0^2}.
\end{equation}
Therefore, the flavor potential is extremized in the $\tau$ direction for either $\tau_0=0$ or $\tau_0 = \infty$.
Moreover, it can be explicitly checked that when the dilaton is zero there is no location in the parameter space $(x_f,\tau_0)$ for which the combination $V_f^0\kappa_0$ diverges. Accordingly, one finds that in this V-QCD setup DBI black hole solutions exist at all $x_f$ so long as $\lambda_0=0$ and $\tau_0 = 0$ or $\infty$.

\subsection{\ads{2} Solution}
There exists another simple solution to the constant scalar warp factor equation of motion (\ref{eq:Aeq}). This is the constant solution $A=A_0$.  In this case, the Maxwell equation is satisfied by a constant electric field of the form
\begin{equation}\label{eq:AdS2E}
\mathcal{E} = -\frac{\hat{n}\,e^{-A_0}}{w_0^2 V_f^0\sqrt{1+e^{-6A_0}\frac{\hat{n}^2}{V_f^0{}^2w_0^2}}}
\end{equation}
giving a potential\footnote{Note that we have anticipated the fact that
the bulk geometry will be different from the DBI black hole
by employing a new radial variable $r$. For the solution of this section
the IR limit is $r\rightarrow 0$ and the UV limit is $r \rightarrow \infty$.}
\begin{equation}
\Phi(r)=\mu + \mathcal{E}\, r
\end{equation}
which is the correct form for a gauge field in \ads{2}. 
The equation for the blackening function is
\begin{equation}
{f^{\prime\prime}}(r) =\frac{e^{-4A_0}}{w_0^2V_f^0}\frac{\hat{n}^2}{\sqrt{1+e^{-6A_0}\frac{\hat{n}^2}{V_f^0{}^2w_0^2 }}}
\end{equation}
and has the general solution
\begin{equation}
f(r) = C_1+C_2 r+\frac{1}{2}\frac{e^{-4A_0}}{w_0^2V_f^0}\frac{\hat{n}^2}{\sqrt{1+e^{-6A_0}\frac{\hat{n}^2}{V_f^0{}^2w_0^2 }}}r^2
\label{418}
\end{equation}
The \ads{2} solution is simply the one in which $C_1=C_2=0$, and we identify the \ads{2} radius, $L_2$, as
\begin{equation}\label{eq:L2}
L_2^2 = 2e^{6A_0}w_0^2\,V_f^0\frac{\sqrt{1+e^{-6A_0}\frac{\hat{n}^2}{V_f^0{}^2w_0^2 }}}{\hat{n}^2}
\end{equation}
All the rest of the equations simply give constraints that determine when this solution can be realized. The ``zero energy" constraint says that
\begin{equation}
0= V_g^0 - V_f^0 \sqrt{1+\frac{\hat{n}^2}{(e^{3A_0}V_f^{0}w_0)^2}}
\end{equation}
while the dilaton equation of motion requires
\begin{equation}
0= \partial_\lambda V_g^0-\frac{1}{\sqrt{1+\frac{\hat{n}^2}{(e^{3A_0}V_f^{0}w_0)^2}}}\left(\partial_\lambda V_f^0 -\frac{V_f^0}{w_0}\frac{\hat{n}^2}{(e^{3A_0}V_f^{0}w_0)^2}\,\partial_\lambda w_0\right)
\end{equation}
and the tachyon equation forces
\begin{equation}
0= \partial_\tau V_f^0 -\frac{V_f^0}{w_0}\frac{\hat{n}^2}{(e^{3A_0}V_f^{0}w_0)^2}\,\partial_\tau w_0.
\end{equation}
In the following section we will investigate these constraints in more detail, to determine whether or not they can be realized in V-QCD models of interest.

\subsection{A closer look at the \ads{2} solution} 
We can summarize the \ads{2} requirements succinctly by recalling the definition of the 
effective potential, Eq. (\ref{1stdefveff}), in the language of this section:
\begin{equation}\label{eq:Veff}
V_{\mathrm{eff}}(\lambda,\tau)=V_g(\lambda)-V_f(\lambda,\tau)\sqrt{1+\frac{\hat{n}^2}{e^{6A}\,V_f(\lambda,\tau)^2w(\lambda)^2}}
\end{equation}
in which case the \ads{2} constraints are simply 
\be
V_{\mathrm{eff}}^0 = \partial_\lambda V_{\mathrm{eff}}^0=\partial_\tau V_{\mathrm{eff}}^0=0\;.
\ee
The zero energy constraint $V_{\mathrm{eff}}^0=0$ shows that the volume form on the $\mathbb{R}^3$ factor is just
\begin{equation}
\mathrm{Vol}_{\mathbb{R}^3} = e^{3A_0} = \frac{|\hat{n}|}{ w_0}\frac{1}{\sqrt{V_g^0\,^2-V_f^0\,^2}}
\end{equation}
so one can think of this condition as an expression describing the size of the $\mathbb{R}^3$, as determined by the values of the potentials at the fixed point. For the class of potentials of immediate interest, this relationship fixes the volume of  $\mathbb{R}^3$ in terms of $(\lambda_0,\tau_0,x_f,\hat{n})$.

Solving the zero energy constraint for $\hat{n}$  allows one to rewrite the extremization conditions like
\begin{align}
0 = &\,\partial_\lambda\log\Big[ w_0^2\left(V_g^0\,^2-V_f^0\,^2\right)\Big]\label{eq:lambdaExt}\\
0 = &\,\partial_\tau\log\Big[ w_0^2\left(V_g^0\,^2-V_f^0\,^2\right)\Big]
\end{align}
Note that these expressions depend only on $x_f$, $\lambda_0$, and $\tau_0$, and that the notation asks one to differentiate the potentials first, then evaluate the result at the constant scalar solution.

\begin{figure}\centering
\includegraphics[scale=0.84]{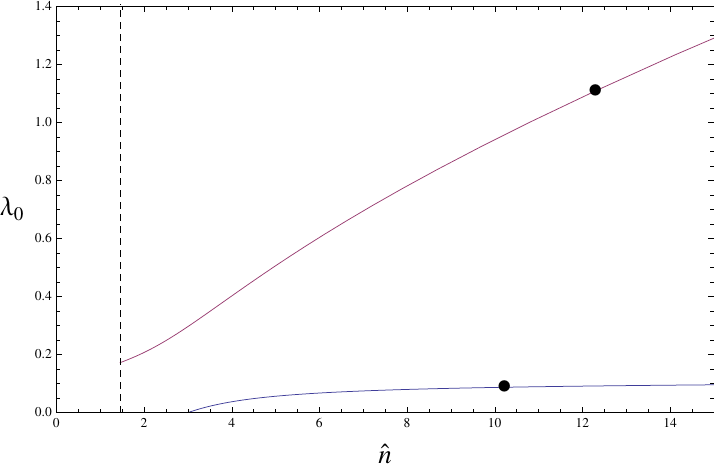}
\includegraphics[scale=0.84]{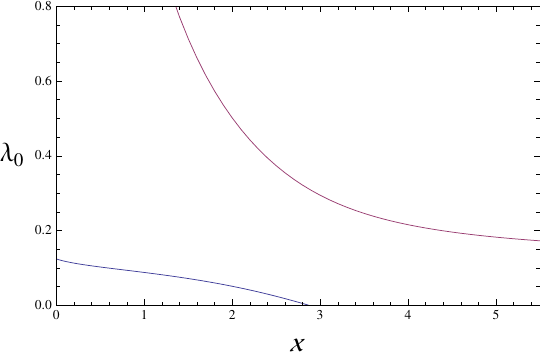}
\caption{\label{fig:AdS2notau} Allowed $(\hat{n},\lambda_0)$  (left) and $(x,\lambda_0)$ (right) values for the \ads{2} solution with vanishing tachyon. At left, the black dots mark the location of $x_f=1$ along each branch, and correspond precisely to the \ads{2} solutions found numerically and shown in figure \ref{phreg}. The black dashed line marks the Banks-Zaks limit at $x_f = 11/2$. From the right plot we find that  when $x_f\gtrsim 2.865 $ the constant dilaton solution becomes negative and is thus excluded as a fixed point candidate.}
\end{figure}

Finding simultaneous solutions to these equations provides the parameter space on a two-parameter plane in which the \ads{2} solution can be realized. For the class of potentials used in V-QCD (\ref{eq:Vfnw}), this constraint is again trivially satisfied for $\tau_0 = 0$ or $\tau_0 = \infty.$  For vanishing $\tau_0$, it is easy to find solutions to the constraint numerically for the V-QCD potentials in Section \ref{act}. They appear in figure \ref{fig:AdS2notau}. Interestingly, there is a region at low $x_f$ where there are two solutions for constant (positive) dilaton. This behavior  may be an artifact of the parametrization of the potential $w(\lambda)$. The second fixed point is not expected, but we also find that it plays no role in the phase diagram.

\begin{figure}\centering
\includegraphics[scale=0.83]{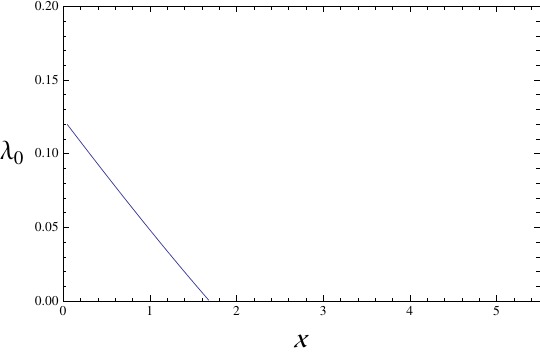}
\includegraphics[scale=0.83]{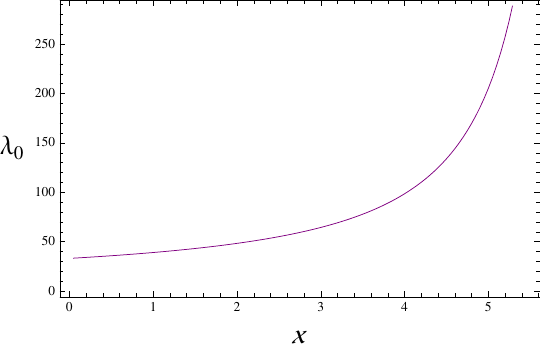}
\caption{\label{fig:AdS2inftau} Allowed $(x,\lambda_0)$  values for the \ads{2} solution with divergent tachyon.  The left plot shows the small branch of solutions, which cease to exist for  $x_f\gtrsim 1.685 $. On the right are the large branch solutions which extend to the Banks-Zaks limit at $x_f=11/2$.}
\end{figure}

In the case of the divergent tachyon, $\tau_0=\infty$ it is clear that $V_f^0 = 0$. One can carry out the same analysis as in the $\tau_0 =0$ case to search for allowed \ads{2} solutions in V-QCD, carefully navigating the somewhat subtle limits implied by this solution. For finite $\hat{n}$ but vanishing $V_f^0$ one finds that a divergent tachyon implies an electric field (\ref{eq:AdS2E}) and \ads{2} radius of the form
\begin{equation}\label{eq:AdS2EtauInf}
\mathcal{E} = -\frac{\hn}{|\hn|}\frac{e^{2A_0}}{w_0}\qquad \mathrm{and} \qquad L_2^2 = \frac{2w_0}{|\hat{n}|}e^{3A_0}
\end{equation}
The extremization condition (\ref{eq:lambdaExt}) becomes
\begin{equation}
0 = \partial_\lambda\log(w_0 V_g^0)
\end{equation}
and the numerical results for the potentials in Section \ref{act} are shown in figure \ref{fig:AdS2inftau}. As before there are two branches of solutions---the smaller of which terminates at some finite value of $x_f$ within the Banks-Zaks limit at $x_f = 11/2$.

\subsection{Stability of the \ads{2} Region\label{stab}} \label{sec:ads2stability}

The \ads{2} solutions can be a priori endpoints or starting points of RG flows. To determine exactly what happens we must do a scaling analysis of the perturbations around them.
  
We perturb the background \ads{2} metric like
\begin{equation}
\dd s^2 = -D(r)\dd t^2 + B(r)\dd r^2 + C(r)\dd \vec{x}^2
\end{equation}
where
\begin{align}
D(r) = & \frac{r^2}{L_2^2}\,\Big(1+D_1\, r^{d_1}\Big)\\
B(r) = & \frac{L_2^2}{r^2}\,\Big(1+B_1\,r^{b_1}\Big)\\
C(r) = & C_0+C_1\,r^{c_1}
\end{align}
In this background, the IR is approached as $r\to 0$ while the UV as $r\to \infty$. Here $L_2$ is the \ads{2} radius as given by (\ref{eq:L2}), $C_0$ controls the volume of the $\mathbb{R}^3$ factor, and the other constants parametrize the fluctuations in the obvious way. Without loss of generality, we set $C_0 = 1$ in what follows. All fluctuation amplitudes are taken to be small.

The background fields are perturbed as well, 
\begin{align}
\lambda(r) = & \lambda_0 + \lambda_1\,r^{a_1}\\
\tau(r) = & \tau_0 + \tau_1\,r^{t_1}\\
\Phi(r) = & \mu +r\,\Big(\mathcal{E} +\Phi_1\,r^{f_1}\Big)
\end{align}
The program is to insert these perturbation Ans\"atze into the equations of motion, linearize the equations about the fluctuations,  and subsequently determine the scaling exponents and the fluctuation amplitudes that describe a given perturbation.

Operationally, one first sets all the fluctuations above the background proportional to the same power, which is to say
\begin{equation}
\alpha = d_1=b_1=c_1=a_1=t_1=f_1
\end{equation}
The linearized fluctuation equations then reduce to a coupled set of homogeneous linear equations in the amplitudes of the fluctuations $F_i=\{D_1,B_1,C_1,\lambda_1,\tau_1,\Phi_1\}$. Importantly, the radial AnsŠtze under investigation leaves a residual gauge freedom related to reparametrizations of $r$. Practically, this means that fixing $B_1$  constitutes a gauge choice, and the linear system consists of 5 independent equations. Requiring that the system have a non-trivial solution is equivalent to  requiring that the determinant of the matrix of coefficients, $M$ vanish for all $r$.

In this case, one finds that the determinant is of the form
\begin{equation}
\det M =\alpha^2(\alpha-1)(\alpha+1)^2(\alpha+2)\,g(\alpha,\lambda_0,\mathcal{E}) 
\end{equation}
which vanishes for $\alpha^* =\{ 0,-2,-1,1\}$ and for the $\alpha=\alpha^*$ such that $g(\alpha^*,\lambda_0,\mathcal{E})=0$. The former correspond to ``universal" modes, while the latter are ``non-universal" in the sense that they depend on the details of the various potentials. Of the universal modes, we find that there are two types of IR relevant ($\alpha<0$) modes in the fluctuation spectrum, with exponents $\alpha^*=\{-2,-1\}$. That they correspond to relevant operators in the IR is clear from the fact that when $\alpha^*<0$ these modes grow as $r\to0$.

To better understand the non-universal modes, it is useful to write them in terms of the effective potential (\ref{eq:Veff}). Note that $\hat{n}$ can be easily related to the boundary value of the electric field ($\mathcal{E}$ in this section) via (\ref{eq:AdS2E}). The effective potential also turns out to govern the properties of two of the four non-universal exponents, 
\begin{equation}\label{eq:sebf}
\alpha_{\pm}^\lambda = -\frac{1}{2}\left[1\pm\sqrt{1-\frac{3}{2}\lambda_0^2\,L_2^2\,\partial_\lambda^2V_{\mathrm{eff}}^0}\right]
\end{equation}
while the other two are
\begin{equation}\label{eq:se2}
\alpha_{\pm}^\tau = -\frac{1}{2}\left[1\pm\sqrt{1+2\,\mathcal{E}^2 w_0^2\frac{\sqrt{1-\mathcal{E}^2 w_0^2}}{\kappa_0}L_2^4\,\partial_\tau^2 V_f^0}\right]
\end{equation}
The superscripts signify the fact that these modes correspond to perturbations of the appropriate scalars as we will see below.

\begin{figure}
\begin{center}
\includegraphics[scale=0.75]{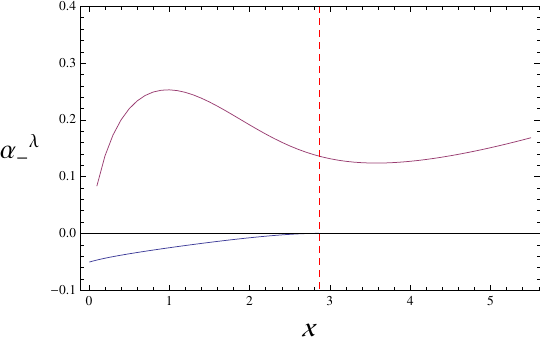}
\includegraphics[scale=0.75]{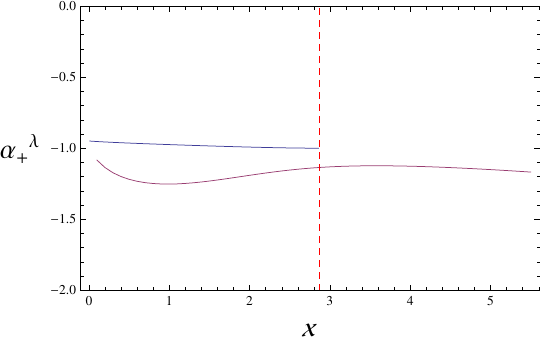}
\caption{
 The numerical values of non-universal exponents $\alpha^{\lambda}_\pm$ from (\ref{eq:sebf}), for the solutions with vanishing tachyon. Relevant operators have negative exponents in this analysis. The large $\lambda_0$ branch of solutions is colored purple. The BF-like bound mentioned in the text is never exceeded. The red dashed line indicates $x_f\approx2.865$ , beyond which the small branch of constant dilaton \ads{2} solutions vanishes. The domain of $x_f$ terminates at the Banks-Zaks limit $x_f=11/2$.}
\label{fig:alpha1}
\end{center}
\end{figure}

\begin{figure}
\begin{center}
\includegraphics[scale=0.75]{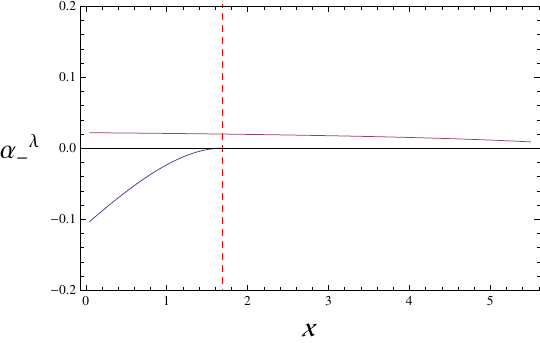}
\includegraphics[scale=0.75]{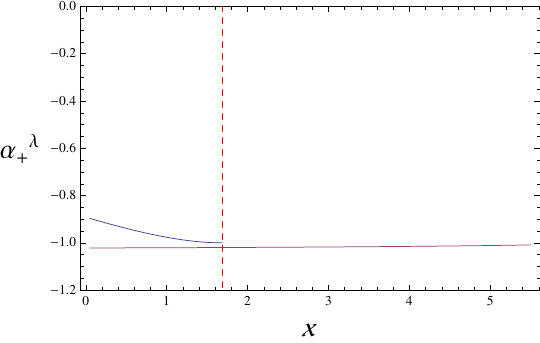}
\caption{
 The numerical values of non-universal exponents $\alpha^{\lambda}_\pm$ from (\ref{eq:sebf}), for the solutions with divergent tachyon. Again, the large $\lambda_0$ branch of solutions is colored purple and the BF-like bound  is never exceeded. The red dashed line indicates $x_f\approx1.685$ , beyond which the small branch of constant dilaton \ads{2} solutions vanishes.}
\label{fig:alpha1tauInf}
\end{center}
\end{figure}

\begin{figure}\centering
\includegraphics[scale=0.75]{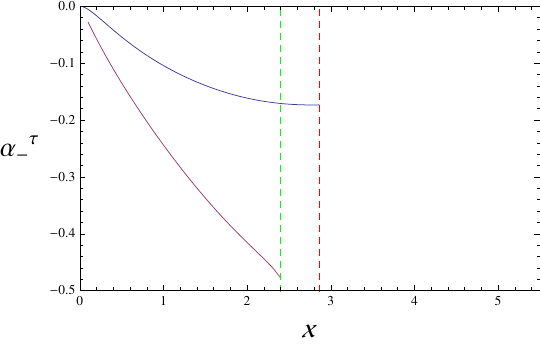}
\includegraphics[scale=0.75]{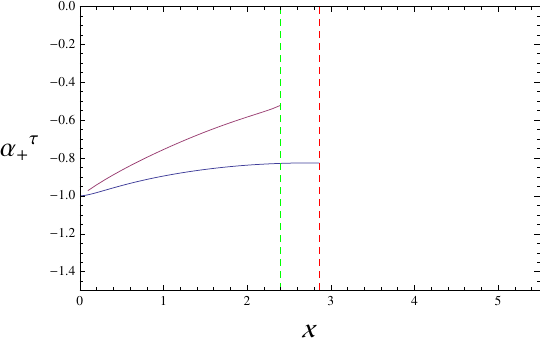}
\caption{The numerical values of non-universal exponents $\alpha^{\tau}_\pm$ from (\ref{eq:se2}), for the solutions with vanishing tachyon. Relevant operators have negative exponents in this analysis. The large $\lambda_0$ branch of solutions is colored purple. Note that for $x_f$ larger than approximately 2.4 the dual operator in the large branch fixed point has complex dimension, signaling an instability (marked by green dashed line). The red dashed line indicates $x_f$ for these potentials beyond which the small branch has $\lambda_0 <0$. Again, the domain of $x_f$ terminates at the Banks-Zaks limit $x_f=11/2$.}
\label{fig:alpha2}
\end{figure}

These exponents have a few noteworthy features. First, all of the exponents --- universal or not --- can be 
pairwise summed to give $\alpha_{+} + \alpha_{-}=-1$, which is the correct structure for modes in \ads{2}, in 
these coordinates. Moreover, we see from (\ref{eq:sebf}) that there is a BF-like bound signaling the onset of 
an instability when $\partial_\lambda^2 V_{\mathrm{eff}}^0 > \frac{2}{3}\frac{1}{L_2^2\lambda_0^2}$. 
For the V-QCD potentials employed for numerical studies, these non-universal exponents are plotted in 
figures \ref{fig:alpha1}, \ref{fig:alpha1tauInf} and \ref{fig:alpha2} as functions of $x_f$ for both branches of 
the \ads{2} fixed point. Evidently, while the BF-like bound is never exceeded in the fluctuations corresponding to $\alpha^\lambda$, the fluctuation characterized by $\alpha^\tau$ realizes an analogous instability around 
$x_f\sim2.4$ in the vanishing $\tau_0$ case. When the tachyon is divergent, the equations of motion require 
$\mathcal{E}^2w_0^2 = 1$ and thus $\alpha_{\pm}^\tau$ 
saturates to $\{-1,0\}$. It will turn out that the fluctuations described by $\alpha^\tau$ are appropriately named, as they correspond to fluctuations of the tachyon alone.

The full description of the perturbation is given by the exponent $\alpha^*$, which contains information about the dimension of the dual IR operator, and the amplitudes of the various modes that are activated by this fluctuation.
  The following cases are pertinent for the two conjugate solutions:
  \begin{itemize}
  
  \item If the operator is UV relevant then both perturbations vanish in the UV boundary.
   
    \item If the operator is IR relevant then both perturbations blow-up in the IR regime.
        
   \item If the operator is UV irrelevant then one perturbation vanishes and one blows up  in the UV boundary.
  
\item If the operator is IR irrelevant then  one perturbation vanishes and one blows up  in the IR regime.
  \end{itemize}
  
The amplitudes are easily obtained by solving the linear system provided by a given $\alpha^*$, and in general depend on one undetermined (but non-vanishing) amplitude and a choice of radial gauge which can be fixed via $B_1$. The results are listed in Appendix~\ref{app:ads2flucts}.  

We conclude this section by assessing the RG stability of \ads{2} solutions.
The one that appears at small values of $\lambda$, denoted by a blue line in figure~\ref{fig:AdS2notau} has dilaton and tachyon perturbations that render it IR unstable. This explains the fact that it plays no role in the phase diagram we describe in this work.

The other \ads{2} solution that corresponds to the purple line in figure~\ref{fig:AdS2notau} has dilaton and other perturbations that are IR irrelevant but the tachyon perturbation is IR relevant in the non-tachyonic black-holes. This is as expected as we need to tune $m_q=0$ to reach this solution in the IR. Once we turn on $m_q\not= 0$ we will avoid it and end up in the tachyonic black hole. On the other hand in the tachyonic case, the dilaton perturbation is IR irrelevant and the tachyon one is marginal. However it does not correspond to an extra parameter in the theory as $\tau=\infty$ is a singular point in field space.

\section{Numerical solution 
}\la{num}
\subsection{Preliminaries}

The equations of motion admit two types of solutions at finite temperature and chemical potential,
which we call black hole and thermal gas solutions. The thermal gas solutions have no horizon in the IR.
In this case the temperature is identified as the inverse of the length of the compactified time coordinate,
while $\Azs=\mathrm{const.}=\mu$. The blackening factor is trivial, $f \equiv 1$, and the
$z$-dependence of the other fields is exactly the same as for the solutions at $T=0=\mu$,
which were constructed in \cite{Jarvinen:2011qe}. When $0<x_f<x_c$, the dominant vacuum was found to have
a nonzero tachyon field (and therefore broken chiral symmetry). The thermodynamics of the corresponding
thermal gas solution is trivial: the pressure is independent of $T$ and $\mu$ and will be normalized to zero here.
Likewise, the condensate, which signals chiral symmetry breaking, will be nonzero but $T$ independent.

The nontrivial task on which we concentrate in this article is the construction of the black hole solutions.
The equations we have to solve numerically are the Einstein's equations \nr{A1} and \nr{A3},
the equations of motion for $\l$, equation \nr{A2},
and the equation of motion for $\tau$, equation \nr{A4}.
Their solution for $\hn=0$ has been discussed in detail in
\cite{Alho:2012mh}. The numerical solving with given initial conditions
as such is very simple using NDSolve of
Mathematica. The main issue is the correct initialization and
subsequent processing of the solutions via the scalings described in section \ref{sectscalings}.

An important general feature is that there are two types of black hole solutions:
\bi
\item
The solutions with $\tau=0$ which describe the hot and dense matter in a chirally
symmetric phase; these are expected to dominate the free energy at large $T$ or $\mu$.
\item
The solutions with $\tau(A)\not=0$. These will describe a chirally broken phase, expected
to dominate at small $T$ or $\mu$. These solutions are parametrized by the value of
the quark mass
\be \label{mqextra}
m_q=\lim_{A\to\infty}\CL_\rmi{UV}^{-2}\,\tau(A)\,e^{A}(A-\ln(\Lambda\CL_\rmi{UV}))^{\gamma_0/b_0}.
\ee
\ei
Since we are interested in solutions with exact chiral symmetry, we need to restrict to $m_q=0$.
This is a technically very demanding task (see Appendix \ref{B}) and necessitates going to
very small values of $z\approx e^{-A}$, up to $A\sim$ hundreds. This is one of the
reasons for using $A$ as a coordinate. The details of the numerical solution and the associated scaling
 properties are discussed in detail in Appendix \ref{scalings}.

In the numerical computations we choose the unit of number density so that $\CL_A=1$.
In section \ref{appD} we shall actually fit that $\CL_A\approx0.97$. 

\subsection{Initialization: expansion around horizon}
For thermodynamics one needs solutions with a black hole. To generate them numerically,
one has to start the integration at the horizon, which we place at $A=A_h$ 
such that
$f(A_h)=0$. Because of the singularities due to the $1/f$ terms in Eqs. \nr{A1}-\nr{A4}
one cannot start the integration precisely at the horizon. Instead, one first writes
the values of the fields at a small distance $\e$ from the horizon by expanding in $\e$ as
\ba
q&=&q_h+\epsilon q'_h + \CO(\e^2),\la{horexpq}\\
\lambda&=&\lambda_h+\epsilon\lambda'_h + \fr12\e^2\l''_h+\CO(\e^3),\\
f&=&\epsilon f'_h +\fra12\e^2 f''_h +\CO(\e^3),\label{horexpf}\\
\tau&=&\tau_h+\epsilon\tau'_h+\fra12\epsilon^2\tau''_h + \CO(\e^3),
\label{horexptau}
\ea
which are then inserted to the equations of motion.
Here and in the following the subscript $h$ denotes quantities evaluated at the horizon.
Then one expands in $\e$ and demands that
the divergences and the constant term
vanish.
Note that the input here is that in \nr{horexpf} $f_h=f(A_h)=0$.

Out of the leading terms in \nr{horexpq}-\nr{horexptau} one can choose $f'_h=+1$ as the magnitude of
$f(A)$ will anyway be fixed by the scaling \nr{fscalA} to the boundary value $f(A\to\infty)=1$. The
dilaton value at the horizon $\l_h$ will remain as a parameter, closely associated with
temperature. The second parameter, closely related to the chemical potential, is $\hn$.
However, in the numerics it turns out to be more practical to use instead
\be \la{defnt}
 \nt = e^{-3A_h}\hn = \frac{\hn}{b_h^3}
\ee
which is invariant in the scaling of~\eqref{Ascaling} and~\eqref{Lascaling}. 
The tachyon value at the horizon will be fixed by the quark mass, $\tau_h=\tau_h(\l_h,\nt;m_q)$.
Including the terms up to $\CO(\e)$ for $q$ and up to $\CO(\e^2)$ for the other fields 
in \eqref{horexpq}-\eqref{horexptau} is sufficient to ensure that the values of these parameters 
in the resulting numerical solution match with their input values to a high precision.

The remaining first-order derivative terms will be fixed by demanding that the $1/A$ (i.e. $1/\epsilon$) singularities
cancel. Canceling the divergent $1/A$ term of \nr{A1}
gives
\be
q_h=-{\sqrt{3f'_h}\over\sqrt{V_g-V_f\sqrt{1+\Ks_h}}},\quad \Ks_h={\nt^2\over \ws^2V_f^2}.
\la{qh}
\ee
with the understanding that the potentials $V_g$, $V_f$, and $\ws$ are evaluated at the horizon.
Canceling the divergent $1/A$ term of the $\l$ equation \nr{A2} gives
\ba
\lambda'_h&=&-{3\lambda_h^2q_h^2\over 8f'_h\sqrt{1+\Ks_h}}
\biggl(\sqrt{1+\Ks_h}\partial_\lambda V_g-\partial_\lambda V_f+\Ks_hV_f{\f'_h\over\f_h}\biggr)
\nn
&=&-{3\lambda_h^2q_h^2\over 8f'_h}\partial_\l \veff(\l_h,\tau_h,\nt)
\la{laprh}
\ea
and canceling the $1/A$ term of the $\tau$ equation gives
\be
\tau'_h={q_h^2\partial_\tau \ln V_f\over f'_h\kappa_h(1+\Ks_h)}.
\ee
In \nr{qh} and \nr{laprh} we again have the important quantity,
\be
\veff=V_g(\l)-V_f(\l,\tau)\sqrt{1+{\hn^2\over b^6\ws^2V_f^2}}=V_g(\l)-V_f(\l,\tau)\sqrt{1+{\nt^2 b_h^6\over b^6\ws^2V_f^2}},
\la{defveff}
\ee
evaluated at the horizon.

This leaves us with the four quantities $f''_h,\, q'_h,\, \l''_h,\,\tau''_h$ to
be determined by requiring that the constant terms of the four equations vanish.
The constant term of equation \nr{A3} gives a simple relation between $f''_h,\,q'_h$:
\be
f''_h+f'_h\biggl(4-{q'_h\over q_h}\biggr)-{\Ks_h\over\sqrt{1+\Ks_h}}q_h^2V_f=0.
\ee
The remaining expressions are too complicated to be reproduced here but can be found in \cite{TSAcode}.
From the algebraic derivation of the initial conditions
to the numerical integration of the system of differential equations \nr{A1}-\nr{A4},
we treat the whole problem in Mathematica.

\subsection{Observables}

It is thus easy to produce some numerical solutions for the functions $q(A)$, $\l(A)$, $f(A)$, and $\tau(A)$
with Mathematica, given $\l_h$ and $\nt$, but an essential and nontrivial part of the numerical work is to
transform the solutions to a standard form satisfying in $z$ coordinates $f(0)=1$ 
and that the scale of the UV expansions equals one (see Appendix~\ref{scalings}). 
In $A$ coordinates these conditions become
\ba
\lim_{A\to\infty}f(A)&=&1 \nn
\lim_{A\to\infty}\biggl({1\over b_0\l(A)}+{b_1\over b_0^2}\ln(b_0\l(A))-A\biggr)&=&-\ln\CL_\rmi{UV}.
\la{Alim}
\ea
The former is implemented by scaling $f$ as in \nr{fscalA}, the latter by scaling $A$ as in
\nr{Ascaling}. To achieve this one determines the scaling factor $\Lambda(\l_h,\nt)$ so that the
asymptotic limit \nr{Alim} holds. 
We start from a numerical solution having $A_h=0$, 
then according to \nr{Ascaling}
the value of $b$ at the horizon in the scaled solutions is simply given in terms of the scaling factor by $b_h=\exp(A_h)/\Lambda=1/\Lambda$.
From the standard configurations so obtained one then computes the temperature
as the black hole temperature and
the chemical potential using \nr{A0value} and \nr{A0cond},
otherwise the configurations as such are not of interest for this calculation.
The procedure is described in detail in Appendix \ref{scalings}.

Summarising, from the numerical integration of equations of motion, for given $(\l_h,\nt)$,
one obtains the following quantities:
\be
b_h(\l_h,\nt),
\quad T(\l_h,\nt),\quad \mu(\l_h,\nt).
\la{quantities}
\ee
From these we obtain the entropy density using the basic formula
\be
s(\l_h,\nt)={A\over4G_5}={b_h^3\over 4G_5}.
\ee
To obtain the 4d physical quark number density note first that,
when deriving the $\Azs$ equation of motion
from the fermionic part of the action, one has, for solutions of equations of motion,
\be
\delta S_f={1\over16\pi G_5}\,{V\over T}\int_\epsilon^{z_h}dz{d\over dz}\biggl(
{\p L_f\over\p \dot \Azs}\delta \Azs\biggr).
\ee
At $z_h$ one has to keep the value $\Azs(z_h)$ fixed to zero so that $\delta \Azs(z_h)=0$ and the
upper limit does not contribute.
Since $S=-\Omega/T$ and $\delta \Azs=d\mu$ the fermionic contribution given by the above integral
is the $n\,d\mu$ term in the free energy, and therefore the correct
normalization of $n$ is
\be \la{defn}
 n = {\hn\over16\pi G_5} = {\nt\,b_h^3\over16\pi G_5} = {\nt\over 4\pi}{b_h^3\over4 G_5}= s\,{\nt\over4\pi},
\ee
where we used the definition~\eqref{defnt}.
This expression also gives a physical interpretation of the parameter $\nt$ of the
integration of the equations of motion:
\be
\nt
=4\pi\, {n\over s}.
\la{physn}
\ee

Next we discuss what values of $(\l_h,\nt)$ are possible and how the
pressure is integrated from $dp=s\,dT+n\,d\mu$.
To compute the pressure we have to integrate over $T$ and $\mu$ and these
one-dimensional integrals are most simply carried out by converting them
into 
integrals over $\l_h$ at fixed $\nt$ or vice versa,
see Section~\ref{sectp}.

\begin{figure}[!tb]

\centering

\includegraphics[width=0.8\textwidth]{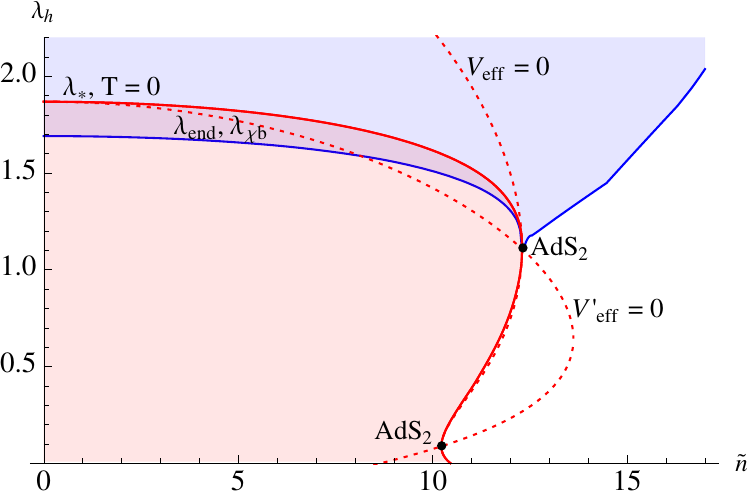}%

\caption{\small
The physical region on the $\l_h,\nt$ plane for chirally symmetric
(red region) and chirally broken (blue region, unbounded above) solutions. Chirally
symmetric region is bounded from above by the curve $\l_*(\nt)$ along which $T=0$ up
to the point \ads{2} at $\nt=12.295,\,\l_h=1.108$, then from the right by a
segment of the curve $V_\rmi{eff}=0$ up to the second \ads{2} point
at $\nt=10.223,\,\l_h=0.0873$ and finally by a segment to $\nt=10.457,\,\l_h=0$.
Tachyonic chiral symmetry breaking solutions exist only above the blue curve
$\l_\rmi{end}(\nt)\equiv\l_{\chi b}$. 
The dashed lines are $\veff=0$ and $\veff'(\l_h)=0$ at $\tau=0$
(see \protect\nr{defveff}).
 }
\label{phreg}
\end{figure}

\subsection{Physical region in the $\l_h,\,\nt$ plane}
For $\nt=0$ one found (see, e.g., \cite{Alho:2012mh}, Fig. 7) that chirally symmetric
solutions, i.e. the ones with zero tachyon,
existed only for $0<\l_h<\l^*$, with $\l^*$ given by the extremum of the
effective potential in~\eqref{defveff}, and chirally broken solutions with nonzero tachyon existed
only for $\l_h>\lend$ with $0<\lend<\l^*$. This is in harmony with the expectation that
large $T$ and chiral symmetry are associated with small coupling, small $\l_h$, and strong
coupling leads to chirality breaking.
The introduction of $\nt$ extends this pattern in an interesting and subtle
way, exhibited in Fig.~\ref{phreg}.

For small $\nt$ the above pattern remains unchanged, only the curves $\lend(\nt)$ and
$\l_*(\nt)$ slowly decrease. Also thermodynamically the situation is only slightly
modified: as we shall soon see along $\lend$ the symmetric and broken phase
pressures are equal and there is a continuous chiral phase transition.
This pattern continues all the way to the point labeled $\mathrm{AdS}_2 $ in Fig.~\ref{phreg}, where $\nt = \nt_\rmi{max} \approx 12.295$. 
On approach to that point along $\lend$ from the small $\nt$ -side, temperature goes to zero. 
When $\nt > \nt_\rmi{max}$, since there is no solution with $\tau_h = 0$, the chiral symmetry breaking solution at the limit $\lend$ cannot, and does not, have $\tau_h \rightarrow 0$, and cannot therefore be a part of the second order transition line. 
The temperature, when approaching that line, does decrease to very small values, but unfortunately the numerics is not stable enough to be confident that it indeed converges to zero. 
The hypothesis is, however, that it indeed does, and that the $\nt > \nt_\rmi{max}$ section of the $\lend$ curve consists of $T = 0$ chiral symmetry breaking solutions of various $\mu$.

The computation of the curves in Fig.~\ref{phreg} 
is mostly numerical, but parts of
the boundary of the symmetric phase can be found analytically.
Note first that \nr{qh} implies that 
\be
\veff(\l_h,\tau_h)=V_g(\l_h)-V_f(\l_h,\tau_h)\sqrt{1+\Ks(A_h)}
=V_g(\l_h)-\sqrt{V_f^2(\l_h,\tau_h)+{\nt^2\over \CL_A^4\f_h^2}}>0,
\la{veff0}
\ee
where we used \nr{defnt} and inserted $\ws$ from~\eqref{defellA}.
For the symmetric phase $\tau_h=0$ one can solve from here the upper boundary for
values of $\nt$:
\be
\nt\le\nt_\rmi{max}=\CL_A^2\f_h\sqrt{V_g^2(\l_h)-V_f^2(\l_h,0)}.
\la{nhatmaxval}
\ee
This with $\CL_A=1$ is the curve $\veff=0$ in Fig.~\ref{phreg}.
Further, due to the interpretation $\beta(\l)=\l'(A)$ one usually expects that $\l(A)$
monotonically decreases from its value $\l_h=\l(A_h)$
towards $\l(A=\infty)=0$, and in particular that $\l'_h<0$. From \nr{laprh} this would imply
\be
\partial_\l\veff(\l_h,\tau_h(\l_h,m_q),\nt)\ge0.
\la{lahpveff}
\ee
However,
deep in the IR 
the interpretation of
$\l'(A)$ as a negative beta function need not be valid and solutions with
signs opposite to those in \nr{lahpveff} are also possible. This is confirmed
by numerical computation and the real boundary is given by finding where
$T=0$ or where the scale factor $\Lambda(\l_h,\nt)$ diverges.
Requiring that both $\veff$ and $\veff'$ vanish has two solutions marked
\ads{2}, since actually
the geometry at these points is asymptotically 
AdS$_2\times {\mathbbm{R}}^3$
in the IR.
The lower \ads{2} point disappears at larger $x_f$.

The boundary of the broken phase marked $\lend$ in Fig.~\ref{phreg} is discussed
in some detail in Appendix \ref{AppNumResults}. It is a lower limit for possible values of
$\l_h$. For $\nt$ there is an upper limit, but there is no upper limit for $\l_h$.

\begin{figure}[!tb]

\centering

\includegraphics[width=0.49\textwidth]{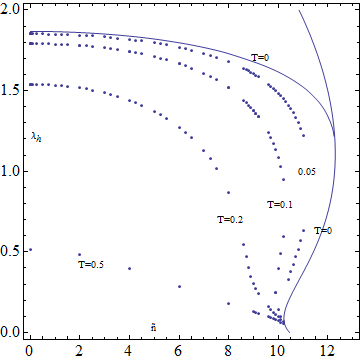}\hfill
\includegraphics[width=0.49\textwidth]{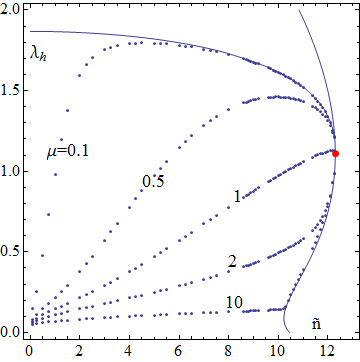}

\caption{\small Constant values of $T$ and $\mu$ on the $\nt,\l_h$ plane for tachyonless
solutions for the discrete values of $\nt$ used in the pressure integration, other
boundary curves as in Fig.~\protect\ref{phreg}. Small $\l_h$ corresponds to large
$T$ as expected. The special role played by the red \ads{2}
point is seen: above it along the boundary curve $\mu=0$, below it $\mu=\infty$.
Thus effectively at the \ads{2} point all positive values of $\mu$ are obtained.
}
\label{Tmuconst}
\end{figure}

The significance of various parts of the physical region is also described by
plotting curves of constant $T$ and $\mu$ as in Fig.~\ref{Tmuconst}. Actually we
show there the result only for the discrete values of $\nt$ used in the pressure
integration. Particularly interesting is the behavior of the $\mu$ = constant
curves. Extrapolating them one sees that clearly asymptotically $\mu=0$ in the
upper part of the $T=0$ curve. States here have $T=\mu=0$ and thus represent vacuum.
In the vertical part of the $T=0$ curve correspondingly $\mu=\infty$. This is also
some special state. All the $\mu=$ constant curves end at the \ads{2} point, where
thus all the exactly $T=0$, $\mu$ finite symmetric phase thermodynamics resides.

\begin{figure}[!tb]

\centering

\includegraphics[width=0.49\textwidth]{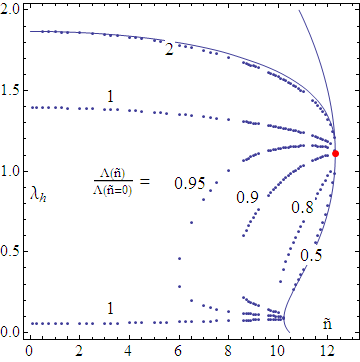}

\caption{\small Constant values of the scale factor $\Lambda$,
normalised to its value at $\tn=0$, on the $\nt,\l_h$ plane for tachyonless
solutions for the discrete values of $\nt$ used in the pressure integration, other
boundary curves as in Fig.~\protect\ref{phreg}. Curves of constant $s$ and $n$ can be inferred
from this (see text).
}
\label{Laconst}
\end{figure}

A third important quantity is the dimensionless scale factor $\Lambda(\l_h,\tn)$.
It varies a lot as a function of $\l_h$. The main part of the variation is contained
in $\Lambda(\l_h,0)$ which at the boundaries of the phase space,
$\l_h\to0$ and $\l_h\to\l_*(0)$, can be fitted by
\ba
\Lambda(\l_h,0)&=&0.714\,e^{-1/(b_0\l_h)}(b_0\l_h)^{b_1/b_0^2}\,(1+2.42\l_h+\cdots )={z_h\over\LUV},
\,\,({\textrm{as}}\,\,\lambda_h\rightarrow 0)\nn
&=&{1.3\over\sqrt{\l_*(0)-\l_h}}+2+\cdots \,\,\,({\textrm{as}}\,\,\lambda_h\rightarrow\lambda_\ast(0)).
\ea
Fig.~\ref{Laconst} shows curves of constant $\Lambda(\l_h,\tn)/\Lambda(\l_h,0)$. Since
$b_h=1/\Lambda$, $s\sim b_h^3$, $n=s\tn/(4\pi)$, also curves of constant entropy and
number density are contained in this figure.  One sees that the dependence on $\tn$
is surprisingly weak except at the $T=0$ boundaries. From the figure one can extrapolate
that
\bi
\item
On the upper $T=0$ boundary from $\nt=0$ to the \ads{2} point: $T=\mu=s=n=0$,
$\Lambda=\infty$. So this really is the vacuum. All the $T=0$, finite $\mu$ chirally
symmetric matter is exactly at the \ads{2} point

\item
 On the vertical $T=0$ boundary between the two \ads{2} points $T=\Lambda=0$,
$\mu=\infty,\,s=\infty,\,n=\infty$, $n/s=\tn/(4\pi)$, $10.223<\tn<12.295$.

\ei
\label{sec:physreg}

\subsection{Constant parameter curves}

A straightforward way to generate the data necessary for solving the thermodynamics would be to
compute black hole solutions in both the symmetric and non-symmetric branch of the solutions
on a sufficiently dense lattice in the physical region of the $(\nt, \l_h)$ -plane.
In order to carry out the pressure integrals without accumulating large cumulative errors,
a reasonably accurate continuum interpolation of the observables is needed.
Since the observables as a function of $(\nt, \l_h)$ are mostly smooth but have
lines of zeroes and divergences, detailed in the following section, it would be
necessary to use at least a somewhat sophisticated interpolation algorithm with an
adaptive local grid size in two dimensions, or alternatively a very large amount
of computing power for the brute force approach of simply a very dense uniform lattice.

However, we have been able to avoid constructing a full 2D interpolation of the solutions
by considering a grid of 1D interpolations, for which well-established adaptive algorithms
are readily available. The two primary interpolations are curves with $\nt$ as a constant
and $\l_h$ as the variable, and those with $\l_h$ as a constant and $\nt$ as the variable.
We compute the interpolations for a number of values of $\nt$ and a number of values of $\l_h$.
Figures \ref{looptest} and \ref{gridfig}  in appendix \ref{AppNumResults} show images of
these curves of the $(\mu, T)$ -plane.We can then compute the pressure integrals along each
of these, for both the symmetric and non-symmetric branches, fixing the constants as
described in the next section.

At least in the specific case handled in this paper, the constant parameter curve method
has allowed us to extract all the thermodynamic features of interest without resorting
to full 2D interpolation. However, in the case of a transition between two regions of
the same branch of solutions, such as happens at small $x_f$ for some of the potentials
explored in \cite{Alho:2012mh}, a complete 2D interpolation may become necessary to extract
the phase transition line.

\section{Results on thermodynamics}

\subsection{Computation of pressure}\la{sectp}

According to the holographic dictionary, the pressure can in principle be computed by evaluating 
the on-shell action. However, this is numerically very challenging in this kind of model with
corrections decaying only logarithmically near the boundary. Therefore, we use instead the
usual thermodynamic formulas. 

\begin{figure}[!tb]

\centering

\includegraphics[width=0.45\textwidth]{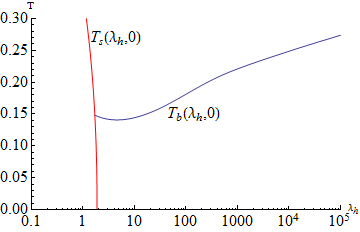}
\includegraphics[width=0.45\textwidth]{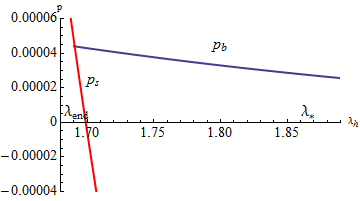}

\caption{\small Left: The temperature as a function of $\lambda_h$
for the symmetric tachyonless and broken tachyonic solutions at $\nt=0$.
Right: The pressures computed from \protect\nr{peeb} and \protect\nr{pees}
near the chiral transition region
with relative normalisation fixed so that they are equal at $\lend$. The broken
phase pressure vanishes outside the figure at $\l_h=3.19$ at the temperature $T_b(3.19,0)=0.14$.
}
\label{figTlah}
\end{figure}

We first review how the pressure is computed by integrating $s(T)=p'(T)$ for
$\nt=0$ since this is how the constant of integration is fixed in \cite{Alho:2012mh}
and will be fixed here, too. One has, see
Fig.~\ref{figTlah},
\ba
4G_5p_b(T)&=&\int^\infty_{\lambda_h(T)}d\lambda_h(-T_b'(\lambda_h))\,b_{h b}^3(\lambda_h)
+p_b(\infty),\label{peeb}\\
4G_5p_s(T)&=&\int^{\lambda_*}_{\lambda_h(T)}d\lambda_h(-T_s'(\lambda_h))\,b_{hs}^3(\lambda_h)
+p_s(\lambda_*),\label{pees}
\ea
where the subscripts $b$ and $s$ denotes quantities in chirally broken and symmetric phases, respectively.
What matters for the phase structure is the
difference 
of the integration
constants $p_b(\infty)$ and $p_s(\l_*)$. This is simply fixed by requiring that
pressure be the same for the two phases at $\l_h=\lend$ \cite{Alho:2012mh}. The outcome is plotted
in Fig.~\ref{figTlah}. At this temperature there is a second order (both $p$ and $s\sim p'$ are
continuous) chiral phase transition. The broken
phase pressure vanishes at $\l_h=3.19$ at the temperature $T_h=T_b(3.19,0)=0.14$.
This is the deconfining transition. At higher $\l_h$
or smaller $T$ the dominant phase is the thermal gas phase with vanishing thermal pressure.

For quantitative correctly normalised results one will need both the energy unit
$\Lambda_0$, which is implicit in formulas involving $T$ and $\mu$, and the
constant $4G_5$. The former is fitted by the value of the critical temperature $T_\chi(0)=
0.148\Lambda_0$. Taking $T_\chi=0.15$ GeV, we fix
\be
\Lambda_0=1 \,{\rm GeV}.
\ee
For $4G_5$ normalisation to the $T^4$ Stefan-Boltzmann term at $T\to\infty$ gives \cite{Alho:2012mh},
see Eq.~\nr{4G5},
\be
{1\over4G_5}={4\over45\pi}\,{1+\fra74 x_f\over\CL_\rmi{UV}^3}N_c^2={4\over45\pi}\,N_c^2.
\la{SBnorm}
\ee

In general, we wish to obtain the pressure by integrating $dp=sdT+nd\mu$. All the quantities on the
RHS are numerically known as functions of $\l_h,\,\nt$, see Appendix \ref{B}.
Note that
we can write $dp$ as
\be
dp={b_h^3\over 4G_5}dT+{b^3_h\,\nt\over16\pi G_5}d\mu
={b^3_h\over4G_5}\left(dT+{\nt\over4\pi}d\mu\right),
\la{dp}
\ee
where all quantities are functions of $\l_h,\nt$.

The differential $dp$ can now be integrated either over curves
of constant $\nt$ from $\l_b$ to $\l_t$,
\be
4G_5p_{\nt}(\l_t)=\int_{\l_b}^{\l_t} \,d\l_h\, b_h^3(\l_h;\nt)\biggl[T'(\l_h;\nt)+{\nt\over4\pi}
\mu'(\l_h;\nt)\biggr],
\la{plah}
\ee
or over curves of constant $\l_h$ from $\nt_b$ to $\nt_t$,
\be
4G_5p_{\l_h}(\nt_t)=\int_{\nt_b}^{\nt_t} \,d\nt \,b_h^3(\nt;\l_h)\biggl[T'(\nt;\l_h)+{\nt\over4\pi}
\mu'(\nt;\l_h)\biggr].
\la{pnt}
\ee

To test the path dependence of the integral, one can choose an arbitrary rectangle
within the physical region in Fig.~\ref{phreg} for either of the phases. This
is mapped to a four-sided region on the grid in Fig.~\ref{gridfig}.
One now integrates numerically around it using Eqs.~\nr{plah} and \nr{pnt}
and checks whether the integral is zero.
This is indeed what we find to a great accuracy.

This proof of path independence is a very impressive confirmation of the validity of
our numerical computations. 
All the quantities included in \nr{plah} and
\nr{pnt} are the results of lengthy numerical solutions of Einstein's equations and it is
striking to see that when they are put together as above, the outcome is path
independent to a very good numerical precision.

The pressures of the two phases $p_s(\l_h,\nt)$ and $p_b(\l_h,\nt)$ can now be computed by
fixing the relative integration constant by demanding that $p_s(\lend,0)=p_b(\lend,0)$
and integrating to the point $(\l_h,\nt)$ along any convenient path. These can trivially
be converted to $p_s(T,\mu)$ and $p_b(T,\mu)$. Three-dimensional plots of pressure vs $T,\mu$ are numerically
rather noisy and we shall focus on the main question: phase structure and phase transition lines.

\subsection{Phase structure}\la{phases}
We have discussed thoroughly the chirally symmetric and broken phases with pressures
$p_s(T,\mu)$ and $p_b(T,\mu)$. Furthermore, as a model for the low $T$ system we shall
use the thermal gas phase, for which the metric Ansatz is like that in \nr{bg} but with
$f(z)=1$ and $T$ is introduced by compactifying the imaginary time region, otherwise
the equations of motion are as before. Note that from $f=1$ and the equation of motion \nr{eq3}
it follows that one must have $\nt=0$ so that also $n=0$. Thus also $\dot \Azs=0$ so that
$\Azs=\mu$ is constant. The property $n=0$ for a low $T$ chirally broken phase is consistent 
with the fact that this model contains no baryons in the spectrum of singlet states.

The pressure of the thermal gas phase 
is $p_\rmi{low}=0$. We identify this with the hadron gas phase of the field theory. 
This is well justified in the case of
pure SU($N_c$) gauge theory, for which the pressure of the plasma phase is $\sim N_c^2$.
In the case here (V-QCD) the degrees of freedom in the low temperature phase are the $N_f^2$ Goldstone
bosons of the spontaneously broken chiral symmetry, while the high energy degrees of freedom are the
deconfined partons, $2 N_c+7/2 N_f N_c$. The ratio of the number of degrees of freedom at low and
high temperature is then $x_f^2/(2+7/2 x_f)$ which at $x_f=1$ is 0.18, and we expect that taking
$p_\rmi{low}=0$ provides still a useful guide towards the location of the deconfinement phase boundary.
However, as $x_f$ increases, the uncertainty associated with this approximation grows.
At $x_f\simeq 4$ the ratio becomes unity signalling the transition to a different vacuum
phase as one enters the conformal window.

The stable phase has the smallest free energy or, equivalently, the largest pressure. For phase
equilibrium one needs both kinetic, thermal and chemical equilibrium, i.e., the same
pressure, temperature and chemical potential for the two phases. The outcome of the
analysis has already been shown in Fig.~\ref{Ttrans}.

Consider first the most reliable prediction of the model: the chiral transition at
$T_\chi(\mu)$. At $\mu=0$ or $\nt=0$ this took place at the point $\lend$ in Fig.~\ref{figTlah}.
As $\nt$ is increased toward $\nt_\rmi{max}$, the $\lend$ -curve decreases monotonically and at $\lend(\nt_\rmi{max})$ hits the \ads{2} point.
We find that the pressure along $\lend$ is positive for the whole interval from $\nt = 0$ to $\nt_\rmi{max}$.
Since the pressure decreases on the $\tau_h \neq 0$ -branch for increasing $\l_h$, and is equal  for the $\tau_h = 0$ and $\tau_h \neq 0$ -branches at $\lend$, the resulting chiral transition is between two stable phases.
Along this transition line, the temperature also decreases monotonically, reaching $T = 0$ at the \ads{2} point.
 It therefore divides the $(\mu, T)$ -plane into a region near the origin, where chiral symmetry is broken, and an outside region with $\mu$ or $T$ large, where chiral symmetry is restored.

At $\nt > \nt_\rmi{max}$, $\lend(\nt)$ starts increasing again.
The temperature along the curve is very low and is consistent with being $0$, but unfortunately the numerics is not quite stable enough to state this with certainty.
The most likely option is that this is an unstable branch of chiral symmetry breaking $T = 0$ solutions, in analogy with the $T = 0$ -boundary of the chirally symmetric branch, as discussed in section \ref{sec:physreg}.

\begin{figure}[!tb]

\centering

\includegraphics[width=0.49\textwidth]{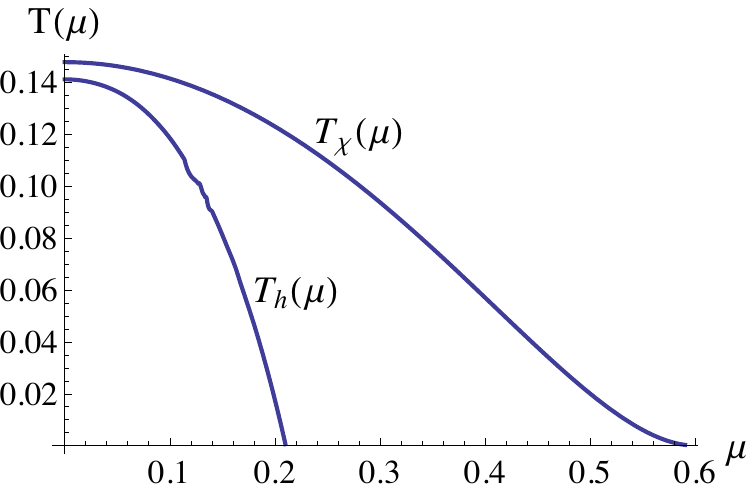}\hfill
\includegraphics[width=0.49\textwidth]{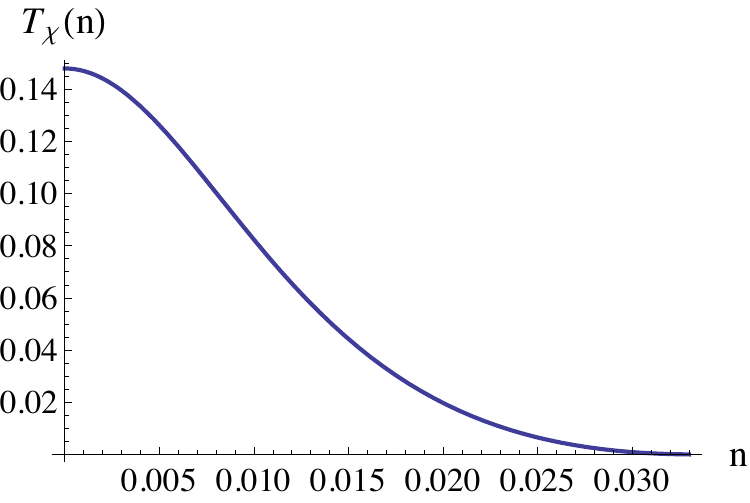}

\caption{\small Dependence of the chiral transition temperature on chemical potential or on
quark number density. 
In the left panel $p_b>0$ between $T_\chi$ and $T_h$ and
vanishes along $T_h(\mu)$. On it $n$ jumps to zero.
}
\label{Tchi}
\end{figure}

Below $T_\chi(\mu)$ the chirally broken phase is the stable one. Its pressure is positive
but starts decreasing when $T$ is further lowered, $\l_h$ increased, just as happens
in Fig.~\ref{figTlah} at $\mu=0$.
Computing the pressure of the broken phase for arbitrary values of $\nt$, one
finds that it vanishes for the values of $\l_h$ plotted in Fig.~\ref{lendTh}. The corresponding
temperature $T_h(\mu)$ is plotted in Fig.~\ref{Tchi}, see also Fig.~\ref{Ttrans}.
We interpret this as a first order deconfining transition between the chirally broken phase and
the zero-pressure low $T$ thermal gas phase.
The transition temperature $T_h(\mu)$ decreases monotonically with increasing $\mu$; our
numerical accuracy does not permit to make definite statements about the limit $T\to0$. Note
that this corresponds to very large values of $\l_h$, see Fig.~\ref{lendTh}.

\begin{figure}[!tb]

\centering

\includegraphics[width=0.6\textwidth]{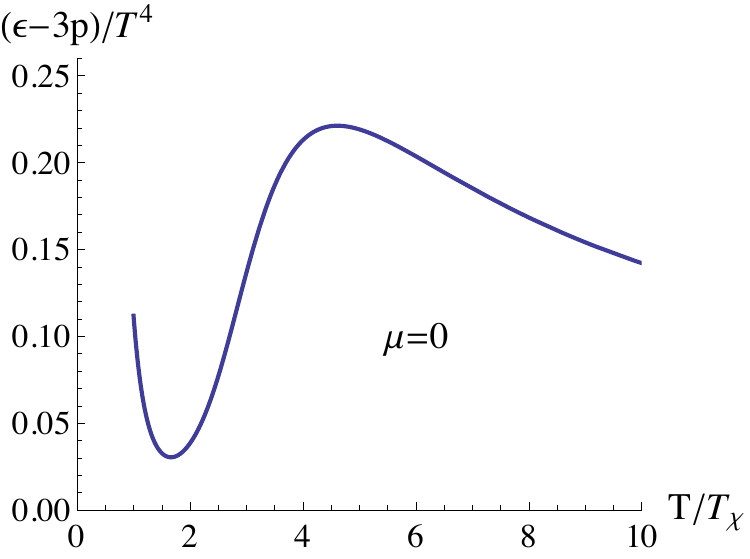}

\caption{\small The interaction measure $\e-3p=Ts+\mu\,n-4p$ scaled by the ideal gas pressure
\protect\nr{pideal} in the region $T>T_\chi(\mu)$ as a function of $T/T_\chi(0)$ for 
$\mu=0$. The curve refers to the symmetric phase and therefore starts
at $T/T_\chi(\mu)=1$ (see also Fig.~\ref{Ttrans}).
}
\label{im}
\end{figure}

One observes that at $\mu=0$ the temperatures $T_\chi(0)$ and $T_h(0)=0.95T_\chi(0)$ are very
close to each other. For reference, one may note that a very similar situation with
$T_h(0)=0.94T_\chi(0)$ was observed in \cite{Qin:2013ufa} in a completely different
Schwinger-Dyson equation model for QCD thermodynamics. There the conclusion was that
the chiral and deconfinement transitions probably coincide. These behaviors are most transparently
understood on the basis of underlying exact and approximate symmetries and related order parameters
\cite{Mocsy:2003qw,Kahara:2012yr}.

From the computed pressure, we can determine the interaction measure
$\e-3p=Ts+\mu\,n-4p=(T\partial_T+\mu\partial_\mu-4)p$, which is shown close to the chiral transition
region in the symmetric phase in Fig.~\ref{im}. Consider the curve for $\mu=0$, the structure
of which is described in Fig.~9 of \cite{Alho:2012mh}. The analogous curve for QCD is plotted, e.g.,
in Fig.~3 of \cite{Borsanyi:2013bia}. The V-QCD curve plotted in Fig.~\ref{im}
starts by decreasing above $T=T_\chi(0)$ but then changes direction and passes through
a maximum at $T\sim 4T_\chi(0)$ with a QCD-like decay above that. This large $T$ maximum can be
interpreted \cite{Alho:2012mh} as a crossover transition. When $x_f$ is increased into the conformal
region at $x_f>x_c\approx4$, this crossover is the only structure in $p/T^4$ which remains.
It is now apparent that increasing $\mu$ does not change this overall pattern qualitatively. In
particular, the large $T$ decay is independent of the chemical potential.

\subsection{Order of transition}
The chiral transition was above numerically observed to be of second order at $\mu=0$, for the potentials
used here. There are also potentials which lead to a 1st order transition, as concretely
shown in \cite{Alho:2012mh}. It is commonly accepted that the chiral QCD transition at
$N_f\ge3$ is of first order \cite{Pisarski:1983ms},
even though this has not been conclusively established with lattice Monte Carlo computations,
say, for $N_f=N_c=3$, $x_f=1$ \cite{deForcrand:2006pv}. It is useful to see how our gauge/gravity
duality model, valid, in principle, for $N_c\gg1$ relates to the general effective theory
arguments.

The order parameter for the effective theory of the QCD chiral transition is a complex
$N_f\times N_f$ matrix $M_{ij}({\bf x})=\langle q_L^i\bar q_R^j\rangle$, $i,j=1,\dots,N_f$,
${\bf x}$ is the $d=3$ dimensional spatial coordinate. The
potential term in the action is
\be
V(M)=m^2\tr M^\dagger M+ g_1\, (\tr M^\dagger M)^2 + g_2 \,\tr M^\dagger M M^\dagger M.
\la{effact}
\ee
To study the phase transition one should compute the effective potential of the theory. In
the 1-loop approximation this was carried out, for $m=0$, in \cite{Paterson:1980fc}. Much information
can already be obtained from the beta functions of the couplings in $d=4-\e$ dimensions: if there
is an infrared stable fixed point, zero of the beta function away from $g_1=g_2=0$, the transition
probably is of second order. If the couplings run to infinity, the transition is of first order.
In the computation of \cite{Paterson:1980fc}, the color and hence the value of $N_c$ is hidden in the
color contraction in $\langle \bar q q\rangle$.
Opening up these color interactions in the 1-loop computation in
full is an impossible task, but in the large $N_c$ limit a single $\tr$ is always one quark
loop and thus suppressed by a factor $1/N_c$. Thus we expect that in the above effective potential
$g_1\sim 1/N_c^2$ and $g_2\sim 1/N_c$.

According to \cite{Paterson:1980fc} the $\beta$-functions of the two couplings in \nr{effact}
(scaled by a factor $\pi^2/3$) have a fixed point at
\be
N_c^2g_1^*={3\e\over x_f^2},\quad g_2^* =0
\ee
with the eigenvalues $\e,-\e$ so that the fixed point is unstable, the flows are plotted in
\cite{Paterson:1980fc}. This is true also at large $N_c,\,N_f$, indicating a first order transition
in this limit, too. However, one may argue that when $N_c=\infty$, the term with $g_1$ in
\nr{effact} should be entirely neglected. Then only the $\beta$ function for $g_2$ remains and
it has an infrared stable fixed point at
\be
N_cg_2^*={3\e\over 2x_f}.
\ee
This indicates a 2nd order transition. The two arguments are compatible if the latent heat
of the 1st order transition is $\sim 1/N_c$. Another way to say this is  that as $N_c\to\infty$ and $g_1/g_2\to 0$, the Hermitian model becomes equivalent to the $O(2N_f^2)$ model that is known to have a second order phase transition.

One should also remember that the $\e$ expansion cannot give any definite answer. A good
example of this is another standard model transition, the electroweak phase transition.
There the $\e$ expansion method also leads to a first order transition \cite{Arnold:1993bq} while
a numerical computation leads to a first order transition for small Higgs masses, $m_H \lesssim 75$ GeV,
while at larger Higgs masses there is only a cross over \cite{Kajantie:1996mn}.

\begin{figure}[!b]

\centering

\includegraphics[width=0.49\textwidth]{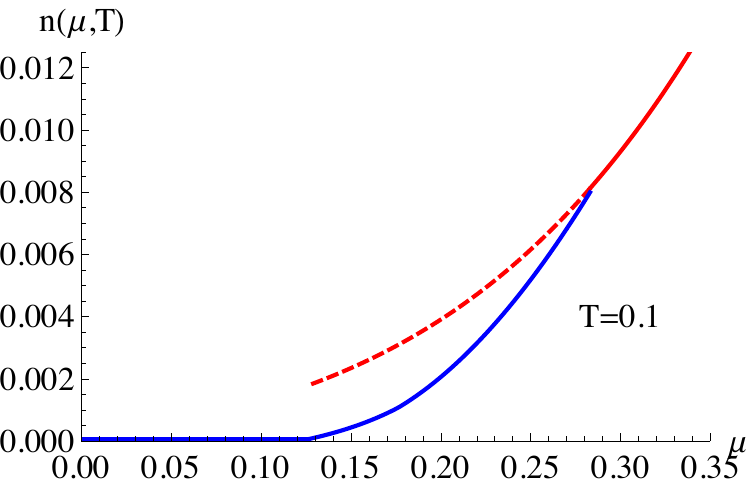}\hfill
\includegraphics[width=0.49\textwidth]{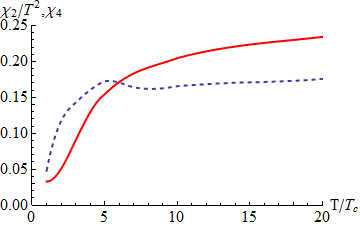}

\caption{\small Left: A plot of the quark number density $n(\mu;T)$
(without the normalisation factor $4/(45\pi)$ in \protect\nr{SBnorm}) for
$T=0.1$. 
At large $\mu$ the stable phase is always the symmetric phase (solid red) which
can exist as a metastable phase (dotted red) even below the transition.
Below the chiral transition the broken phase is stable (blue) and at
the lowest $\mu$s the stable phase is the thermal gas phase with $n=0$.
Above $T_c=T_\chi(0)$ only the symmetric phase exists.
Right: The quark number susceptibilities $\chi_2$ (continuous) and $\chi_4$ (dashed) per $N_c^2$
including the normalisation factor $4/(45\pi)$.
The limit of $\chi_2/T^2$ at large $\mu$ is $\fra13$, that of $\chi_4$ is $2/\pi^2$.
}
\label{nmu}
\end{figure}

\subsection{Quark number density}
The quark number density as a function of $\mu$ is plotted in Fig.~\ref{nmu}
for $T=0.1$, as a representative value.
As expected, 
$n(\mu)$ is continuous as the transition is of 2nd order.
When plotted for
$T>T_\chi(0)$, the fixed-$T$ curves contain just the symmetric phase
with monotonically increasing $n(\mu)$.

It is common to characterize the $\mu$ dependence by the susceptibilities at
$\mu=0$:
\be
\chi_2(T)={\partial^2 p(\mu,T)\over\partial\mu^2}\vert_{\mu=0},\quad
\chi_4(T)={\partial^4 p(\mu,T)\over\partial\mu^4}\vert_{\mu=0}
\la{susk}
\ee
In the ideal gas limit and for $\CL_A=1$, $\chi_2\to\fra{x_f}3N_c^2T^2$, $\chi_4\to\fra{2x_f}{\pi^2}N_c^2$ .
These (per $N_c^2$) are also plotted in
Fig.~\ref{nmu}, now also including the normalisation term (see Appendix~\ref{appD}).
One is approaching
the ideal gas limit but rather slowly.


\subsection{Polyakov line}
The basic difficulty in the study of thermodynamic deconfinement is that there is no symmetry
and thus no order parameter associated with deconfinement. For $N_f=0$ the Polyakov line,
trace of path ordered exponential of $A_0$ over the periodicity range $0,\,1/T$ in imaginary
time, signals breaking of Z($N_c$) symmetry and separates low and high $T$ phases. Even though
it is not an order parameter for finite $N_f$, it is a gauge invariant measurable observable,
which in lattice Monte Carlo studies varies together with the chiral condensate $\langle \bar q q\rangle$.

Constructing the gravity dual of the Polyakov line is very complicated, but we can model it
in a simple way following \cite{Noronha:2009ud}. The idea is to start from a duality determination
of a string tension in a thermal ensemble, interpret this as $dF/dz=dF/dT\times dT/dz$, compute from
here $dF/dT$, integrate $F(T)$ by choosing the integration constant by physical arguments and
finally plotting $L=\exp(-F(T)/T)$. Here one can start from the determination of the spatial
string tension $\sigma_s$ \cite{Alanen:2009ej}, determined from Wilson loops with sides in spatial directions,
\be
\sigma_s={1\over 2\pi \alpha'}\,{b_h^2\over\f_h}={dF\over dz_h}={dT\over dz_h}\,{dF\over dT}.
\ee
Apart from the string tension all quantities here are known and the evaluation, using \nr{qh}, gives
\be
F'(T,\mu)=-{2\over\alpha'}\,{1\over \f(\l_h)\veff(\l_h,\tau_h)c_s^2(\l_h,\tau_h)}.
\la{pol}
\ee
On the RHS $\l_h,\tau_h$ are functions of $T,\mu$ on the LHS, for the symmetric
phase $\tau_h=0$. In the symmetric phase at large $T$,
\be
F'(T)\to-{\LUV^2\over 2\alpha'}.
\ee

\begin{figure}[!tb]

\centering

\includegraphics[width=0.49\textwidth]{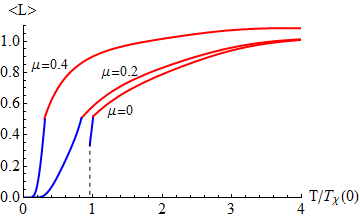}

\caption{\small The model Polyakov line plotted using \protect\nr{pol} for $\mu=0,\,0.2,\,0.4$. The red curves
are for the symmetric and the blue ones (at smaller $T$) for the broken phase. The dashed line
indicates the discontinuity at $T_h$.
}
\label{polfig}
\end{figure}

In the symmetric phase we shall, somewhat arbitrarily, fix the integration constant
in the integration of $F'(T)$ so that $F_s(T_\chi(\mu))=0$. Normalising $\langle L\rangle$ to 1 at
large $T$ (actually, due to limitations of numerics, at $T=10T_c$) one obtains the red large $T$ curves
in Fig.~\ref{polfig} for $\mu=0,\,0.2,\,0.4$. For the broken phase we shall
enforce continuity at $T_\chi(\mu)$ by demanding that also $F_b(T_\chi(\mu))=0$. This is
reasonable for the 2nd order transition which is the case here.
The discontinuity shown for $\mu=0$ is due to the transition at $T_h$.

It would be very valuable to derive a theoretically better founded gravity dual for
the Polyakov line. There is no order parameter for deconfinement but this operator
anyway is and will be used in lattice Monte Carlo studies.

\subsection{Sound speed}
Now that we have two thermodynamic variables we have three second derivatives of
$p(T,\mu)$. Out of the standard quantities $C_V$ and $C_p$ are complicated to
compute in the present framework, but it so happens that the formula for
the sound speed squared
\be
c_s^2={dp\over d\epsilon}={s\,dT+n\,d\mu\over T\,ds+\mu\,dn}={b_h\,[T'(\l_h)+\fra{\nt}{4\pi}\mu'(\l_h)]
\over 3b_h'(\l_h)(T+\fra{\nt}{4\pi}\mu)},
\la{svel}
\ee
where all quantities are to be taken at fixed $\nt$, can be directly evaluated.
Rather fortunately, our method
of computation makes it trivial to take into account the extra condition among the fluctuations
of pressure and energy density in \nr{svel}, they are to be taken at fixed $n/s$ and, due
to \nr{physn} this is just fixed $\nt$. In particular, for $\nt=0$ the derivative should be
taken in the direction of $T$, as is usually done, though in this direction the volume
density of entropy varies.

\begin{figure}[!tb]

\centering

\includegraphics[width=0.49\textwidth]{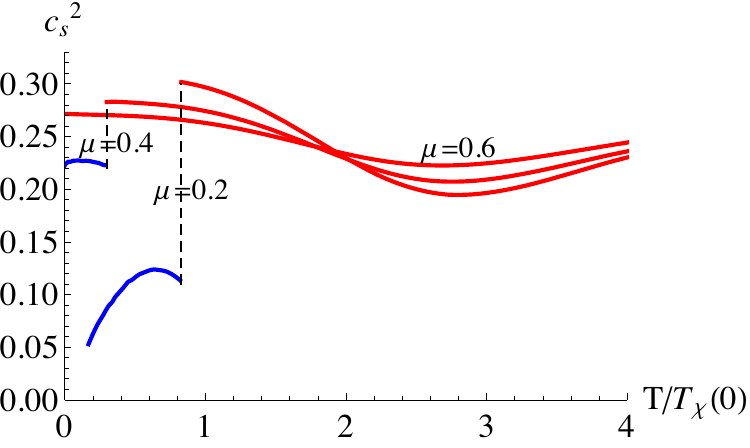}

\caption{\small Sound speed squared plotted vs $T/T_\chi$ at fixed $\mu$.
At small $T$ and small $\mu$ 
sound velocity drops markedly, the more the smaller $\mu$ is.
At large $T$ in  the symmetric case $c_s^2$ at large $T$ approaches $1/3$.
}
\label{soundvel}
\end{figure}

The most interesting region is that near the phase transitions.
Fig.~\ref{soundvel} shows $c_s^2$ plotted vs $T/T_\chi(0)$
(numerically $T_\chi(0)=0.1484$) at $\mu=0.2$, $0.4$ and $0.6$, i.e. as
one moves vertically in the $T$ direction in Fig.~\ref{Tchi}. At very large $T$, outside
the figure, $c_s^2$ approaches the conformal value $1/3$. For $\mu=0.6$ one does not
cross any phase transition and $c_s^2$ approaches some fixed value at $T=0$.


\subsection{The $T\to0$ limit}
The $T\to 0$ limit would be particularly interesting. The end point, $T=0$,
can be reached by putting $f=1$ in the metric Ansatz \cite{Arean:2013tja}. 
In this limit one can determine, for example, the lowest states of the mass
spectrum of the theory. We have tried to approach $T=0$ by using the
thermal Ansatz by trying to make $T \sim f^\prime (z_h)$ as small as
possible. This clearly leads to large numerical fluctuations. The
overall structure of the phase diagram along the $T=0$ axis
should nevertheless be as shown in Fig.~\ref{Tchi}, for the potentials
used here. We are still, for example, missing the
constant $T$ and $\mu$ curves on the $(\lambda_h, \nt)$-plane in the chirally
broken phase. With these one could more reliable conclude what
$p(T\to0,\mu)$ is in both symmetric and broken phases.

\section{Conclusions}

We have analysed  in this paper a non-fine-tuned gauge/gravity duality model for hot and dense QCD in the
limit of large number of colors and flavors. The model contains 5-dimensional gravity with
\ads{5} symmetry on the boundary,
a scalar dilaton for confinement and asymptotic freedom, a
scalar tachyon for quark mass and condensate  and the zeroth component of a bulk 4-vector for chemical
potential and quark number density. The potentials of the model are constructed so that one obtains the correct
QCD beta function and mass running in the weak coupling region and color confinement in the strong
coupling region.

Tuning the quark mass to
zero, the main result of this paper is the phase diagram and a description of dynamical chiral symmetry breaking when
temperature or density is decreased. Chiral symmetry corresponds to solutions with vanishing
tachyon, which automatically leads to the vanishing of both $m_q$ and the condensate. Spontaneous chiral symmetry breaking
corresponds to solutions with non-zero tachyon, which are constructed such that $m_q=0$ but
nevertheless the condensate is nonvanishing.
By explicit calculation of the pressures of chirally symmetric
and broken phases we find that the broken one dominates at small temperature and chemical
potential, $T<T_\chi(\mu)$. The transition in between is of second order.

When $T$ is further decreased below $T_\chi(\mu)$,  the system ultimately
goes to another phase at some $T_h(\mu)$ with a non-zero tachyon but without black holes: the thermal gas phase.
We use this as a model for the low $T$ hadron phase.
In lattice Monte Carlo simulations one normally finds that the chiral transition (as identified
by variation of the quark condensate) and deconfinement transition (as identified by the Polyakov
line or energy density discontinuities) coincide;
there is effectively just one transition line between a quark-gluon plasma phase and a
hadronic phase. Here we find that these lines are separated.

The numerical effort needed to obtain the results presented here is extensive and we have thus
limited ourselves to a quantitative study of one set of potentials and the case $N_f=N_c$.
With improved numerical techniques many further questions can be addressed. 
A set of potentials 
which describes all $T=0$ 
QCD physics in quantitative detail has been identified \cite{Arean:2012mq,Arean:2013tja}. 
Computing also its thermodynamics with good accuracy would make it possible to correlate
zero and finite $T$ properties reliably. For example, how does the requirement of linear
Regge trajectories in particle spectra affect the thermodynamics? How does the quark
condensate behave as a function of density? Further, the approach
to the conformal limit at $x_f=x_c\approx4$ has been computed for $\mu=0$ in \cite{Alho:2012mh}
and it will be interesting to study also the $T,\mu$ phase diagram in this limit.

There is a surprise in the phase diagram at low temperatures. Our analysis suggests that 
there is a new quantum critical 
regime with exotic properties at $T=0$ which realizes the symmetries 
of the associated geometry, AdS$_2\times {\mathbbm{R}}^3$.
This exists both on the $T=0$  segment of the chirality breaking plasma as well as 
the $T=0$ line of the chirally symmetric plasma.

The presence of the AdS$_2\times {\mathbbm{R}}^3$ geometry in the holographic solution 
indicates that there is a scaling symmetry of the time direction which does not act in the spatial directions. 
Such symmetries have been called semilocal.
This is an unexpected symmetry in a theory at finite density, but it is natural and 
generic in the holographic context \cite{Gouteraux:2012yr} and appears even for simple 
black holes like the Reissner-Nordstr\"om black hole \cite{Liu:2009dm}.
The physics in this critical regime is similar to that of a theory with zero speed of light: 
all spatial points decouple in the IR.

The local RG pattern of such \ads{2} solutions is fully compatible with the phase diagram we derived. 
It is an interesting question to determine physical implications of this scaling regime as 
well as its potential experimental signatures.

\section{Acknowledgments}

We thank Blaise Gout\'eraux, Misha Stephanov and Aleksi Vuorinen for discussions.
This work was partially supported by European Union's Seventh Framework Programme 
under grant agreements (FP7-REGPOT-2012-2013-1) No 316165, PIF-GA-
2011-300984, the ERC Advanced Grant BSMOXFORD 228169, the EU program
Thales MIS 375734 and was also co-financed by the European Union (European Social
Fund, ESF) and Greek national funds through the Operational Program ``Education
and Lifelong Learning'' of the National Strategic Reference Framework (NSRF)
under ``Funding of proposals that have received a positive evaluation in the 3rd and
4th Call of ERC Grant Schemes''. TA thanks the Vaisala foundation for financial support. 
KT acknowledges financial support from the Academy of Finland project 267842.

\appendix

\vspace{2cm}
\section{Fluctuation modes around the \ads{2} point} \label{app:ads2flucts}

In this appendix we compute explicitly the amplitudes for the fluctuations of the \ads{2} region discussed in Sec.~\ref{sec:ads2stability}.
The amplitudes are easily obtained by solving the linear system provided by a given $\alpha^*$, and in general depend on one undetermined (but non-vanishing) amplitude and a choice of radial gauge which can be fixed via $B_1$. For the V-QCD \ads{2} fixed point, the fluctuations are as follows:

\subsubsection*{\underline{$\alpha^* = -2$}}
This is a relevant perturbation that couples the metric and the gauge field like
\begin{equation}
D_1 \ne 0 \qquad \qquad \Phi_1 = -\frac{\mathcal{E}}{2}(D_1+B_1)\qquad \mathrm{and} \qquad \lambda_1 = \tau_1 = C_1 = 0
\end{equation}
This mode corresponds to one sort of finite temperature perturbation to \ads{2}.  It is the \ads{2} black hole studied in \cite{Liu:2009dm}.

\subsubsection*{\underline{$\alpha^* = -1$}}
There exists another finite temperature perturbation, which is again characterized by a relevant mode. In this case the perturbation has
\begin{equation}
\Phi_1,\,D_1 \ne 0 \qquad \mathrm{and}\qquad  B_1 = -D_1\qquad \mathrm{and} \qquad \lambda_1 = \tau_1 = C_1 = 0
\end{equation}
which corresponds to a shift in the chemical potential
\begin{equation}
\Phi(r) = \mu + r\left(\mathcal{E}+\frac{\Phi_1}{r}\right)
\end{equation}
and a metric of the familiar form
\begin{equation}\label{eq:AdS2BH2}
\dd s^2 = -\frac{r^2}{L_2^2}\left(1-\frac{D_1}{r}\right)\dd t^2+ L_2^2\frac{\dd r^2}{r^2\left(1-\frac{D_1}{r}\right)} + C_0\,\dd \vec{x}^2
\end{equation}
Again, this is a black hole in \ads{2},  related to the one obtained from the $\alpha^*=-2$ fluctuation by a radial coordinate transformation.

More specifically, under the transformation
\begin{equation}\label{eq:AdS2BHtrans}
r\to r' = \frac{1}{2}(D_1+\rho) \qquad \mathrm{and} \qquad t\to t' =2\tau
\end{equation}
the metric (\ref{eq:AdS2BH2}) becomes
\begin{equation}\label{eq:AdS2BH1}
\dd s^2 = -\frac{\rho^2}{L_2^2}\left(1-\frac{D_1'}{\rho^2}\right)\dd \tau^2+ L_2^2\frac{\dd \rho^2}{\rho^2\left(1-\frac{D_1'}{\rho^2}\right)} + C_0\,\dd \vec{x}^2
\end{equation}
where $D_1' = D_1{}^2$, which is the metric implied by the $\alpha^*=-2$ perturbation in the gauge $B_1' = -D_1'$. Note the rescaling of the time coordinate in the transformation (\ref{eq:AdS2BHtrans}). Because black holes in \ads{2} are coordinate equivalent to the vacuum \ads{2} solution \cite{Spradlin:1999bn}, the black hole (\ref{eq:AdS2BH1}) obtained via this coordinate change  lives at a different temperature $T$ than its parent solution (\ref{eq:AdS2BH2}), $T_\rho/T_r = 2$.

\subsubsection*{\underline{$\alpha^* = 0$}}
These are marginal modes corresponding to rescalings of space and time. They are described by
\begin{equation}
C_1,\, D_1\ne 0 \qquad \mathrm{and} \qquad \Phi_1 = \frac{\mathcal{E}}{2}D_1\qquad \mathrm{and} \qquad \lambda_1 = \tau_1 = 0
\end{equation}
 When $C_1\ne 0$, then the volume form on the $\mathbb{R}^3$ changes by a factor of
\begin{equation}
\mathrm{Vol}_{\mathbb{R}^3} \approx \mathrm{Vol}_{\mathbb{R}^3}^0\left(1+\frac{3}{2}\frac{C_1}{C_0}\right)
\end{equation}
and when  $D_1\ne 0$ then one obtains a shift in the time coordinate
\begin{equation}
t\to t'=\sqrt{1+D_1}\,t\qquad \mathrm{so} \qquad \dd t'\approx \left(1+\frac{1}{2}D_1\right)\dd t
\end{equation}
These are the conjugate modes to the $\alpha^*=-1$ finite temperature fluctuation above.

\subsubsection*{\underline{$\alpha^* = 1$}}
The last universal mode is the irrelevant perturbation conjugate to the $\alpha^* = -2$ mode. This perturbation couples all of the fields except for the tachyon, and is partially responsible for driving the system away from the \ads{2} fixed point. The amplitudes are somewhat complicated, but take the form
\begin{equation}\label{eq:irrVQCD}
\lambda_1 \ne 0 \qquad  C_1 = \Gamma_1 \qquad  D_1=\Delta_1\qquad\Phi_1 =  (\Gamma_1+\Delta_1-\frac{1}{2}B_1)\mathcal{E}\qquad \tau_1 = 0
\end{equation}
where
\begin{equation}\label{eq:Gamma1}
\Gamma_1 = \frac{2\lambda_1\left(16+3L_2^2\,\lambda_0^2\,\partial_\lambda^2V_{\mathrm{eff}}^0\right) w_0}{9\lambda_0^2\left[ 2(2-\mathcal{E}^2 w_0^2)\partial_\lambda w_0+ L_2^2\,\mathcal{E}^2\, w_0^3\sqrt{1-\mathcal{E}^2 w_0^2}\,\partial_\lambda V_f^0\right]}
\end{equation}
and
\begin{equation}\label{eq:Delta1}
\Delta_1 = B_1-\frac{1}{27}\left[\frac{2}{\lambda_0^2}\left(16+3L_2^2\,\lambda_0^2\,
\partial_\lambda^2V_{\mathrm{eff}}^0\right)\frac{\lambda_1^2}{\Gamma_1}+63\Gamma_1\right]+\mathcal{E}^2 w_0^2\,\Gamma_1
\end{equation}
That this mode interpolates between the IR and UV solutions is suggested by the fact that for $\lambda_1\ne0$ the spatial part of the metric acquires a non-trivial radial dependence as per (\ref{eq:Gamma1}).

 For the non-universal exponents, one finds a simple perturbation
 \subsubsection*{\underline{$\alpha^* = \alpha^\tau$}}
\begin{equation}
\tau_1 \ne 0 \qquad D_1 = \frac{2}{\mathcal{E}}\Phi_1 \qquad B_1 =\alpha^\tau\frac{2}{\mathcal{E}}\Phi_1\qquad  \mathrm{and} \qquad \lambda_1 =  C_1=0
\end{equation}
This is gauge equivalent to a mode consisting of only a tachyon fluctuation.

Finally, there exists a somewhat more complicated perturbation
\subsubsection*{\underline{$\alpha^* = \alpha^\lambda$}}
\begin{equation}
\lambda_1 \ne 0 \qquad  \Phi_1 = \frac{\mathcal{E}}{2}\left(\Delta_1'(1+\alpha^\lambda)-B_1\right)\qquad D_1 = \Delta_1' \qquad \mathrm{and} \qquad \tau_1 = C_1  = 0
\end{equation}
where
\begin{equation}
\Delta_1' = \frac{1}{\alpha^\lambda}B_1-\lambda_1\frac{(4-2\mathcal{E}^2w_0^2)\partial_\lambda w_0+ L_2^2\,\mathcal{E}^2\,w_0^3\sqrt{1-\mathcal{E}^2 w_0^2}\,\partial_\lambda V_f^0}{\alpha^\lambda(2+\alpha^\lambda) w_0}
\end{equation}
where $\alpha^\lambda$ is given by (\ref{eq:sebf}). This is a perturbation which couples the dilaton fluctuations to the metric fluctuations, leaving the spatial part of the metric unchanged.

Note that the above expressions for the perturbation amplitudes hold for generic values of the constant scalars $\lambda_0$ and $\tau_0$. In the special case of the divergent tachyon, many of the amplitudes simplify as in this case $\mathcal{E}^2w_0^2 = 1$. To wit, (\ref{eq:Gamma1}) becomes
\begin{equation}
\Gamma_1 = \frac{\lambda_1\left(16+3L_2^2\,\lambda_0^2\,\partial_\lambda^2V_{\mathrm{eff}}^0\right) w_0}
{9\lambda_0^2\partial_\lambda w_0}.
\end{equation}

\section{Numerical solution of the equations of motion}\la{scalings}
\subsection{Definitions}

The purpose of this appendix is to clarify the technical details of the numerical solutions 
to the equations of motion, the scaling properties of the results,
and their dimensional analysis in the $A = \ln b$ -coordinates. This is
essential for extracting the physics out of the numerics. All the details
are built in the numerical code \texttt{SolveFiniteTTachyons} deposited
in \cite{TSAcode}.

In this treatment,
the solutions in the A-coordinate system are considered primary, and the
z-system is just an auxiliary coordinate system used to relate the results
to known holographic formulae. The treatment extends \cite{Alho:2012mh},
but we shall rewrite it explicitly in the form that the actual numerical code
\cite{TSAcode} uses, and keep each stage of the equations dimensionally consistent.

We first define the notation: the fields $q_1(A_1), f_1(A_1), \lambda_1(A_1)$ and
$\tau_1(A_1)$ are the fields produced by numerical equation solving, expressed as a
function of the coordinate $A_1$. These will be referred to as \emph{level 1} solutions.
The level 1 coordinate $A_1$ is the one in which the numerics is defined, and the horizon
sits at $A_{1,h} = 0$.
{\emph Level 2} solutions are obtained after $f$-scaling (see next section) and \emph{level 3} solutions
\cite{Alho:2012mh} are the final ones with
the fields, observables and the coordinate in the units corresponding to the desired
UV boundary conditions.

We shall consider at first only V-QCD at $\mu = 0$, and then devote a separate section to
the $\mu \neq 0$ case.

We define the coordinate $z$ by
\be
\frac{dA}{dz} = \frac{e^A}{q(A)},
\ee
with the boundary condition $z(A = \infty) = 0$. Notice that this is defined with the
final scaled level 3 fields, and so we have precisely one system of $z$-coordinates,
which we never scale.

\subsection{The $f$-scaling}

We generally want the function $f$ to asymptote to 1 in the UV ($z\rightarrow 0$ or equivalently
$A_1 \rightarrow \infty$) in order to have the standard Minkowski coordinate system with
$c=1$ on the boundary. When the boundary conditions are set at the horizon, this is not
in general guaranteed. Fortunately the equations of motion are invariant under a
combined scaling of $f$ and $q$, such that if $f_1, q_1$ are solutions, then also the pair
\ba
f_2 &=& f_{\rm scale}^2 f_1\\
q_2 &=& f_{\rm scale} q_1,
\ea
with no change to the other fields or the coordinate $A_1 = A_2$, is a solution for
any value of $f_{\rm scale}$, although with different boundary conditions. Choosing
$f_{\rm scale} = 1/\sqrt{f(A_1 = \infty)}$ gives us the desired solution. From now on,
fields and coordinates with the subscript 2 denote the numerical solutions scaled in
such a way. We will call these the \emph{level 2} solutions.

In \cite{TSAcode}, this scale factor appears as \texttt{fscale} and is explicitly
used in generating the scaled solutions. The solutions produced by
\texttt{SolveFiniteTTachyons} are level 1 in this notation, whereas
\texttt{SolveAndScaleFiniteTTachyons} produces func\-tions that are level 2
in this notation. The scaling itself is car\-ried out in \texttt{ScaleSolution},
which can also be used to convert the level 1 solutions produced by
\texttt{SolveFiniteTTachyons} to level 2 solutions.

The solution produced by this scaling no longer corresponds to the initial conditions
set in the numerics. Specifically, if the original equation solver was started with the condition
\ba
q_1(A_{1,h}) &=& q_{1,h}\\
f_1'(A_{1,h}) &=& f'_{1,h}
\ea
then the new solution corresponds to
\ba
q_2(A_{2,h}) &=& q_{2,h} f_{\rm scale}\label{eq:qhorizon}\\
f'_2(A_{2,h}) &=& f'_{2,h} f_{\rm scale}^2\label{eq:fphorizon},
\ea
with the initial conditions for the other fields unchanged.

The code \cite{TSAcode} sets the initial conditions
\footnote{We present here already the formula with the chemical potential included,
for $\mu = 0$ set $\nt = 0$. See section \ref{sec:chempot} for further explanation.}
\ba
f'_{1,h} &=& 1\\
q_{1,h} &=& -\frac{\sqrt{3}}{\sqrt{V_g(\lambda_h) - V_f(\lambda_h, \tau_h)
\sqrt{1 + \frac{\tilde{n}^2}{\CL_A^4\kappa(\lambda_h)^2 V_f(\lambda_h, \tau_h)^2}}}},
\ea
where the first is chosen arbitrarily, since the magnitude of $f$ is anyway set by $f$-scaling,
and the second was derived in \nr{qh}.
The post scaling boundary condition then simply is
\ba
f'_{2,h} &=& f_{\rm scale}^2\\
q_{2,h} &=& -\frac{\sqrt{3}f_{\rm scale}}{\sqrt{V_g(\lambda_h) -
V_f(\lambda_h, \tau_h) \sqrt{1 + \frac{\tilde{n}^2}{\CL_A^4\kappa(\lambda_h)^2 V_f(\lambda_h, \tau_h)^2}}}},
\ea
with the rest of the fields unchanged.

\subsection{The $\Lambda$-scaling}

The UV -expansion (see \nr{Alim}  and Appendix A in \cite{Alho:2012mh}) is
\ba
A_2 &=& \hat{A}_0 +  \frac{1}{b_0 \lambda_2(A_2)} + \frac{b_1}{b_0^2}
\ln(b_0 \lambda_2(A_2)) + \CO(\lambda),\label{eq:AUV}\\
\lambda_2(z) &=& -\frac{1}{b_0 \ln(z \Lambda)} + \CO\left(\frac{\ln(-\ln(z \Lambda))}
{\ln(z \Lambda)^2}\right),\label{eq:lambdaUV}\\
A_2(z) &=& -\ln\frac{z}{\LUV} + \CO\left(\frac{1}{\ln(z \Lambda)}\right),\label{eq:AUVz}
\ea
where $\hat{A}_0$ is a constant of integration. Here $\LUV$ is the asymptotic value of
$-q_2(A)$ at large A (or equivalently $A_2$).
Using these, we find
\be
\hat{A}_0 = \ln(\LUV \Lambda) = \lim_{A_2 \rightarrow \infty} (A_2 - \frac{1}{b_0 \lambda_2(A_2)}
+ \frac{b_1}{b_0^2} \ln(b_0 \lambda_2(A_2))).
\ee
Since we want to find a solution where $\Lambda = \Lambda_0$, we write this in the form
\be
\hat{A} = \hat{A}_0 - \ln{\LUV \Lambda_0} = \ln(\Lambda/\Lambda_0) =
\lim_{A_2 \rightarrow \infty} (A_2 - \ln(\LUV\Lambda_0) - \frac{1}{b_0 \lambda_2(A_2)} +
\frac{b_1}{b_0^2} \ln(b_0 \lambda_2(A_2))) \la{Alimit}
\ee
and observe that the solution with the shifted coordinate $A = A_2 - \hat{A}$ has
the required asymptotics.
Since the equations of motion are invariant with respect to shifts in $A$, this is also a
solution of the equations, although with different boundary conditions.
These are the level 3 solutions. We further denote $e^{\hat{A}} = \Lambda/\Lambda_0 \equiv \Lambda_{\rm scale}$.
This is the factor that appears in \cite{TSAcode} as $\Lambda$\texttt{scale},
although in some places it is (inaccurately) denoted as simply as $\Lambda$. This factor is dimensionless.

Using \nr{Alimit} converges somewhat slowly for practical purposes due to the $\CO(\lambda) = \CO(A^{-1})$ corrections. We speed up that convergence by considering $\hat{A}$ as a function of $A_{\rm max}$ as given by \nr{Alimit}, where $A_{\rm max}$ is the limit up to which the numerical solution has been computed. From the numerical process, we know the derivatives of the fields, so we can compute $\hat{A}'(A_{\rm max})$ and derive the formula
\be
\hat{A} = \hat{A}(A_{\rm max}) - \hat{A}'(A_{\rm max}) \frac{\lambda(A_{\rm max})}{\lambda'(A_{\rm max})},
\ee
which cancels the $\CO(\lambda)$ corrections. The value of $\hat{A}$ computed by this method is returned by \texttt{SolveAndScaleFiniteTTachyons}, which uses \texttt{ScaleSolution}
to scale the level 1 solutions to level 2 and to derive $\Lambda_{\rm scale}$.


We now write the transformation equations explicitly:
\ba
A(A_2) &=& A_2 - \hat{A},\\
A_2(A) &=& A + \hat{A},\\
A_h &=& A(A_{2,h}) =  -\hat{A}.
\ea
Especially note that the horizon value of, for example,
$\lambda(A_h) = \lambda_2(A_2(A_h))= \lambda_2(A_h + \hat{A}) = \lambda_2(-\hat{A} + \hat{A}) = \lambda_2(0)$.
In other words, the horizon value of any field $h$ in the solution with the correct asymptotics,
is the same as the value of the original function $h_1$ coming from the numerics, evaluated at $A_{1,h} = 0$.

The conformal factor of the metric appears in several physical observables. In level 3 coordinates it is simply
\be
b(A) = e^A = e^{A_2 - \hat{A}} = \frac{e^{A_2}}{\Lambda_{\rm scale}}.
\ee
In addition, the derivatives of fields in the $z$-coordinate system often play a role,
and we observe that for example
\ba
\frac{d(f(z))}{dz} &=& \frac{dA}{dz} \frac{df}{dA}\Big|_{A=A(z)} = \frac{e^A}{q(A)}
\frac{df}{dA}\Big|_{A=A(z)} = \frac{e^{A_2 - \hat{A}}}{q_2(A_2)}\frac{df_2}{dA_2}\Big|_{A_2 = A_2(A(z))}\\
&=& \frac{e^{A_2}}{\Lambda_{\rm scale} q_2(A_2)} \frac{df}{dA_2}\Big|_{A_2 = A(z) + \hat{A}}
\ea
and especially at the horizon
\be
\frac{d(f(z))}{dz}\Big|_{z=z_h} =  \frac{1}{\Lambda_{\rm scale} q_2(0)}
\frac{df}{dA_2}\Big|_{A_2 = 0}\label{eq:dzhorizon}.
\ee
An identical result holds for any field.

Since it is possible in this way to eliminate the need to explicitly shift the fields,
and thus the need to keep track of one extra variable, the actual numerical code
\cite{TSAcode} does precisely this. In the code $A$ always refers to $A_2 = A_1$,
the horizon is always at $A_2 = 0$, and the fields used to compute the physical observables are level 2.

Since the equations of motion are invariant under shifts of $A$ without any corresponding
change in the fields, the initial conditions for the fields themselves at horizon when
expressed in terms of the $A$ -coordinates are not changed by this scaling.
The relation between $A$ and $z$ is what changes. Note however that once we introduce
the gauge field $\Azs$ corresponding to a chemical potential, this changes since $\Azs$
explicitly breaks this shift invariance. We will return to that later.

\subsection{Physical observables at $\mu = 0$}

Using the previous results, we can work out the formulas for physical quantities used in the code.
The temperature is
\ba
4\pi T &=& -\frac{df}{dz}\Big|_{z = z_h} =
-\frac{1}{\Lambda_{\rm scale} q_2(A_2)} \frac{df}{dA_2}\Big|_{A_2 = 0} =
-\frac{f'_{2,h}}{\Lambda_{\rm scale} q_2(0)} \nonumber\\
&=& \frac{f_{\rm scale}}{\sqrt{3}\Lambda_{\rm scale}} \sqrt{V_g(\lambda_h) -
V_f(\lambda_h, \tau_h) \sqrt{1 + \frac{\tilde{n}^2}{\CL_A^4\kappa(\lambda_h)^2 V_f(\lambda_h, \tau_h)^2}}},
\ea
where we used (\ref{eq:dzhorizon}), (\ref{eq:fphorizon}) and (\ref{eq:qhorizon}) . In the code, $T$ is returned by
\texttt{TemperatureFromSols}.
Note that both $f_{\rm scale}$ and $\Lambda_{\rm scale}$ are dimensionless, with the
function $q$ carrying one dimension of length, giving the correct unit of 1/length = energy.

Also note that from (\ref{eq:qhorizon}) one sees that the unit of length in $q$ ultimately
comes from the potential, which is proportional to $1/\LUV^2$. This shows that
$\Lambda_{\rm scale}$ is the dimensionless factor which tells the relation between
the 4D boundary units and the units of the potential.

The entropy density comes from
\be
4 G_5 s = b(A_h)^3 = e^{3 A_h} = e^{-3 \hat{A}} = \frac{1}{\Lambda_{\rm scale}^3}.\label{eq:entropy}
\ee
This is returned in the code by \texttt{s4G5FromSols}.
Note that $b(A_h)$ is dimensionless, so the entropy density picks up its units
from the gravitational constant $G_5$.

The quark mass is expressed as
\ba
\tau(z)/\CL_{\rmi UV} &=& m_q(-\ln(\Lambda_0 z))^{-\gamma_0/b_0} z (1 + \CO(1/\ln z )) \nonumber \\
&=& - m_q (A_2 - \ln(\Lambda_0 \LUV))^{-\gamma_0/b_0} q_2(A_2) \Lambda_{\rm scale} e^{-A_2}(1 + \CO(A^{-1})) \nonumber \\
\Rightarrow m_q &=& \lim_{A_2 \rightarrow \infty}\CL_{\rmi UV}^{-2} \tau(A_2) e^{A_2} (A_2 - \ln(\Lambda_0 \LUV))^{\gamma_0/b_0}\frac{1}{ \Lambda_{\rm scale}} \nonumber \\
&\approx& \CL_{\rmi UV}^{-2} \tau(A_{\rm max}) e^{A_{\rm max}} (A_{\rm max}- \ln(\Lambda_0 \LUV))^{\gamma_0/b_0}\frac{1}{ \Lambda_{\rm scale}},
\ea
where $A_{\rm max}$ is the maximum $A$ to which the equations of motion have been solved. Except for the appearance of  $\Lambda_{\rm scale}$, the shift between $A$ and $A_2$ is $\CO(A^{-1}) = \CO(A_2^{-1})$ for large $A$.

Similarly as with the determination of $\hat{A}$, the $A^{-1}$ corrections to $m_q$ are rather large at easily reachable values of $A_{\rm max}$. As before, we can consider $m_q$ as a function of $A_{\rm max}$ and take its value at another point $A_b < A_{\rm max}$. Using from the above that $m_q(A) = m_q(1 + kA^{-1})$ for some unknown coefficient $k$, we can cancel the $\CO(A^{-1})$ corrections:
\be
m_q = \frac{m_q(A_{\rm max}) A_{\rm max} - m_q(A_b) A_b}{A_{\rm max} - A_b}(1 + \CO(A^{-2})).\la{mqfinitediff}
\ee
Since we know the derivatives of the fields from the numerical process, we can go further and take the limit $A_b \rightarrow A_{\rm max}$, yielding
\be
m_q = m_q(A_{\rm max}) + m_q'(A_{\rm max}) A_{\rm max}.
\ee
In practice the finite difference method of \nr{mqfinitediff} is slightly more stable and converges only very slightly slower, and that method is therefore used in the code by default. The function \texttt{QuarkMass} computes the mass from the solutions with this method.

\subsection{Chemical potential}
\label{sec:chempot}

In \nr{lagrang}, we introduce a zero component $\Azs$ of a gauge vector field in the bulk
to model a chemical potential
in the boundary theory. It turns out that the UV asymptotics are not affected by this addition,
and so we will want to do similar scalings as in the zero chemical potential case.

However, the full structure of the solution with respect to scaling the UV-variables does change,
since there are new terms in the equations of motion, of the form
\be
f^{\prime\prime}+ (4-\frac{q^\prime}{q})f^{\prime} - V_f \frac{\CL_A^4 \kappa^2 e^{-2A} \Azs^{\prime 2}}
{\sqrt{1+\frac{f \kappa}{q^2}\tau'^2 - \frac{\kappa^2}{q^2}e^{-2A}\CL_A^4 \Azs^{\prime 2} }} = 0,
\ee
where we have written \nr{A3} without substituting the solution of the $\Azs$ equation of motion.

If we have a solution with subscripts 1, including $\Azs_{1}$ which has as yet undefined
transformation properties, we have
\ba
0 &=& f_1^{\prime\prime} + (4-\frac{q_1^\prime}{q_1})f_1^{\prime} - V_f \frac{\CL_A^4 \kappa^2 e^{-2A_1}
\Azs_{1}^{\prime 2}}{\sqrt{1+\frac{f_1 \kappa}{q_1^2}\tau_1^{\prime 2} -
\frac{\kappa^2}{q_1^2}e^{-2A_1}\CL_A^4 \Azs_{ 1}^{\prime 2} }}\\
&=& f_{\rm scale}^{-2} f_2^{\prime\prime} + (4-\frac{q_2^\prime}{q_2})f_{\rm scale}^{-2} f_2^{\prime} -
V_f \frac{\CL_A^4 \kappa^2 e^{-2A_2} \Azs_{1}^{\prime 2}}{\sqrt{1+\frac{f_2 \kappa}{q_2^2}\tau_2^{\prime 2} -
\frac{\kappa^2 f_{\rm scale}^2}{ q_2^2}e^{-2A_2}\CL_A^4 \Azs_{ 1}^{\prime 2} }} \\
&=& f_{\rm scale}^{-2} f^{\prime\prime} + (4-\frac{q^\prime}{q})f_{\rm scale}^{-2} f^{\prime} -
V_f \frac{\CL_A^4 \kappa^2 e^{-2A - 2\hat{A}} \Azs_{1}^{\prime 2}}{\sqrt{1+\frac{f\kappa}{q^2}\tau^{\prime 2} -
\frac{\kappa^2 f_{\rm scale}^2 }{ q^2}e^{-2A - 2\hat{A}}\CL_A^4 \Azs_{1}^{\prime 2} }}.
\ea
From this we conclude that if $f_1, q_1, \lambda_1, \tau_1, \Azs_{1}$ solve the equations
of motion with UV asymptotics corresponding to the level 1 fields, then the corresponding
level 3 functions solve the equations of motion with the correct UV asymptotics, if the
function $\Azs_{1}$ is replaced with $\Azs$, such that
\be
\Azs(A) = e^{-\hat{A}}f_{\rm scale} \Azs_{ 1}(A_1) =
\frac{f_{\rm scale}}{\Lambda_{\rm scale}} \Azs_{1}(A_1).\label{eq:A0scale}
\ee
It is apparent by inspection that the rest of the equations of motion are also
invariant under this substitution. We naturally call $\Azs(A)$ a level 3 gauge field.

The addition of a new field of course also adds a new pair of initial conditions.
The fundamental physical constraint \nr{A0cond} in $A$-coordinates requires that
we set $\Azs_{h} = 0$, so the remaining initial condition is
determined by the derivative of $\Azs$ at the horizon, $\Azs_{h}^\prime$.
Now given a level 1 solution, corresponding to the initial condition $\Azs_{1,h}$,
the scaled solution clearly corresponds to the initial condition
\be
\Azs_{h} = \frac{ f_{\rm scale}}{\Lambda_{\rm scale}} \Azs_{1,h}.
\ee

The field $\Azs$ is a cyclic coordinate: its equation of motion is
\be
\frac{d}{dA}{\partial L_f\over\partial \Azs^\prime}= \frac{d}{dA}{-\CL_A^4 \frac{e^{2A}}{q} V_f\f^2 \Azs^\prime\over
\sqrt{1+{f\f\over q^2} \tau^{\prime 2}-{\f^2\over e^{2 A} q^2}\CL_A^4 \Azs^{\prime 2}}}=0, \label{eq:Adef}
\ee
which we can immediately integrate to the form \nr{A0eom}:
\be
{-\CL_A^4 e^{2A} V_f\f^2 \Azs^\prime\over q
\sqrt{1+{f\f\over q^2} \tau^{\prime 2}-{\kappa^2\over e^{2 A} q^2}\CL_A^4 \Azs^{\prime 2}}}
=\hat n.\label{eq:Aintdef}
\ee

Different values of $\hat n$ correspond to different initial conditions for the $\Azs$ field.
Evaluating this at the horizon for a given solution or a set of initial conditions gives
us the value of $\hat n$ corresponding to that solution. Specifically, using the standard
boundary conditions for starting the numerics we have in terms of the level 1 solution
\be
{-\CL_A^4 V_{f,h} \kappa_h \Azs_{1,h}^\prime \over q_{1,h}
\sqrt{1 - \frac{\kappa_h^2}{q_{1,h}^2} \CL_A^4 \Azs_{1,h}^{\prime 2}}} = \hat{n}_1.
\ee
On the other hand, applying known scaling properties of the fields to the lhs
of the same expression for the level 3 solution leads to
\ba
\hat{n} &=& {-\CL_A^4 e^{2A_h} V_f\f^2 \Phi^\prime_{h}\over q_h
\sqrt{1-{\kappa_h^2\over e^{2 A_h} q_h^2}\CL_A^4 \Phi_h^{\prime 2}}} \\
&=& {-\CL_A^4 V_{f,h} \kappa_h \frac{1}{\Lambda_{\rm scale}^2} \frac{f_{\rm scale}}{\Lambda_{\rm scale}}
\Azs_{1,h}^\prime\over f_{\rm scale} q_{1,h} \sqrt{1-\frac{\kappa^2}{f_{\rm scale}^2}\Lambda_{\rm scale}^2
\frac{f_{\rm scale}^2}{\Lambda_{\rm scale}^2 }\CL_A^4 \Azs_{1,h}^2}}\\
&=& \frac{\hat{n}_1}{\Lambda_{\rm scale}^3}.
\ea
Since the level 3 solutions were the final ones, this gives us the scaling property of $\hat{n}$.

We can solve (\ref{eq:Aintdef}) to yield an explicit expression for $\Azs^\prime$ in terms of
$\hat{n}$ and the other fields (for $\dot \Azs$, see \nr{A0value}):
\be
\CL_A^2 \Azs^\prime(A) = - \frac{e^A q}{\kappa} \sqrt{\left(1 + \frac{f \kappa}{q^2} \tau^{\prime 2}\right)
\left[\frac{\frac{\hat n^2}{\CL_A^4}}{\frac{\hat n^2}{\CL_A^4} + e^{6 A} V_f^2 \kappa^2}\right]}.
\ee
Since $\Azs$ appears in the equations of motion always in the combination $\CL_A^2 \Azs^\prime$,
we could 
entirely eliminate the choice of $\CL_A$ at this stage by
rescaling $\hn \to \hn \CL_A^2$. Therefore we can set $\CL_A=1$ without loss of generality in the numerics.
Plugging the resulting formula into the equations of motion gives us the equations
\nr{A1}-\nr{A4} on which \cite{TSAcode} is based on.
Solving the highest derivatives from those leads to the form in \texttt{TachyonEquationsOfMotion}.
Since $A_{1,h}=0$, $\hn_1$ matches with the scale invariant quantity $\nt$ of~\eqref{defnt}:
\be
 \hn_1 = \nt = \nt_1.
\ee

With this substitution and using the scaling properties it is apparent  that, once the
equations have been solved and subjected to the $f$ -scaling to yield
$\lambda_2, f_2, \tau_2$ and $q_2$, we can write $\Azs^\prime$ as
\ba
&&\CL_A^2 \Azs_{2}^\prime(A_2) = -\frac{e^{A_2}  q_2(A_2)}{\Lambda_{\rm scale} \kappa(\lambda_2(A_2))} \\\nonumber
&&\times \sqrt{\left(1 + \frac{f_2(A_2) \kappa(\lambda_2(A_2))}{q_2(A_2)^2}
\tau^{\prime 2}_2(A_2)\right)\left[\frac{\nt^2}{\nt^2 + \CL_A^4e^{6 A_2} V_f(\lambda_2(A_2), \tau_2(A_2))^2
\kappa(\lambda_2(A_2))^2}\right]}.
\ea
This function is returned
in the code by \texttt{APrimeFromSols}.
It is the fully scaled form,
but expressed as a function of the coordinate $A_2$, which has not been shifted, i.e.
it is in the same coordinate system as all the other functions returned by the code.
It is level 2 in the same sense as the rest of the level 2 functions: the coordinate system
is such that the horizon is at zero, but the units are such that it needs no further factors
of $\Lambda_{\rm scale}$. It can be used fully consistently with all the other output functions,
but note that if for some reason the coordinate system would be shifted again, $\Azs^\prime$
would then be scaled again according to (\ref{eq:A0scale}).

\subsection{Physical observables for $\mu \neq 0$}

When $\mu \neq 0$ we immediately have two new physical observables.
First there is the quark number density,
\be
\hn = \frac{\nt}{\Lambda_{\rm scale}^3 },
\la{enph}
\ee
returned in the code by \texttt{nFromSols}. The relation to the physical quark number density $n$ is
given in the text in Eq.~\nr{defn}. 

An interesting point is that the dependence on the actual numerical solution is precisely the same as
for the entropy density $s$ in (\ref{eq:entropy}). Thus one has a physical interpretation
for the input parameter $\nt$, it is simply $\nt = 4\pi n(\lambda_h; \nt)/s(\lambda_h; \nt)$,
where one also inserted the constants given in \nr{physn}.

The other observable is of course the chemical potential itself. The holographic formula for it is
\be
\mu = \lim_{A\rightarrow \infty} \Azs(A),
\ee
for which we need to integrate (\ref{eq:Aintdef}). The correct boundary condition is that $\Azs(A_h) = 0$, yielding
\ba
\mu &=& \int_{A_h}^\infty \Azs^\prime(A) dA 
= \int_0^\infty \Azs_{2}^\prime(A_2) dA_2
\ea
This, and also the function $\Azs(A_2)$, is returned in the code by \texttt{AAndMuFromSols},
with $\infty$ replaced by the upper limit $A_{\rm max}$ of the range for which the equations have been solved. In addition, the code uses $\CL_A = 1$, but any other choice can be implemented by simply scaling $n$ and $\mu$.

\begin{figure}[!tb]

\centering

\includegraphics[width=0.49\textwidth]{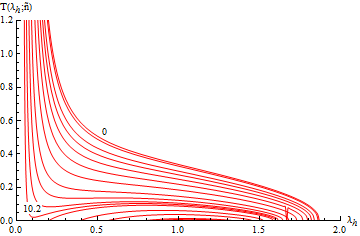}
\hfill
\includegraphics[width=0.49\textwidth]{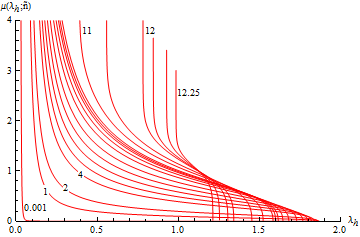}

\caption{\small $T=T(\l_h;\nt)$ and $\mu=\mu(\l_h;\nt)$
for tachyonless solutions for the
PotILogMod potential with $\bar\mu=-\fra12\,x_f=1$ and for
$\nt = 0, 2, 4, 5, 6, 7, 8, 9, 9.6, 10, 10.2, 10.4, 11,
11.5, 12, 12.1, 12.2, 12.25$. For $\mu$ the smallest values of $\nt$
are $0.001,1$.
Note that $T$ develops
a minimum around $\l_h=0.3$ for $\nt>9.5$. The $T,\mu$ derived from here is in Fig.
\protect\ref{looptest}.
 }
\label{ntfix}
\end{figure}

\begin{figure}[!tb]

\centering

\includegraphics[width=0.49\textwidth]{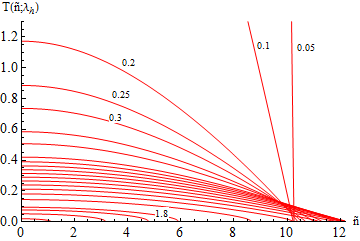}
\hfill
\includegraphics[width=0.49\textwidth]{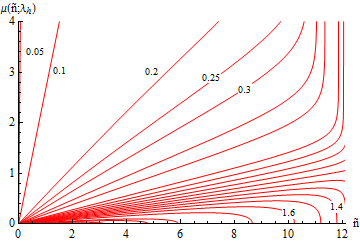}

\caption{\small $T=T(\nt;\l_h)$ and $\mu=\mu(\nt;\l_h)$
for PotILogMod potential with $\bar\mu=-\fra12\,x_f=1$ and for some values
$\l_h$.
 }
\label{lahfix}
\end{figure}

\section{Determination of $\tau_h(\l_h; m_q)$\label{B}}

The quark mass $m_q$ can be computed from the formulas presented in Appendix \ref{scalings} given the initial conditions at horizon. However, for computing physical results, we are interested rather in finding a class of solutions corresponding to a predetermined value of $m_q$, in this paper specifically $m_q = 0$. This is in principle a simple problem of numerical function inversion, but it is complicated by the fact that the inverse is multivalued and that computing values of $m_q(\l_h, \nt, \tau_h)$ takes a considerable amount of time (of the order of 1 second per point on a single core of an Intel i7 level processor).

The main task involves determining $\tau_h(\l_h, \nt; m_q)$, given a fixed pair $(\l_h, \nt)$. We will denote $m_q(\tau_h) \equiv m_q(\l_h, \nt, \tau_h)$. As discussed in previous papers \cite{Jarvinen:2011qe, Alho:2012mh}, $m_q(\tau_h)$ may have several zeroes, corresponding to different Efimov vacuums, of which the most stable is the one with the largest $\tau_h$. This means that it is not enough to find a zero, but rather we have to be able to bracket an interval containing the last zero before the asymptotic rise of $m_q(\tau_h)$ at large $\tau_h$, or alternatively deduce that no zero at finite $\tau_h$ exists. In addition, since this needs to be done in at least tens of thousands of points on the $(\l_h, \nt)$ -plane, the search must be fully automated and reliable enough to not need manual checking of the solutions.

The function \texttt{$\tau$hFromQuarkMass} in \cite{TSAcode} does this with a heuristic method that will be briefly described here. We omit some details, for which we invite the interested reader to look into the the code itself.
\begin{enumerate}
\item First we find a point where the solution exists and $m_q < 0$, starting from an initial guess, by default $\tau_h = 1$. If the solution does not exist at all at the initial guess, $\tau_h$  is multiplied by 2 to form a new guess. This is repeated until a point $\tau_{h,\rm exist}$ where the solution exists is found. After that, a point where $m_q(\tau_h) < 0$ is found by a binary search between $]0, \tau_{h, \rm exist}]$. We denote that point by $\tau_{h,\rm min}$. If this point is not found in a predetermined number of bisections (default is 80), we conclude that a chiral symmetry breaking solution does not exist for this pair $(\l_h, \nt)$.
\item We look for $\tau_{h,\rm max}$ such that $m_q(\tau_{h,\rm max}) > 0$ by progressively doubling $\tau_h$ until $\tau_{h, \rm max}$ is found.
\item We now have two points $\tau_{h,\rm min} < \tau_{h,\rm max}$ such that at least one root of
$m_q$ lies between them. Starting a numerical root finder in this bracket with
Brent's method would be guaranteed to find a root, but unfortunately there is no control over
which root. We need to start looking for zeros of $m_q$ in this interval. This is complicated by the
fact that the distances between the zeros in $\tau_h$ become exponentially larger toward increasing
$\tau_h$. The heuristic we use determines an initial step length by the formula $\Delta \tau_{h} = \tau_{h, \rm min}((\frac{\tau_{h,\rm max}}{\tau_{h, \rm min}})^{\frac{1}{N}} - 1)$ (default $N = 10^4$).
Then $m_q(\tau_{h,i})$ is computed at points $\tau_{h, \rm min} + n \Delta \tau_{h, \rm min}$, $n = 0, 1, 2$, and we form the unique parabola
passing through all of these points. The distance between its two roots is used to provide a local estimate of the distances between zeroes, which is used to determine a new step length $\Delta \tau_h$. We then compute $m_q$ at intervals of $\Delta \tau_h$ until we find a zero, that is, $m_q$ changes sign during a step.
\item Once a zero is found, we update $\Delta \tau_h$ to be the distance between $\tau_{h, \rm min}$ and the zero divided by small safety factor (default = 5). 
\item We continue to iterate with step length $\Delta \tau_h$, and whenever a new zero (a change of sign in $m_q$) is found, we update the step length $\Delta \tau_h$ to the distance between the two latest zeroes divided by the safety factor. This iteration is terminated, and we take the last zero found, $\tau_{h, \rm last}$, as the correct root, when the following conditions hold simultaneously:
\begin{enumerate}
\item $m_q$ has not decreased from the last step, since we know that asymptotically $m_q$ grows.
\item $m_q > 0$
\item $m_q > k \max(m_q(\tau_h);  \tau_h < \tau_{h,\rm last}))$, where $k$ is a heuristically determined number, typically a few hundred. This is the main condition used to ensure that the search goes on for long enough to reach the region of asymptotic growth.
\end{enumerate}
\end{enumerate}

Once the iteration described above completes, we are reasonably confident that $\tau_{h,\rm last}$ and $\tau_{h, \rm last} - \Delta \tau_h$ bracket the largest zero, and simply use a standard root finder implementing Brent's method to find the precise location of that root.

\section{Numerical results for $T$ and $\mu$}
\la{AppNumResults}

As everywhere in this paper,
numerical results in this appendix are all computed for the potentials \nr{VfSB}-\nr{akappa}
with $\bar\mu=-\fra12$ and for $x_f=1$.

\begin{figure}[!tb]
\centering
\includegraphics[width=0.49\textwidth]{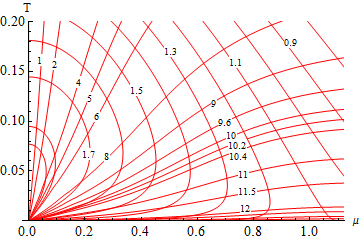}
\caption{\small
Formation of the symmetric phase grid on the $T,\mu$ plane. The curves
marked $1.7,\,1.5,\dots,0.9$ are those for constant $\l_h$, those marked
$1,\,2,\dots 11.5,\,12$ are those for constant $\tn$.
 }
\label{looptest}
\end{figure}

First, Fig. \ref{ntfix}  plots $T$ and $\mu$ as functions of $\l_h$
for fixed values of $\nt$ for the tachyonfree chirally symmetric solutions.
In Fig.~\ref{lahfix} the roles of $\l_h$ and $\nt$
are interchanged.
From these one then determines the two families of curves
\be
T=T(\mu;\nt),\quad T=T(\mu;\l_h),
\la{Tmulahnt}
\ee
which, when plotted on the $T,\mu$ plane, form a grid, see Fig. \ref{looptest}.

\begin{figure}[!tb]
\centering
\includegraphics[width=0.32\textwidth]{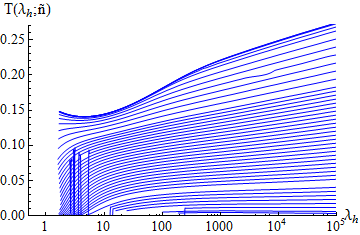}
\hfill
\includegraphics[width=0.32\textwidth]{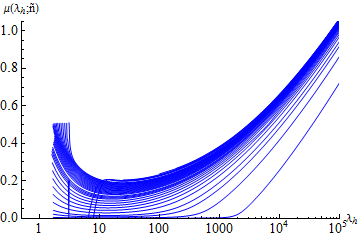}
\hfill
\includegraphics[width=0.32\textwidth]{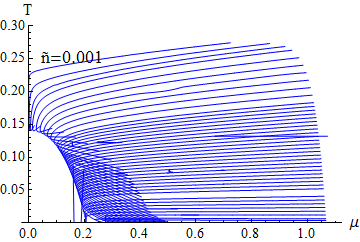}
\caption{\small Plots of $T(\l_h;\nt)$ and $\mu(\l_h;\nt)$ in the broken phase as
functions of $\l_h$ for values of $\nt$ varying from
$0$ to $23.5$ in steps of $0.5$, see the physical region in Fig. \protect\ref{phreg}.
From these one derives
curves of constant $\nt$ on the $(T,\mu)$ plane (rightmost panel) for the broken
phase.
 }
\label{Tmutauhntconst}
\end{figure}

\begin{figure}[!tb]
\centering
\includegraphics[width=0.32\textwidth]{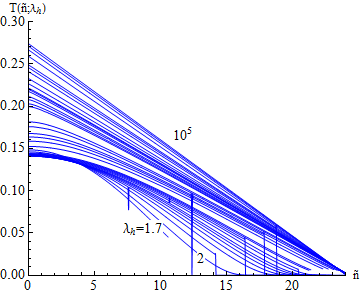}
\hfill
\includegraphics[width=0.32\textwidth]{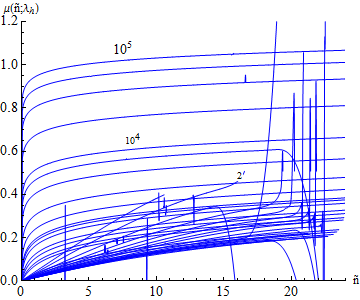}
\hfill
\includegraphics[width=0.32\textwidth]{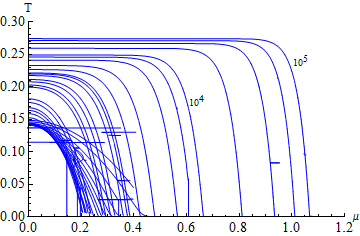}
\caption{\small
Plots of $T(\l_h;\nt)$ and $\mu(\l_h;\nt)$ in the broken phase as
functions of $\nt$ for values of $\l_h$ varying from
$1.7$ to $10^5$, see the physical region in Fig. \protect\ref{phreg}.
From these one derives
curves of constant $\l_h$ on the $(T,\mu)$ plane (rightmost panel) for the broken
phase.
 }
\label{Tmutauhlahconst}
\end{figure}

\begin{figure}[!tb]
\centering
\includegraphics[width=0.6\textwidth]{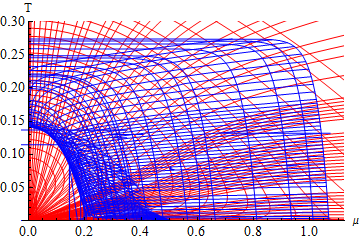}
\caption{\small The symmetric and broken phase grids.}
\label{gridfig}
\end{figure}

Further, Figs.~\ref{Tmutauhntconst} and \ref{Tmutauhlahconst} show the same for the
solutions with a nonzero tachyon. Several of the curves have numerical fluctuations.
Note that there is a region near the origin where broken phase solutions do not exist.
Putting the symmetric phase and broken phase grids together one obtains the grid in Fig.~\ref{gridfig}, on points of which the pressure $p(T,\mu)$ is numerically
computed as discussed in Section \ref{sectp}.

\begin{figure}[!b]

\centering
\includegraphics[width=0.49\textwidth]{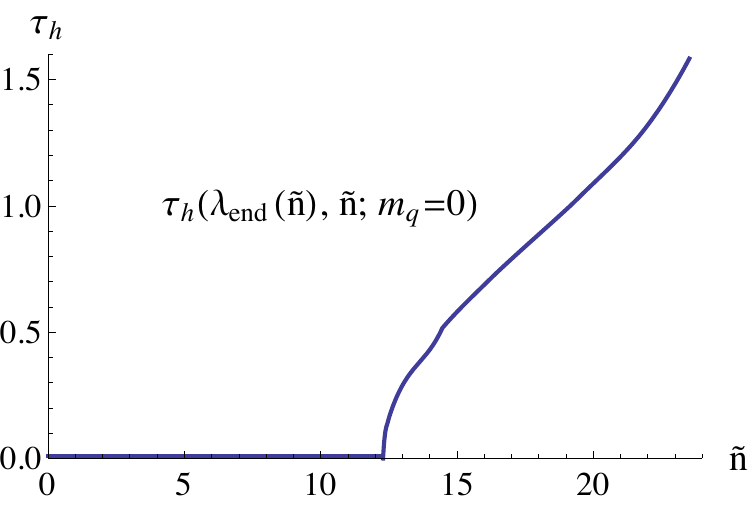}\hfill
\includegraphics[width=0.49\textwidth]{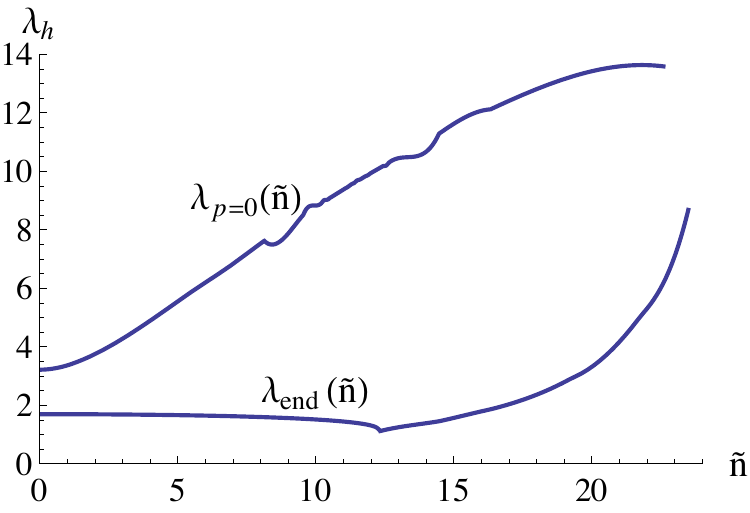}
\caption{\small Left:
Behavior of $\tau_h(\l_h,\nt, m_q=0)$ for as a function of $\nt$. 
Right: The physical region of tachyonic solutions $\l_h>\lend$ compared with the
values of $\l_h$ at which the broken phase pressure vanishes.
This is where the deconfining transition takes place, the corresponding
temperatures $T_h(\mu)$ are plotted in Fig.~\ref{Ttrans}. The chiral transition takes place along
$\lend$ (see Section~\ref{sectp}). }
\label{lendTh}
\end{figure}

\begin{figure}[!b]
\centering
\includegraphics[width=0.49\textwidth]{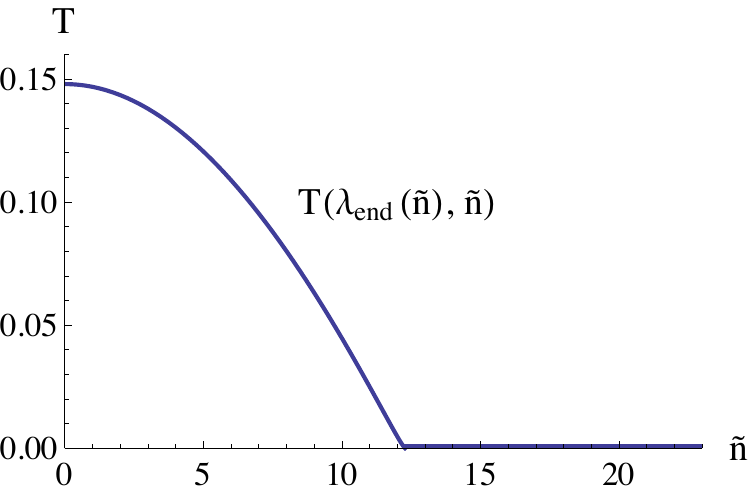}\hfill
\includegraphics[width=0.49\textwidth]{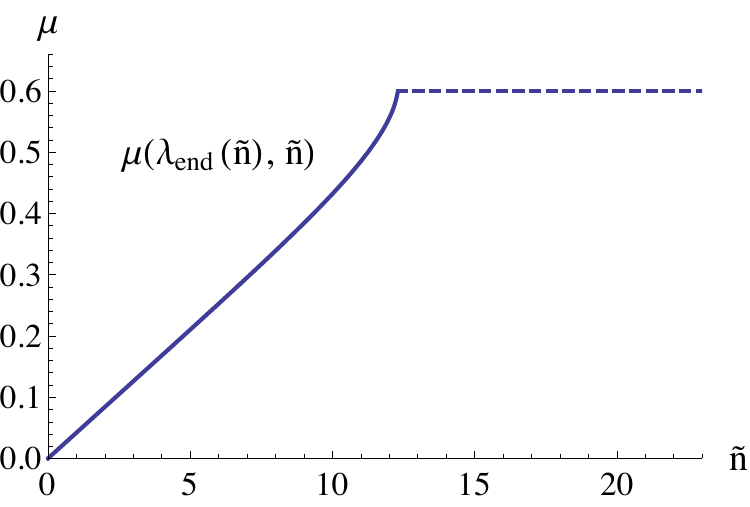}
\caption{\small Temperature and chemical potential along $\lend(\nt)$.
 }
\label{lendTandmu}
\end{figure}

\begin{figure}[!b]

\centering

\includegraphics[width=0.49\textwidth]{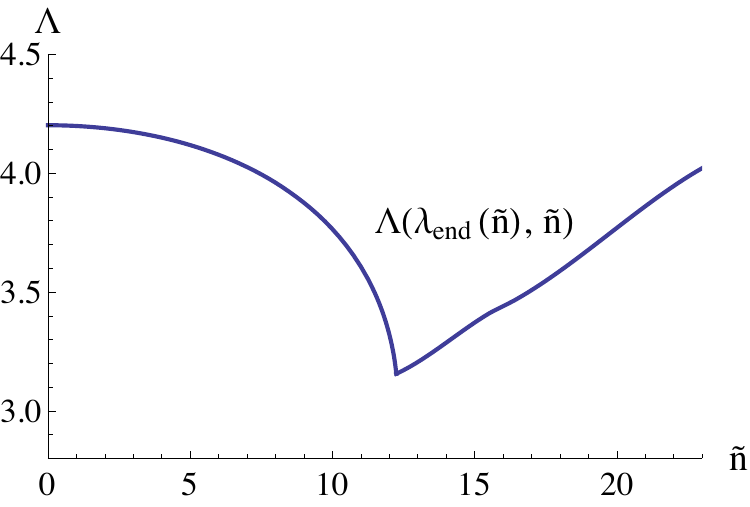}\hfill
\includegraphics[width=0.49\textwidth]{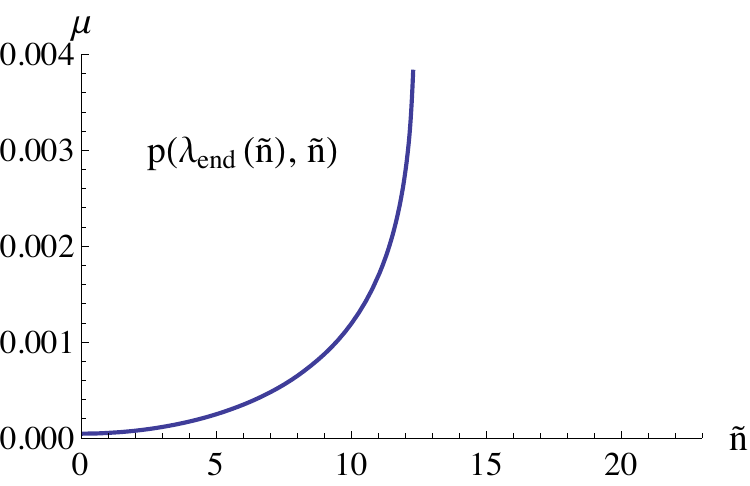}

\caption{\small The scale factor $\Lambda=1/b_h$ and the pressure along $\lend(\nt)$. Note that
$s\sim b_h^3$ varies only very little along the curve.
 }
\label{lendLa}
\end{figure}

\section{Thermodynamics along $\lend$}\la{sectlaend}
The lower limit $\lend(\nt)$ of the physical region of the tachyonic solutions on the
$\nt,\l_h$ plane plays an important role in the thermodynamics. We shall here analyse
its properties.

The lower limit $\lend(\nt)$ arises when one tries to determine what values of $\tau_h$
are possible so that after integrating towards the boundary $m_q=0$ is obtained.
One finds that $\tau_h=\tau_h(\l_h,m_q=0,\nt)$ is
a monotonically growing function of $\l_h$ (see, e.g., Fig. 5 of \cite{Alho:2012mh}) which
starts at some $\l_h=\l_\rmi{end}(\nt)$.
As long as $\nt \lesssim 8$
one has the normal situation in which $\tau_h(\lend)=0$ but if $\nt\gtrsim8$ the value $\tau_h(\lend)>0$,
(see Fig.\ref{lendTh}). An accurate plot of $\lend(\nt)$ has been presented in Fig.~\ref{phreg}, in
Fig.~\ref{lendTh} it is shown together with the location of the deconfining transition. As discussed
in Section~\ref{sectp}, the chiral transition takes place along $\lend$.

The extreme situation is that when
$\tau_h$ is so large that $V_f$ decouples due to the $e^{-a\tau^2}$
factor. The large $\nt$ limit of \nr{nhatmaxval} is then simply 
\be
\nt_\rmi{max}=V_g(\l_h)\f(\l_h).
\la{vefflargetau}
\ee
The upper limit of $\l_h$ moves to infinity and the large $\l_h$ limit
of \nr{vefflargetau} is
\be
\nt_\rmi{max}=23.99+{12.34\over\l_h^{1/3}\sqrt{\ln\l_h+1}}+\CO\left({1\over\l_h}\right).
\la{nmaxnotau}
\ee

In the $T=0$ limit it seems
that $s\sim b_h^3$ goes to a finite limit even though $T=0$. In fact, $s_b$ varies only
very little along the chiral equilibrium curve.

\section{Large scale behavior}\la{appD}
Since our model has asymptotic freedom built in it, we can at large $T$ fit the magnitude of the
pressure to the ideal gas limit
\be
p=N_c^2\biggl[\fra{\pi^2}{45}(1+\fra74 x_f)T^4+\fra16 x_f\mu^2T^2+\fra1{12\pi^2}x_f\mu^4\biggr].
\la{pideal}
\ee

Large $T$ means small $\l_h$ and
we thus want to compute the pressure $p(\l_h,\tn)$ at very small $\l_h$ and finite $\tn$
in the approximation
\be
b(z)={\CL\over z},\quad {1\over z_h}=e^{1/(b_0\l_h)}(b_0\l_h)^{b_1/b_0^2}.
\la{UVappro}
\ee
In this section, the limit $\l_h\to0$ is always implied and the argument $\l_h$
(equivalent to $z_h$) is often omitted. Thus here $V_g=12$, $\kappa=1$ and, for $x_f=1$, 
$\CL\equiv\LUV=(1+\fra74x_f)^{1/3}=1.401$ and
\be
V_f=x_fW_0=12(1-1/\CL^2)=5.886,\quad {\CL^2\over V_f}={1\over 2.999},\quad \sqrt{V_g^2-V_f^2}=10.457.
\la{numvals}
\ee

The pressure is obtained by first doing the pressure integral \nr{plah} along $\l_h$ at
$\tn=0$ and then at fixed $\l_h$ the pressure integral \nr{pnt} from $\tn=0$ to some
$\tn$. The former is simple and gives
\be
p_{\tn=0}(\l_h)={\CL^3\over 16\pi G_5}\,{1\over z_h^4}.
\ee
The latter becomes in the approximation \nr{UVappro}
\be
4G_5p_{\l_h}(\tn)={\CL^3\over z_h^3}\int_0^{\tn} \,d\tn\biggl[T'(\tn;\l_h)+{1\over4\pi}
\tn\,\mu'(\tn;\l_h)\biggr].
\la{enint1}
\ee
To evaluate this we must work out $A_0(z)$ and $\mu=\mu(z_h,\tn)$ from \nr{A0int} and $T=T(z_h,\tn)$ from
\nr{chargedtemp} in the approximation \nr{UVappro}. Here $z_h$ is equivalent to $\l_h$ due to \nr{UVappro}
and the order of arguments is irrelevant.

Noting that
\be
\int_0^{x^2}\,du{1\over \sqrt{1+y^2 \,u^3}}=x^2\,_2F_1(\fra13,\fra12,\fra43,-x^6y^2)
\ee
one finds
\ba
z_h\,\CL_A^4\mu&=&{\CL^2\over 2V_f}\,_2 F_1(\fra13,\fra12,\fra43,-\fra{\tn^2}{\CL_A^4V_f^2})\,\tn\nn
&=&
{\CL^2\over 2V_f}\biggl(\tn-{\tn^3\over 8\CL_A^4V_f^2}+\cdots \biggr)\la{mures1}
\ea
and
\ba
z_h\,\pi T&=&{\veff(\tn)\over\veff(0)}=1-{\CL^2V_f\over 12}
\biggl(\sqrt{1+\fra{\tn^2}{\CL_A^4V_f^2}}-1\biggr)\nn
&=&
1-{\CL^2\over24\CL_A^4V_f}\tn^2+{\CL^2\over96\CL_A^8V_f^3}\tn^4+\cdots \la{Tres1}
\ea
This form of $T$ shows explicitly that $T$ vanishes at $\tn=\CL_A^2\sqrt{V_g^2-V_f^2}= 10.457\, \CL_A^2$,
i.e., at the physical region boundary in Fig.~\ref{phreg} (where $\CL_A=1$).
By taking the ratio one sees that $\mu/(\pi T)$ is essentially determined by $\tn$ so that
it grows monotonically from $0$ to $\infty$ at the physical region boundary.

Inserting these exact forms to \nr{enint1} and integrating one finds that
\be
4G_5p_{\l_h}(\tn)={\CL^3\over 4\pi z_h^3}\biggl(-{1\over z_h}+\pi T +\fra14 \tn\,\mu\biggr)
\ee
so that the final total pressure in the limit of $\l_h\to0$, $\tn$ finite becomes
\be
p=p_{\tn=0}(\l_h)+p_{\l_h}(\tn)={\CL^3\over16G_5z_h^3}\biggl(T+{\tn\over4\pi}\mu\biggr)
=\fra14 s\biggl(T+{\tn\over4\pi}\mu\biggr)=\fra14(Ts+\mu n),
\la{pUV}
\ee
where we also used \nr{physn}. This further implies that $\e=Ts-p+\mu n=3p$ in this UV corner
of parameter space.
Note the mixed notation, $p=p(z_h,\tn)$ is given directly by the above equations, but
if we want $p(T,\mu)$ we must solve
$z_h=z_h(T,\mu)$ and $\tn=\tn(T,\mu)$ from the exact expressions \nr{mures1} and \nr{Tres1}.

To compare with \nr{pideal}, consider first the limit $T\to\infty$, $\mu$ = constant.
Taking the ratio of \nr{mures1} and \nr{Tres1}, expanding in $\tn$ and inverting the series
one obtains
\be
\tn={2V_f\over\CL^2}\,{\CL_A^4\mu\over\pi T}\biggl[1+{1\over2\CL^2}(1-\fra13\CL^2V_f)
{\CL_A^4\mu^2\over\pi^2T^2}+\cdots \biggr].
\la{muoT}
\ee
and
\be
{1\over z_h^3}=\pi^3 T^3\biggl[1+{V_f\over2\CL^2}\,{\CL_A^4\mu^2\over \pi^2T^2}+
\CO({\mu^6\over T^6})\biggr]
\ee
Inserting this to \nr{pUV} gives
\be
p={\CL^3\over16\pi G_5}\biggl[(\pi T)^4+{V_f\CL_A^4\over\CL^2}\mu^2(\pi T)^2+
\fr16\biggl({V_f\CL_A^4\over\CL^2}\biggr)^2\biggl(1+{3\over 2\CL^2V_f}\biggr)\mu^4+
\CO({\mu^6\over T^2})\biggr].
\la{pexp}
\ee

The two parameters $G_5$ and $\CL_A$ can be fixed by the magnitudes of the $T^4$
and $\mu^2T^2$ terms.
Comparing the $T^4$ terms of \nr{pideal} and \nr{pexp} gives first \cite{Alho:2012mh}
\be
{\CL^3\over 16\pi G_5}=N_c^2\,{1+\fra74 x_f\over 45\pi^2}.
\la{4G5}
\ee
Using this the $\mu^2T^2$ terms agree if, inserting \nr{numvals} and $x_f=1$,
\be
\CL_A^4={\CL^2\over V_f}\,{15x_f\over 2+\fra72 x_f}=
{5x_f(1+\fra74 x_f)^{1/3}\over8((1+\fra74 x_f)^{2/3}-1)}\approx0.9094\qquad (x_f=1).
\la{CLAfit}
\ee
The parameter $\gamma$ then is
\be
{1\over\gamma^2}={V_f\CL_A^4\over\CL^2 }={15x_f\over 2+\fra72 x_f}={30\over11} 
\qquad (x_f=1).
\la{gamsq}
\ee
However, one can also determine $\CL_A$ requiring agreement with the $\mu^4$ term.
The answer is
\be
\CL_A^4=\sqrt{{5x_f(1+\fra74 x_f)^{5/3}\over 32((1+\fra74 x_f)^{2/3}-1)((1+\fra74 x_f)^{2/3}-\fra78)}}
\approx0.8973\qquad (x_f=1).
\la{CLA2}
\ee
The values are automatically remarkably close also for other values of $x_f$, for $x_f=4$
\nr{CLAfit} gives 1.667 and \nr{CLA2} 1.461.
Thus both terms are reproduced almost correctly and without further parameters,
We thus have fitted that $\CL_A^2$ in \nr{defellA} is very close to one. 

\begin{figure}[!b]

\centering

\includegraphics[width=0.6\textwidth]{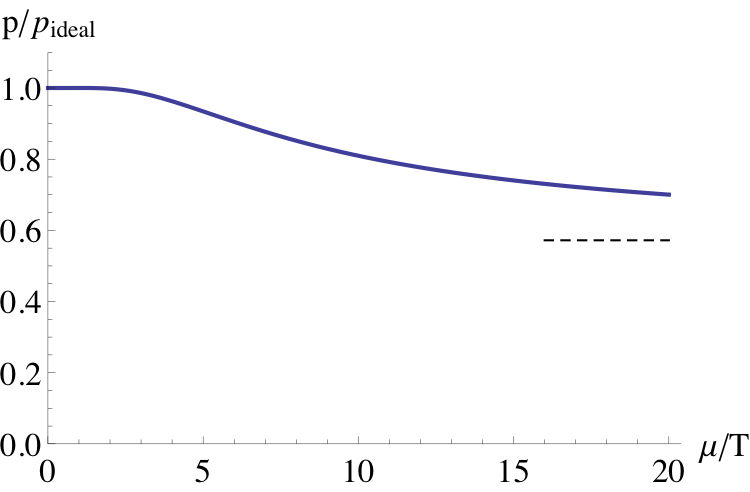}

\caption{\small Computed values of the ratio of the symmetric phase pressure integral \nr{pees}
and the ideal gas approximation \nr{pideal} at $T\to\infty$ and $\mu/T$ = fixed. The dashed
line shows the asymptotic $T\ll\mu$ limit $0.572$.
 }
\label{ppSB}
\end{figure}

For completeness, the $\mu^6/T^2$ term in \nr{pexp} is
\be
-{1\over\gamma^6}\,{1\over216}\,\biggl(29-{54\over\CL^2V_f}+{81\over \CL^4V_f^2}\biggr)\,
{\mu^6\over\pi^2T^2}=-2.34\,{\mu^6\over\pi^2T^2}.
\ee
Thus $p$ starts falling below $p_\rmi{idea}$, the non-expanded result is in Fig.~\ref{ppSB}.

The comparison of \nr{pideal} and \nr{pexp} can also be carried out in the limit
$T\to\infty$, $\mu/T$ = constant. Fixing the normalisation at $\mu=0$ we simply have
\be
{p\over p_\rmi{ideal}}={1+\fra1{4\pi}\tn(\fra{\mu}T)\,\fra\mu T\over
(1+\fra{30}{11\pi^2}\fra{\mu^2}{T^2}+\fra{15}{11\pi^4}\fra{\mu^4}{T^4})
(\veff(\tn(\fra{\mu}T))/\veff(0))^3},
\ee
where $\tn(\fra{\mu}T)$ is to be determined by inverting the ratio of \nr{mures1} and
\nr{Tres1} numerically; the small-$\mu/T$ terms were given in \nr{muoT}. The result is
plotted in Fig.~\ref{ppSB}. The value of $\CL_A$ was fixed so that the ideal gas $\mu^2$ term
was correctly reproduced. Now one sees that the good agreement extends to large values of $\mu$,
at $\mu=4T$ the deviation is 3\%.
One can work out analytically the asymptotic limit at $\mu\gg T$ which corresponds to
$\tn\to\tn_\rmi{max}=\CL_A^2\sqrt{V_g^2-V_f^2}$. It depends on $V_g=12$ and
$V_f$ and its numerical value is $0.572$. There is no obvious constraint leading to the
value $1$, but it is nevertheless rather close to this value.

It may be useful to compare holographic and perturbative QCD predictions for other quantities, too.
At $T=0$ \cite{Arean:2012mq,Arean:2013tja} finds that the holographic and perturbative QCD results for
the correlator of vector flavor currents agree in the UV if
\be
 {\CL_A^4\CL\,W_0\over 16\pi G_5}=N_c^2\,{1\over 6\pi^2},
\la{4G5arean}
\ee
see Eq. (C.10) in~\cite{Arean:2012mq} with $w^2=\CL_A^4\kappa^2=\CL_A^4$ and $W_0= V_f/x_f$. 
This matches exactly with the combination of 
\nr{4G5} and \nr{CLAfit}. Note that both the pressure at large $T,\mu$ and the vector correlator
at large momentum depend only on the combination $\CL_A^2\kappa(\l_h=0)$, here we have
assumed $\kappa(0)=1$. 
These quantities can be fixed separately using the scalar correlator. Combining the result
in Eq. (C.21) of \cite{Arean:2012mq} and \nr{CLAfit} one finds that
\be
\kappa(0)={2\CL^4\over3\CL_A^4}={16(1+\fra74 x_f)\over 15x_f}\biggl[(1+\fra74 x_f)^{2/3}-1)\biggr]
=2.82,\quad (x_f=1).
\ee
This modified value of $\kappa$ would affect the normalisation of $\tau$ and consequently that of 
the chiral condensate, but not the results in this article. As another example of
comparisons of weak coupling and holographic computations  
one may also compare this result with an analogous analysis of the finite temperature  
correlators of the energy momentum tensor in the UV~\cite{Kajantie:2013gab}. 
Using the thermal normalisation \nr{4G5},
the holographic result of the shear correlator is too small by a factor $4/9$
with respect to the perturbation theory one, for the bulk correlator the results agree. 
This result illustrates the fact that this action cannot describe all the phenomena 
in the weak coupling region at the same time, an action with higher derivatives is needed.

\bibliography{kempotFinal}

\providecommand{\href}[2]{#2}\begingroup\raggedright\begin{thebibliography}{10}

\bibitem{Kogut:2004su}
J.~Kogut and M.~Stephanov, {\it {The phases of quantum chromodynamics: From
  confinement to extreme environments}},  {\em
  Camb.Monogr.Part.Phys.Nucl.Phys.Cosmol.} {\bf 21} (2004) 1--364.

\bibitem{Scavenius:2000qd}
O.~Scavenius, A.~Mocsy, I.~Mishustin, and D.~Rischke, {\it {Chiral phase
  transition within effective models with constituent quarks}},  {\em
  Phys.Rev.} {\bf C64} (2001) 045202,
  [\href{http://xxx.lanl.gov/abs/nucl-th/0007030}{{\tt nucl-th/0007030}}].

\bibitem{Kahara:2008yg}
T.~Kahara and K.~Tuominen, {\it {Degrees of freedom and the phase transitions
  of two flavor QCD}},  {\em Phys.Rev.} {\bf D78} (2008) 034015,
  [\href{http://xxx.lanl.gov/abs/0803.2598}{{\tt arXiv:0803.2598}}].

\bibitem{Stephanov:2004wx}
M.~A. Stephanov, {\it {QCD phase diagram and the critical point}},  {\em
  Prog.Theor.Phys.Suppl.} {\bf 153} (2004) 139--156,
  [\href{http://xxx.lanl.gov/abs/hep-ph/0402115}{{\tt hep-ph/0402115}}].

\bibitem{Myers:2009ij}
R.~C. Myers, M.~F. Paulos, and A.~Sinha, {\it {Holographic Hydrodynamics with a
  Chemical Potential}},  {\em JHEP} {\bf 0906} (2009) 006,
  [\href{http://xxx.lanl.gov/abs/0903.2834}{{\tt arXiv:0903.2834}}].

\bibitem{DeWolfe:2010he}
O.~DeWolfe, S.~S. Gubser, and C.~Rosen, {\it {A holographic critical point}},
  {\em Phys.Rev.} {\bf D83} (2011) 086005,
  [\href{http://xxx.lanl.gov/abs/1012.1864}{{\tt arXiv:1012.1864}}].

\bibitem{DeWolfe:2011ts}
O.~DeWolfe, S.~S. Gubser, and C.~Rosen, {\it {Dynamic critical phenomena at a
  holographic critical point}},  {\em Phys.Rev.} {\bf D84} (2011) 126014,
  [\href{http://xxx.lanl.gov/abs/1108.2029}{{\tt arXiv:1108.2029}}].

\bibitem{Kaczmarek:2011zz}
O.~Kaczmarek, F.~Karsch, E.~Laermann, C.~Miao, S.~Mukherjee, et~al., {\it
  {Phase boundary for the chiral transition in (2+1) -flavor QCD at small
  values of the chemical potential}},  {\em Phys.Rev.} {\bf D83} (2011) 014504,
  [\href{http://xxx.lanl.gov/abs/1011.3130}{{\tt arXiv:1011.3130}}].

\bibitem{Karsch:2010hm}
F.~Karsch, B.-J. Schaefer, M.~Wagner, and J.~Wambach, {\it {Towards finite
  density QCD with Taylor expansions}},  {\em Phys.Lett.} {\bf B698} (2011)
  256--264, [\href{http://xxx.lanl.gov/abs/1009.5211}{{\tt arXiv:1009.5211}}].

\bibitem{Endrodi:2011gv}
G.~Endrodi, Z.~Fodor, S.~Katz, and K.~Szabo, {\it {The QCD phase diagram at
  nonzero quark density}},  {\em JHEP} {\bf 1104} (2011) 001,
  [\href{http://xxx.lanl.gov/abs/1102.1356}{{\tt arXiv:1102.1356}}].

\bibitem{deForcrand:2008zi}
P.~de~Forcrand and O.~Philipsen, {\it {The curvature of the critical surface
  (m(u,d),m(s))**crit(mu): A Progress report}},  {\em PoS} {\bf LATTICE2008}
  (2008) 208, [\href{http://xxx.lanl.gov/abs/0811.3858}{{\tt
  arXiv:0811.3858}}].

\bibitem{Maldacena:1997re}
J.~M. Maldacena, {\it {The Large N limit of superconformal field theories and
  supergravity}},  {\em Adv.Theor.Math.Phys.} {\bf 2} (1998) 231--252,
  [\href{http://xxx.lanl.gov/abs/hep-th/9711200}{{\tt hep-th/9711200}}].

\bibitem{Witten:1998qj}
E.~Witten, {\it {Anti-de Sitter space and holography}},  {\em
  Adv.Theor.Math.Phys.} {\bf 2} (1998) 253--291,
  [\href{http://xxx.lanl.gov/abs/hep-th/9802150}{{\tt hep-th/9802150}}].

\bibitem{Gubser:1998bc}
S.~Gubser, I.~R. Klebanov, and A.~M. Polyakov, {\it {Gauge theory correlators
  from noncritical string theory}},  {\em Phys.Lett.} {\bf B428} (1998)
  105--114, [\href{http://xxx.lanl.gov/abs/hep-th/9802109}{{\tt
  hep-th/9802109}}].

\bibitem{Hartnoll:2009sz}
S.~A. Hartnoll, {\it {Lectures on holographic methods for condensed matter
  physics}},  {\em Class.Quant.Grav.} {\bf 26} (2009) 224002,
  [\href{http://xxx.lanl.gov/abs/0903.3246}{{\tt arXiv:0903.3246}}].

\bibitem{Herzog:2009xv}
C.~P. Herzog, {\it {Lectures on Holographic Superfluidity and
  Superconductivity}},  {\em J.Phys.} {\bf A42} (2009) 343001,
  [\href{http://xxx.lanl.gov/abs/0904.1975}{{\tt arXiv:0904.1975}}].

\bibitem{Gursoy:2007cb}
U.~Gursoy and E.~Kiritsis, {\it {Exploring improved holographic theories for
  QCD: Part I}},  {\em JHEP} {\bf 0802} (2008) 032,
  [\href{http://xxx.lanl.gov/abs/0707.1324}{{\tt arXiv:0707.1324}}].

\bibitem{Gursoy:2007er}
U.~Gursoy, E.~Kiritsis, and F.~Nitti, {\it {Exploring improved holographic
  theories for QCD: Part II}},  {\em JHEP} {\bf 0802} (2008) 019,
  [\href{http://xxx.lanl.gov/abs/0707.1349}{{\tt arXiv:0707.1349}}].

\bibitem{Gursoy:2010fj}
U.~Gursoy, E.~Kiritsis, L.~Mazzanti, G.~Michalogiorgakis, and F.~Nitti, {\it
  {Improved Holographic QCD}},  {\em Lect.Notes Phys.} {\bf 828} (2011)
  79--146, [\href{http://xxx.lanl.gov/abs/1006.5461}{{\tt arXiv:1006.5461}}].

\bibitem{Gursoy:2008za}
U.~Gursoy, E.~Kiritsis, L.~Mazzanti, and F.~Nitti, {\it {Holography and
  Thermodynamics of 5D Dilaton-gravity}},  {\em JHEP} {\bf 0905} (2009) 033,
  [\href{http://xxx.lanl.gov/abs/0812.0792}{{\tt arXiv:0812.0792}}].

\bibitem{Gursoy:2008bu}
U.~Gursoy, E.~Kiritsis, L.~Mazzanti, and F.~Nitti, {\it {Deconfinement and
  Gluon Plasma Dynamics in Improved Holographic QCD}},  {\em Phys.Rev.Lett.}
  {\bf 101} (2008) 181601, [\href{http://xxx.lanl.gov/abs/0804.0899}{{\tt
  arXiv:0804.0899}}].

\bibitem{Gursoy:2009jd}
U.~Gursoy, E.~Kiritsis, L.~Mazzanti, and F.~Nitti, {\it {Improved Holographic
  Yang-Mills at Finite Temperature: Comparison with Data}},  {\em Nucl.Phys.}
  {\bf B820} (2009) 148--177, [\href{http://xxx.lanl.gov/abs/0903.2859}{{\tt
  arXiv:0903.2859}}].

\bibitem{Jarvinen:2009fe}
M.~J{\"a}rvinen and F.~Sannino, {\it {Holographic Conformal Window - A Bottom
  Up Approach}},  {\em JHEP} {\bf 1005} (2010) 041,
  [\href{http://xxx.lanl.gov/abs/0911.2462}{{\tt arXiv:0911.2462}}].

\bibitem{Alanen:2009na}
J.~Alanen and K.~Kajantie, {\it {Thermodynamics of a field theory with infrared
  fixed point from gauge/gravity duality}},  {\em Phys.Rev.} {\bf D81} (2010)
  046003, [\href{http://xxx.lanl.gov/abs/0912.4128}{{\tt arXiv:0912.4128}}].

\bibitem{Alanen:2010tg}
J.~Alanen, K.~Kajantie, and K.~Tuominen, {\it {Thermodynamics of Quasi
  Conformal Theories From Gauge/Gravity Duality}},  {\em Phys.Rev.} {\bf D82}
  (2010) 055024, [\href{http://xxx.lanl.gov/abs/1003.5499}{{\tt
  arXiv:1003.5499}}].

\bibitem{Alanen:2011hh}
J.~Alanen, T.~Alho, K.~Kajantie, and K.~Tuominen, {\it {Mass spectrum and
  thermodynamics of quasi-conformal gauge theories from gauge/gravity
  duality}},  {\em Phys.Rev.} {\bf D84} (2011) 086007,
  [\href{http://xxx.lanl.gov/abs/1107.3362}{{\tt arXiv:1107.3362}}].

\bibitem{Bigazzi:2005md}
F.~Bigazzi, R.~Casero, A.~Cotrone, E.~Kiritsis, and A.~Paredes, {\it
  {Non-critical holography and four-dimensional CFT's with fundamentals}},
  {\em JHEP} {\bf 0510} (2005) 012,
  [\href{http://xxx.lanl.gov/abs/hep-th/0505140}{{\tt hep-th/0505140}}].

\bibitem{Casero:2007ae}
R.~Casero, E.~Kiritsis, and A.~Paredes, {\it {Chiral symmetry breaking as open
  string tachyon condensation}},  {\em Nucl.Phys.} {\bf B787} (2007) 98--134,
  [\href{http://xxx.lanl.gov/abs/hep-th/0702155}{{\tt hep-th/0702155}}].

\bibitem{Iatrakis:2010zf}
I.~Iatrakis, E.~Kiritsis, and A.~Paredes, {\it {An AdS/QCD model from Sen's
  tachyon action}},  {\em Phys.Rev.} {\bf D81} (2010) 115004,
  [\href{http://xxx.lanl.gov/abs/1003.2377}{{\tt arXiv:1003.2377}}].

\bibitem{Iatrakis:2010jb}
I.~Iatrakis, E.~Kiritsis, and A.~Paredes, {\it {An AdS/QCD model from tachyon
  condensation: II}},  {\em JHEP} {\bf 1011} (2010) 123,
  [\href{http://xxx.lanl.gov/abs/1010.1364}{{\tt arXiv:1010.1364}}].

\bibitem{Iatrakis:2011ht}
I.~Iatrakis and E.~Kiritsis, {\it {Vector-axial vector correlators in weak
  electric field and the holographic dynamics of the chiral condensate}},  {\em
  JHEP} {\bf 1202} (2012) 064, [\href{http://xxx.lanl.gov/abs/1109.1282}{{\tt
  arXiv:1109.1282}}].

\bibitem{Arean:2013tja}
D.~Arean, I.~Iatrakis, M.~J{\"a}rvinen, and E.~Kiritsis, {\it {The
  discontinuities of conformal transitions and mass spectra of V-QCD}},  {\em
  JHEP} {\bf 1311} (2013) 068, [\href{http://xxx.lanl.gov/abs/1309.2286}{{\tt
  arXiv:1309.2286}}].

\bibitem{Jarvinen:2011qe}
M.~J{\"a}rvinen and E.~Kiritsis, {\it {Holographic Models for QCD in the
  Veneziano Limit}},  {\em JHEP} {\bf 1203} (2012) 002,
  [\href{http://xxx.lanl.gov/abs/1112.1261}{{\tt arXiv:1112.1261}}].

\bibitem{Alho:2012mh}
T.~Alho, M.~J{\"a}rvinen, K.~Kajantie, E.~Kiritsis, and K.~Tuominen, {\it {On
  finite-temperature holographic QCD in the Veneziano limit}},  {\em JHEP} {\bf
  1301} (2013) 093, [\href{http://xxx.lanl.gov/abs/1210.4516}{{\tt
  arXiv:1210.4516}}].

\bibitem{Arean:2012mq}
D.~Arean, I.~Iatrakis, M.~J{\"a}rvinen, and E.~Kiritsis, {\it {V-QCD: Spectra,
  the dilaton and the S-parameter}},  {\em Phys.Lett.} {\bf B720} (2013)
  219--223, [\href{http://xxx.lanl.gov/abs/1211.6125}{{\tt arXiv:1211.6125}}].

\bibitem{Stoffers:2010sp}
A.~Stoffers and I.~Zahed, {\it {Improved AdS/QCD Model with Matter}},  {\em
  Phys.Rev.} {\bf D83} (2011) 055016,
  [\href{http://xxx.lanl.gov/abs/1009.4428}{{\tt arXiv:1009.4428}}].

\bibitem{Gouteraux:2012yr}
B.~Gouteraux and E.~Kiritsis, {\it {Quantum critical lines in holographic
  phases with (un)broken symmetry}},  {\em JHEP} {\bf 1304} (2013) 053,
  [\href{http://xxx.lanl.gov/abs/1212.2625}{{\tt arXiv:1212.2625}}].

\bibitem{Liu:2009dm}
H.~Liu, J.~McGreevy, and D.~Vegh, {\it {Non-Fermi liquids from holography}},
  {\em Phys.Rev.} {\bf D83} (2011) 065029,
  [\href{http://xxx.lanl.gov/abs/0903.2477}{{\tt arXiv:0903.2477}}].

\bibitem{TSAcode}
T.~Alho, {\it {Numerical code for thermodynamics of holographic V-QCD}},
  {\url{https://github.com/timoalho/VQCDThermo}}.

\bibitem{Gouteraux:2011ce}
B.~Gouteraux and E.~Kiritsis, {\it {Generalized Holographic Quantum Criticality
  at Finite Density}},  {\em JHEP} {\bf 1112} (2011) 036,
  [\href{http://xxx.lanl.gov/abs/1107.2116}{{\tt arXiv:1107.2116}}].

\bibitem{Qin:2013ufa}
S.-x. Qin and D.~H. Rischke, {\it {Quark Spectral Function and Deconfinement at
  Nonzero Temperature}},  {\em Phys.Rev.} {\bf D88} (2013) 056007,
  [\href{http://xxx.lanl.gov/abs/1304.6547}{{\tt arXiv:1304.6547}}].

\bibitem{Mocsy:2003qw}
A.~Mocsy, F.~Sannino, and K.~Tuominen, {\it {Confinement versus chiral
  symmetry}},  {\em Phys.Rev.Lett.} {\bf 92} (2004) 182302,
  [\href{http://xxx.lanl.gov/abs/hep-ph/0308135}{{\tt hep-ph/0308135}}].

\bibitem{Kahara:2012yr}
T.~Kahara, M.~Ruggieri, and K.~Tuominen, {\it {Deconfinement vs. chiral
  symmetry and higher representation matter}},  {\em Phys.Rev.} {\bf D85}
  (2012) 094020, [\href{http://xxx.lanl.gov/abs/1202.1769}{{\tt
  arXiv:1202.1769}}].

\bibitem{Borsanyi:2013bia}
S.~Borsanyi, Z.~Fodor, C.~Hoelbling, S.~D. Katz, S.~Krieg, et~al., {\it {Full
  result for the QCD equation of state with 2+1 flavors}},
  \href{http://xxx.lanl.gov/abs/1309.5258}{{\tt arXiv:1309.5258}}.

\bibitem{Pisarski:1983ms}
R.~D. Pisarski and F.~Wilczek, {\it {Remarks on the Chiral Phase Transition in
  Chromodynamics}},  {\em Phys.Rev.} {\bf D29} (1984) 338--341.

\bibitem{deForcrand:2006pv}
P.~de~Forcrand and O.~Philipsen, {\it {The Chiral critical line of N(f) = 2+1
  QCD at zero and non-zero baryon density}},  {\em JHEP} {\bf 0701} (2007) 077,
  [\href{http://xxx.lanl.gov/abs/hep-lat/0607017}{{\tt hep-lat/0607017}}].

\bibitem{Paterson:1980fc}
A.~Paterson, {\it {{Coleman-Weinberg} Symmetry Breaking in the Chiral SU($N$) X
  SU($N$) Linear Sigma Model}},  {\em Nucl.Phys.} {\bf B190} (1981) 188.

\bibitem{Arnold:1993bq}
P.~B. Arnold and L.~G. Yaffe, {\it {The epsilon expansion and the electroweak
  phase transition}},  {\em Phys.Rev.} {\bf D49} (1994) 3003--3032,
  [\href{http://xxx.lanl.gov/abs/hep-ph/9312221}{{\tt hep-ph/9312221}}].

\bibitem{Kajantie:1996mn}
K.~Kajantie, M.~Laine, K.~Rummukainen, and M.~E. Shaposhnikov, {\it {Is there a
  hot electroweak phase transition at m(H) larger or equal to m(W)?}},  {\em
  Phys.Rev.Lett.} {\bf 77} (1996) 2887--2890,
  [\href{http://xxx.lanl.gov/abs/hep-ph/9605288}{{\tt hep-ph/9605288}}].

\bibitem{Noronha:2009ud}
J.~Noronha, {\it {Connecting Polyakov Loops to the Thermodynamics of SU(N(c))
  Gauge Theories Using the Gauge-String Duality}},  {\em Phys.Rev.} {\bf D81}
  (2010) 045011, [\href{http://xxx.lanl.gov/abs/0910.1261}{{\tt
  arXiv:0910.1261}}].

\bibitem{Alanen:2009ej}
J.~Alanen, K.~Kajantie, and V.~Suur-Uski, {\it {Spatial string tension of
  finite temperature QCD matter in gauge/gravity duality}},  {\em Phys.Rev.}
  {\bf D80} (2009) 075017, [\href{http://xxx.lanl.gov/abs/0905.2032}{{\tt
  arXiv:0905.2032}}].

\bibitem{Spradlin:1999bn}
M.~Spradlin and A.~Strominger, {\it {Vacuum states for AdS(2) black holes}},
  {\em JHEP} {\bf 9911} (1999) 021,
  [\href{http://xxx.lanl.gov/abs/hep-th/9904143}{{\tt hep-th/9904143}}].

\bibitem{Kajantie:2013gab}
K.~Kajantie, M.~Krssak, and A.~Vuorinen, {\it {Energy momentum tensor
  correlators in hot Yang-Mills theory: holography confronts lattice and
  perturbation theory}},  {\em JHEP} {\bf 1305} (2013) 140,
  [\href{http://xxx.lanl.gov/abs/1302.1432}{{\tt arXiv:1302.1432}}].

\end{thebibliography}\endgroup
\end{document}